\documentclass[aps,preprint,nofootinbib,superscriptaddress]{revtex4}
\usepackage{amsmath}
\usepackage{amssymb}
\usepackage{amsfonts}
\usepackage{relsize}
\usepackage{enumerate}
\usepackage{booktabs}
\usepackage{xcolor}
\usepackage{float}
\usepackage{placeins}
\usepackage{lscape}
\usepackage{graphicx}
\usepackage{epstopdf}
\usepackage[caption=false]{subfig}
\usepackage{hyperref}
\usepackage{xspace}
\usepackage{bm}
\graphicspath{ {./image/} }

\begin{document}
\title{Entanglement Entropy Inequalities in BCFT by Holography}
\author{Chia-Jui Chou}
\email{agoodmanjerry.ep02g@nctu.edu.tw}
\affiliation{Department of Electrophysics, National Chiao Tung University, Hsinchu, R.O.C.}
\author{Bo-Han Lin}
\email{ejoelin@gmail.com}
\affiliation{Department of Physics, National Tsing Hua University, Hsinchu, R.O.C.}
\author{Bin Wang}
\email{1962373127@qq.com}
\affiliation{Department of Electrophysics, National Chiao Tung University, Hsinchu, R.O.C.}
\author{Yi Yang}
\email{yiyang@mail.nctu.edu.tw}
\affiliation{Department of Electrophysics, National Chiao Tung University, Hsinchu, R.O.C.}
\begin{abstract}
    We study entanglement entropy inequalities in boundary conformal field theory (BCFT) by holographic correspondence. By carefully classifying all the configurations for different phases, we prove the strong subadditiviy and the monogamy of mutual information for holographic entanglement entropy in BCFT at both zero and finite temperatures.
\end{abstract}
\maketitle
\tableofcontents
\section{Introduction}

The concept of entanglement reveals one of the major differences between classical physics and quantum physics that has been pursued by physicists for a long time. We could understand the quantum entanglement from the Von Neumann entropy which is subject to satisfy certain inequalities such as the subadditivity \cite{Araki&Lieb} and the strong subadditivity \cite{Lieb&Ruskai}.

In quantum field theory, it is difficult to compute the entanglement entropy by the usual method. Remarkably, Ryu and Takayanagi proposed a geometric prescription by holographic correspondence \cite{0603001,0605073}. The holographic entanglement entropy (HEE) can be calculated by the area of the minimal entangling surface, also known as the RT surface. The prescription gives a very simple geometric picture to compute the HEE and has been widely studied for the various holographic setups \cite{0606184,0705.0016,1006.0047,1102.0440,1304.4926,1609.01287}.

Using the geometric realization of the entanglement entropy, several HEE inequalities have been proven, including the strong subadditivity \cite{0704.3719,1211.3494} and the monogamy of mutual information \cite{1107.2940,1211.3494}. Moreover, the authors in \cite{1505.07839} have attempted to construct a systematic way to find out all the entanglement inequalities from holography by using the concept of holographic entropy cone, which is lately generalized by proving the $I_n$-theorem from the primitive information quantities for multipartite systems \cite{1612.02437,1808.07871}.

On the other hand, boundary conformal field theory (BCFT) is a conformal field theory defined on a manifold with boundaries where suitable boundary conditions imposed \cite{9505127}. BCFT provides important applications in many physical systems with boundaries, such as  D-branes in string theory and boundary critical behavior in condensed matter physics, including Hall effect, chiral magnetic effect, topological insulator etc. Early studies of holographic dual of defect or interface CFT can be found in \cite{0011156,0111135}. The holographic dual of BCFT by including extra boundaries in the gravity dual was proposed in \cite{1105.5165,1205.1573}. Many interesting developments of holographic BCFT can be found in \cite{1108.5152,1205.1573,1305.2334,1309.4523,1509.02160,1604.07571,1601.06418,1701.04275,1701.07202,1702.00566,1703.04186,1708.05080}.

In this work, we study the strong subadditivity (SSA) and the monogamy of mutual information (MMI) for the tripartite systems in BCFT by holography. In the presence of boundaries, the RT surface could be in different shapes \cite{1701.04275,1701.07202}. We classify all possible configurations of the RT surfaces by using the phase diagrams obtained in \cite{1805.06117}. In each configuration, we verify the SSA and MMI for the tripartite systems at both zero and finite temperatures.

The paper is organized as follows. In the next section, we briefly review the phase diagrams of the HEE in a holographic BCFT. In section III, we apply the phase diagram of HEE to the bipartite and tripartite systems, and prove the SSA and the MMI in the pure AdS background. We then extend our proof to the AdS black hole background in section IV, and summarize our results in Section V.

\section{Holographic Entanglement Entropy in BCFT}

\subsection{Holographic BCFT}

\begin{figure}[h]
  \subfloat[Pure AdS spacetime]{
  \includegraphics[width=.45\linewidth]{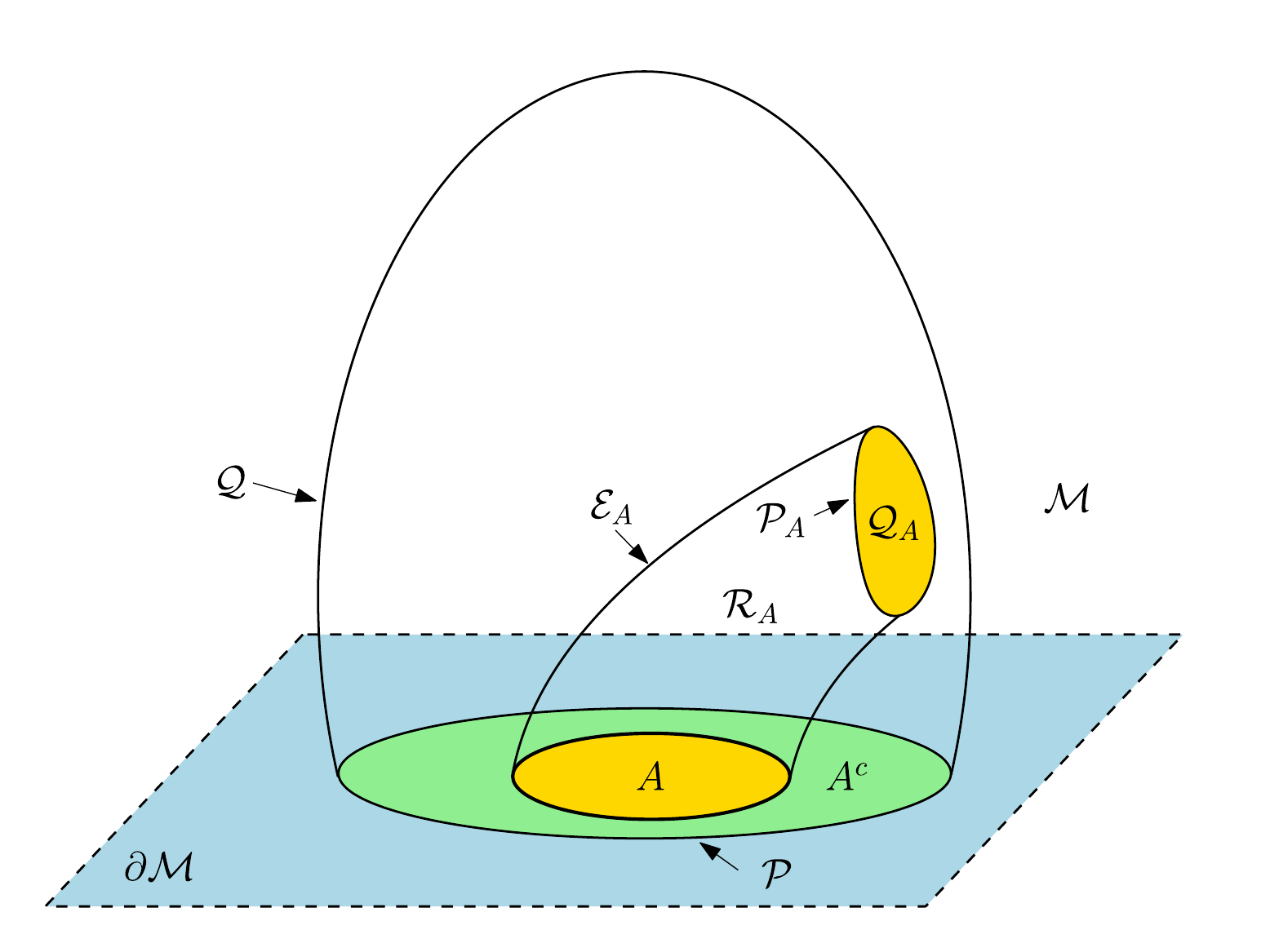} }
  \subfloat[AdS black hole]{
    \includegraphics[width=.45\linewidth]{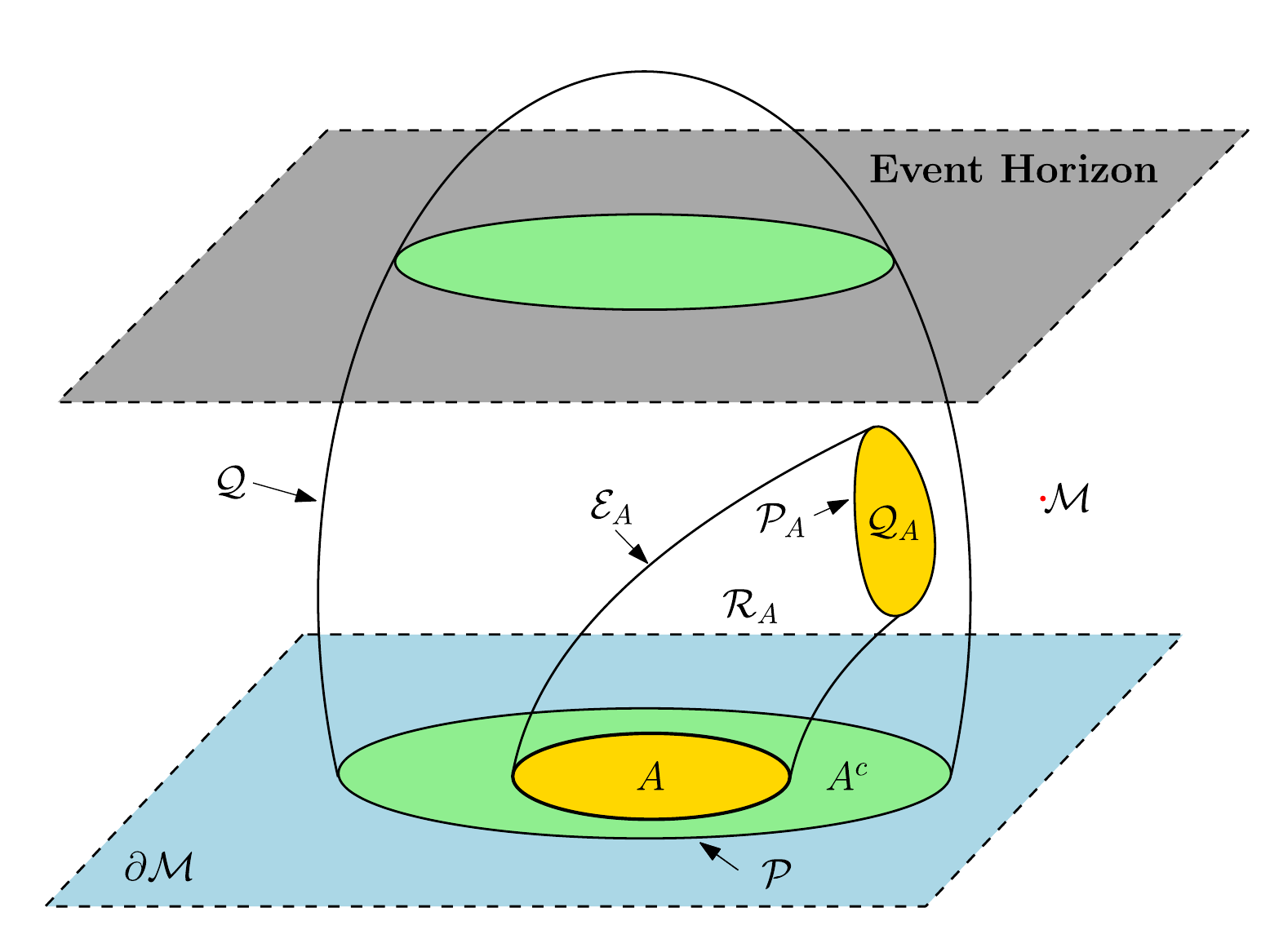} }
  \caption{Spacetime setup for the holographic BCFT with an entangled region $ {A} $.
  The minimal surface $\mathcal{E}_{ {A} }$ could end not only on the conformal boundary $\partial\mathcal{M}$ but also on the geometric boundary $\mathcal{Q}$.
  (a) The bulk manifold is a pure AdS spacetime. The minimal surface for ${ {A} }$ and ${ {A} ^{c}}$ have the same homology.
  (b) The bulk manifold is an asymptotically AdS black hole. The minimal surface for ${ {A} }$ and ${ {A} ^{c}}$ have different homologies due to the black hole horizon.}
  \label{Setup2}
\end{figure}

We consider a $(d+2)$-dimensional bulk manifold $\mathcal{M}$ which has a $(d+1)$-dimensional conformal boundary $\partial\mathcal{M}$ as shown in the Fig.\ref{Setup2}. The bulk manifold $\mathcal{M}$ is either a pure AdS spacetime, as in Fig.\ref{Setup2}(a), or an asymptotic AdS black hole with an event horizon, as in Fig.\ref{Setup2}(b). In addition, there is a $(d+1)$-dimensional hypersurface $\mathcal{Q}$ in $\mathcal{M}$ that intersects the conformal boundary $\partial\mathcal{M}$ at a $d$-dimensional hypersurface $\mathcal{P}$. A BCFT is defined on $\partial\mathcal{M}$ within the boundary $\mathcal{P}$. The hypersurface $\mathcal{Q}$ could be considered as the extension of the boundary $\mathcal{P}$ from $\partial\mathcal{M}$ into the bulk $\mathcal{M}$ and represents a geometric boundary of the bulk. This is our holographic setup for a BCFT living in $\partial\mathcal{M}$ with a boundary $\mathcal{P}$.

The total action of the system is the sum of the actions of the various geometric objects and their boundary terms \cite{1701.07202,1805.06117},
\begin{equation} \label{action}
  \mathcal{S} = \mathcal{S}_{\mathcal{M}_{d+2}} + \mathcal{S}_{\mathcal{Q}_{d+1}} + \mathcal{S}_{\mathcal{\partial M}_{d+1}} + \mathcal{S}_{\mathcal{P}_{d}},
\end{equation}
with
\begin{align}
  \mathcal{S}_{\mathcal{M}_{d+2}} & = \int_{\mathcal{M}} \sqrt{-g} (R-2\Lambda_{\mathcal{M}}), \\
  \mathcal{S}_{\mathcal{Q}_{d+1}} & = \int_{\mathcal{Q}} \sqrt{-h} (\lambda R_{\mathcal{Q}} - 2\Lambda_{\mathcal{Q}}+2K), \\
  \mathcal{S}_{\mathcal{\partial M}_{d+1}} & = 2\int_{\mathcal{\partial M}} \sqrt{-\gamma} K^{\prime}, \\
  \mathcal{S}_{\mathcal{P}_{d}} & = 2\int_{\mathcal{P}} \sqrt{-\sigma} \theta,
\end{align}
where $S_{\mathcal{M}_{d+2}}$ is the action of the bulk manifold $\mathcal{M}$ with $R$ and $\Lambda_{\mathcal{M}}$ being the intrinsic Ricci curvature and the cosmological constant of $\mathcal{M}$.
$S_{\mathcal{Q}_{d+1}}$ is the action of the geometric boundary $\mathcal{Q}$ with $R_{\mathcal{Q}}$, $\Lambda_{\mathcal{Q}}$ and $K$ being the intrinsic Ricci curvature, the cosmological constant and the extrinsic curvatures of $\mathcal{Q}$ embedded in $\mathcal{M}$, and $\lambda$ is a constant carrying the dimension of length. $S_{\partial\mathcal{M}_{d+1}}$ is the action of the conformal boundary of $\partial\mathcal{M}$ with $K^{\prime}$ being the extrinsic curvatures of $\partial\mathcal{M}$ embedded in $\mathcal{M}$.
We remark that the terms of $K$ and $K^{\prime}$ are the Gibbons-Hawking boundary terms for the boundaries $\mathcal{Q}$ and $\partial\mathcal{M}$ of the bulk manifold $\mathcal{M}$, respectively.
Finally, $S_{\mathcal{P}_{d}}$ is the common boundary term of $\mathcal{Q}$ and $\partial\mathcal{M}$ with $\theta=\cos^{-1}{(n^{\mathcal{Q}}\cdot n^{\mathcal{M}})}$ being the
supplementary angle between $\mathcal{Q}$ and $\partial \mathcal{M}$, which makes a well-defined variational principle on $\mathcal{P}$.
Furthermore, $g_{ab}$ denotes the metric of the bulk manifold $\mathcal{M}$, $h_{ab}$ and $\gamma_{ab}$ denote the induced metric of the boundaries $\mathcal{Q}$ and $\partial\mathcal{M}$, $\sigma_{ab}$ denotes the metric of $\mathcal{P}$.

Varying $\mathcal{S}_{\mathcal{M}_{d+2}}$ with $g^{ab}$ gives the equation of motion of the bulk $\mathcal{M}$,
\begin{equation}
0=R_{ab}-\frac{1}{2}Rg_{ab}+\Lambda_{\mathcal{M}}g_{ab}.
\end{equation}
Varying $\mathcal{S}_{\mathcal{Q}_{d+1}}$ with $h^{ab}$ gives the equation of motion of the geometric boundary $\mathcal{Q}$,
\begin{equation}
\lambda {R_{\mathcal{Q}ab}}+2K_{ab}-\left(  \frac{1}{2}{\lambda R_{\mathcal{Q}}}%
+K-\Lambda_{\mathcal{Q}}\right)  h_{ab}=0,\label{NBC}%
\end{equation}
which is just the Neumann boundary condition originally proposed by Takayanagi in \cite{1105.5165}. However, the boundary condition (\ref{NBC}) is too strong to have a solution even in the pure AdS spacetime because there are more constraint equations than the degrees of freedom. In
\cite{1701.04275,1701.07202}, the authors proposed the following reduced
boundary condition,
\begin{equation}
\left(  d-1\right)  \left( \lambda {R_{\mathcal{Q}}}+2K\right)  -2\left(  d+1\right)
\Lambda_{\mathcal{Q}}=0, \label{BC}%
\end{equation}
by taking the trace of the Eq.(\ref{NBC}).

In this work, we first consider the bulk manifold $\mathcal{M}$ as the $(d+2)$-dimensional pure AdS spacetime $AdS_{d+2}$ with the metric,
\begin{equation}
ds_{\mathcal{M}}^{2}=\frac{l_{AdS}^{2}}{z^{2}}\left(  -dt^{2}+dz^{2}%
+\sum_{i=1}^{d}dx_{i}^{2}\right)  , \label{AdS}%
\end{equation}
where $l_{AdS}$ is the $AdS$ radius. The conformal boundary of $AdS_{d+2}$ is
a $(d+1)$-dimensional Minkowski spacetime located at $z=0$.

We propose a simple solution of the geometric boundary $\mathcal{Q}$ as a
$(d+1)$-dimensional hepersurface embedded in the bulk manifold as,
\begin{equation}
ds_{\mathcal{Q}}^{2}=\frac{l_{AdS}^{2}}{z^{2}}\left(  -dt^{2}+dz^{2}%
+\sum_{i=2}^{d}dx_{i}^{2}\right)  , \label{AdS-Q}%
\end{equation}
with a simple embedding $x_{1}=$ constant. The intrinsic curvature, the
extrinsic curvature and the cosmological constant on $\mathcal{Q}$ can be
calculated as,
\begin{equation}
R_{\mathcal{Q}}=-\frac{d(d+1)}{l_{AdS}^{2}}\text{, }K_{ab}=0\text{, }%
\Lambda_{\mathcal{Q}}=-\frac{d(d-1)\lambda}{2l_{AdS}^{2}}.
\end{equation}
It is easy to verify that the mixed boundary condition (\ref{BC}) is satisfied.

Next, to study the holographic BCFT at finite temperature, we consider the bulk manifold $\mathcal{M}$ to be a black hole spacetime. However, even though with the reduced boundary condition (\ref{BC}), it is still very difficult to find a black hole solution for $\mathcal{M}$. So far the only known black hole solution for $\mathcal{M}$ is the $(d+2)$-dimensional Schwarzschild-AdS black hole \cite{1805.06117},
\begin{equation}
ds_{\mathcal{M}}^{2}=\frac{R^{2}}{z^{2}}\left(  -g\left(  z\right)
dt^{2}+\frac{dz^{2}}{g\left(  z\right)  }+\sum_{i=1}^{d}dx_{i}^{2}\right)
\label{AdS-Schw}%
\end{equation}
where
\begin{equation}
g\left(  z\right)  =1-\frac{z^{d+1}}{z_{h}^{d+1}}. \label{blacken}
\end{equation}
The black hole temperature can be calculated as,
\begin{align}
T  & =\left\vert
\dfrac{g^{\prime}\left(  z_{H}\right)  }{4\pi}\right\vert =\dfrac{d+1}{4\pi
z_{H}}.
\end{align}
Similar to the pure AdS case, we propose a  $(d+1)$-dimensional black hole for the geometric boundary $\mathcal{Q}$ by setting $x_{1}=$ constant,
\begin{equation}
ds_{\mathcal{Q}}^{2}=\frac{l_{AdS}^{2}}{z^{2}}\left(  -g\left(  z\right)
dt^{2}+\frac{dz^{2}}{g\left(  z\right)  }+\sum_{i=2}^{d}dx_{i}^{2}\right)
.\label{AdS-Schw-Q}%
\end{equation}
where the blacken factor $g(z)$ is the same .

The intrinsic curvature, the extrinsic curvature and the cosmological constant on $\mathcal{Q}$ are calculated as,
\begin{equation}
R_{\mathcal{Q}}=-\frac{d(d+1)}{l_{AdS}^{2}}\text{, }K_{ab}=0\text{, }%
\Lambda_{\mathcal{Q}}=-\frac{d(d-1)\lambda }{2l_{AdS}^{2}},
\end{equation}
which satisfy the reduced boundary condition (\ref{BC}).

\subsection{Holographic Entanglement Entropy}
The standard HEE is given by the RT formula \cite{0603001,0605073},
\begin{equation}
S_{ {A} }=\min_{X}\frac{Area\left(  \mathcal{E}_{ {A} }\right)
}{4G_{N}^{\left(  d+2\right)  }}\text{, }X=\left\{  \mathcal{E}_{ {A} %
}\Big\vert~\left.  \mathcal{E}_{ {A} }\right\vert _{\mathcal{\partial M}%
}=\partial {A} \text{; }\exists\mathcal{R}_{ {A} }\subset
\mathcal{M}\text{, }\partial\mathcal{R}_{ {A} }=\mathcal{E}%
_{ {A} }\cup {A} \right\}  ,
\end{equation}
where $\mathcal{E}_{ {A} }$ is the RT surface anchored on $\partial {A} $, and $\mathcal{R}_{ {A} }$ is the entanglement wedge of $ {A} $.

The RT formula for the HEE in BCFT is proposed as \cite{1701.04275,1701.07202,1805.06117},
\begin{equation}
S_{ {A} }=\min_{X}\frac{Area\left(  \mathcal{E}_{ {A} }\right)
}{4G_{N}^{\left(  d+2\right)  }}\text{, }X=\left\{  \mathcal{E}_{ {A} %
}\Big\vert~\left.  \mathcal{E}_{ {A} }\right\vert _{\partial \mathcal{M}%
}=\partial {A} \text{, }\left.  \mathcal{E}_{ {A} }\right\vert
_{\mathcal{Q}}=\mathcal{P}_{ {A} }\text{; }\exists\mathcal{R}_{ {A} }%
\subset\mathcal{M}\text{, }\partial\mathcal{R}_{ {A} }=\mathcal{E}%
_{ {A} }\cup {A} \cup \mathcal{Q}_{ {A} }\right\}  ,
\end{equation}
where $\mathcal{P}_{ {A} }$ divides the geometric boundary $\mathcal{Q}$
into two parts, $\mathcal{Q}_{ {A} }$ and $\mathcal{Q}_{ {A} ^{c}%
}$, with $\mathcal{Q}_{ {A} }$/ $\mathcal{Q}_{ {A} ^{c}}$ having the same homology as
$ {A} $/$ {A} ^{c}$, as shown in Fig.\ref{Setup2}. Requiring the boundary
condition (\ref{BC}) to be smooth, $\mathcal{E}_{ {A} }$ should be
orthogonal to $\mathcal{Q}$ when they intersect as shown in
\cite{1701.04275,1701.07202}.

To be concrete, in this work, we consider a bulk spacetime $\mathcal{M}$ with
two boundaries $\mathcal{Q}_{L,R}$, which intersect the conformal boundary
$\partial \mathcal{M}$ at $\mathcal{P}=\pm l/2$ perpendicularly. We choose the region
$ {A} \subset \partial \mathcal{M}$ as an infinite long strip,
\begin{equation}
x_{1}\in\left[  x-\frac{a}{2},x+\frac{a}{2}\right]  \text{, }x_{i}%
\in\mathbb{R}^{d-1}\text{ for }i=2,\cdots d,
\end{equation}
which preserves $(d-1)$-dimensional translational invariance in the directions
$x_{i}$ for $i=2,\cdots d$. Since we only consider the infinite strip in this work, the entangled region effectively lives in a one dimensional space $x_1$. Thus $ {A} $ can be
described by two parameters $\left(  x,a\right)  $, the middle point $x$, and
the width $a$ along the $x_1$ direction.

In the static gauge,
\begin{equation}
z=z\left(  x_{1}\right)  \text{, }z\left(  x\pm\frac{a}{2}\right)  =0\text{,
}z\left(  x\right)  =z_{0}\text{, }z^{\prime}\left(  x\right)  =0,
\end{equation}
where $x_{1}=x$ is the turning point of the minimal surface $\mathcal{E}%
_{ {A} }$.

For a general $(d+2)$-dimensional bulk metric,
\begin{equation}
ds^{2}=-g_{tt}\left(  z\right)  dt^{2}+\sum_{i=1}^{d}g_{ii}\left(  z\right)
dx_{i}^{2}+g_{zz}\left(  z\right)  dz^{2}\text{, }i=1,\cdots d,
\end{equation}
the size $a$ and the HEE $S_{ {A} }$ of the entangled region
$ {A} $ can be calculated as
\begin{align}
a &  =2z_{0}\int_{0}^{1}dv\left[  \frac{g_{11}\left(  z_{0}v\right)  }%
{g_{zz}\left(  z_{0}v\right)  }\left(  \frac{\tilde{g}^{2}\left(
z_{0}v\right)  }{\tilde{g}^{2}\left(  z_{0}\right)  }-1\right)  \right]
^{-1/2},\label{a}\\
S_{ {A} } &  =\frac{l_{AdS}^{d}L^{d-1}}{2G_{N}^{(d+2)}}\int_{0}%
^{1}dv\text{ }z_{0}\tilde{g}\left(  z_{0}v\right)  \left[  \frac{g_{11}\left(
z_{0}v\right)  }{g_{zz}\left(  z_{0}v\right)  }\left(  1-\frac{\tilde{g}%
^{2}\left(  z_{0}\right)  }{\tilde{g}^{2}\left(  z_{0}v\right)  }\right)
\right]  ^{-1/2},\label{SEE}%
\end{align}
where $v=z/z_{0}$ and
\begin{equation}
\tilde{g}\left(  z\right)  = \sqrt{ \prod_{i=1}^{d} g_{ii}\left(  z\right)  },
\end{equation}
and $L$ is the length of the directions in which the translational invariance is preserved,
\begin{equation}
\int_{\mathbb{R}^{d-1}}d^{d-1}\mathbf{x}=L^{d-1}.
\end{equation}
Using Eqs.(\ref{a}) and (\ref{SEE}), the HEE can be solved as a function of the size
$a$.

\section{Pure AdS Background}

We first consider the bulk spacetime $\mathcal{M}$ as a $(d+2)$-dimensional
pure AdS spacetime with the metric (\ref{AdS}), and choose the geometric
boundary $\mathcal{Q}$ as a $(d+1)$-dimensional hypersurface with the metric
(\ref{AdS-Q}). This is dual to BCFT at zero temperature.

In the case of pure AdS spacetime, the size $a$ can be integrated to obtain
\begin{equation}
a=2z_{0}\int_{0}^{1}\frac{v^{d}dv}{\sqrt{1-v^{2d}}}=2z_{0}\sqrt{\pi}%
\frac{\Gamma\left(  \frac{d+1}{2d}\right)  }{\Gamma\left(  \frac{1}%
{2d}\right)  }.
\end{equation}
The HEE is divergent near the boundary at $v\rightarrow0$. We thus need to
regulate the HEE by putting a small cut-off $\epsilon\ll1$. After the
regulation, the HEE can be obtained as
\begin{equation}
S_{ {A} }=\frac{l_{AdS}^{d}}{2\left(  d-1\right)  G_{N}^{(d+2)}}\left[
\left(  \frac{L}{\epsilon}\right)  ^{d-1}-\left(  \frac{L}{z_{0}}\right)
^{d}\frac{a}{2L}\right]  ,
\end{equation}
where the divergent term is proportional to the boundary of the entangled
region $ {A} $, i.e. $S^{ {A} }\sim L^{d-1}\sim\partial
 {A} $, as expected. The remaining term is finite.

\subsection{Phases of HEE}
\begin{figure}
  \subfloat[Sunset]{
    \includegraphics[width=.3\linewidth]{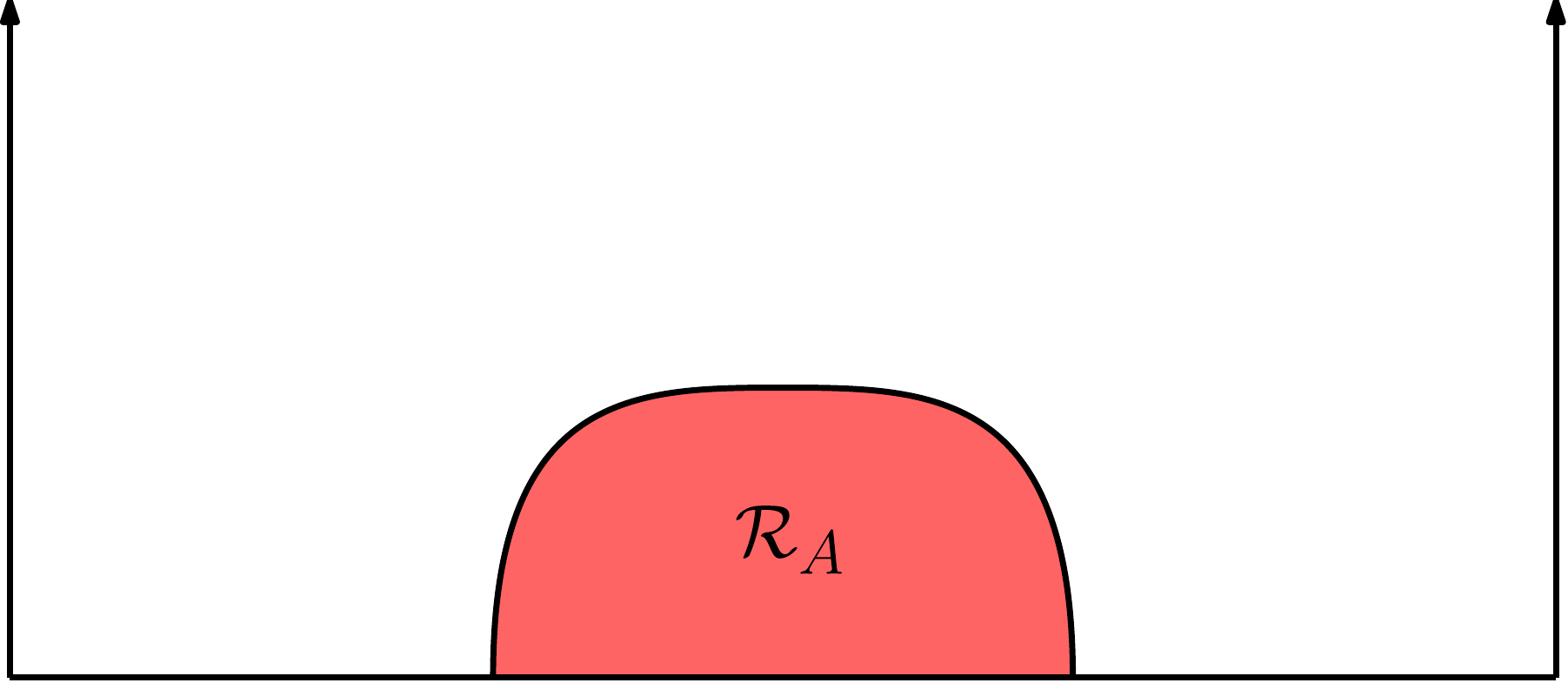} }
  \subfloat[Sky]{
    \includegraphics[width=.3\linewidth]{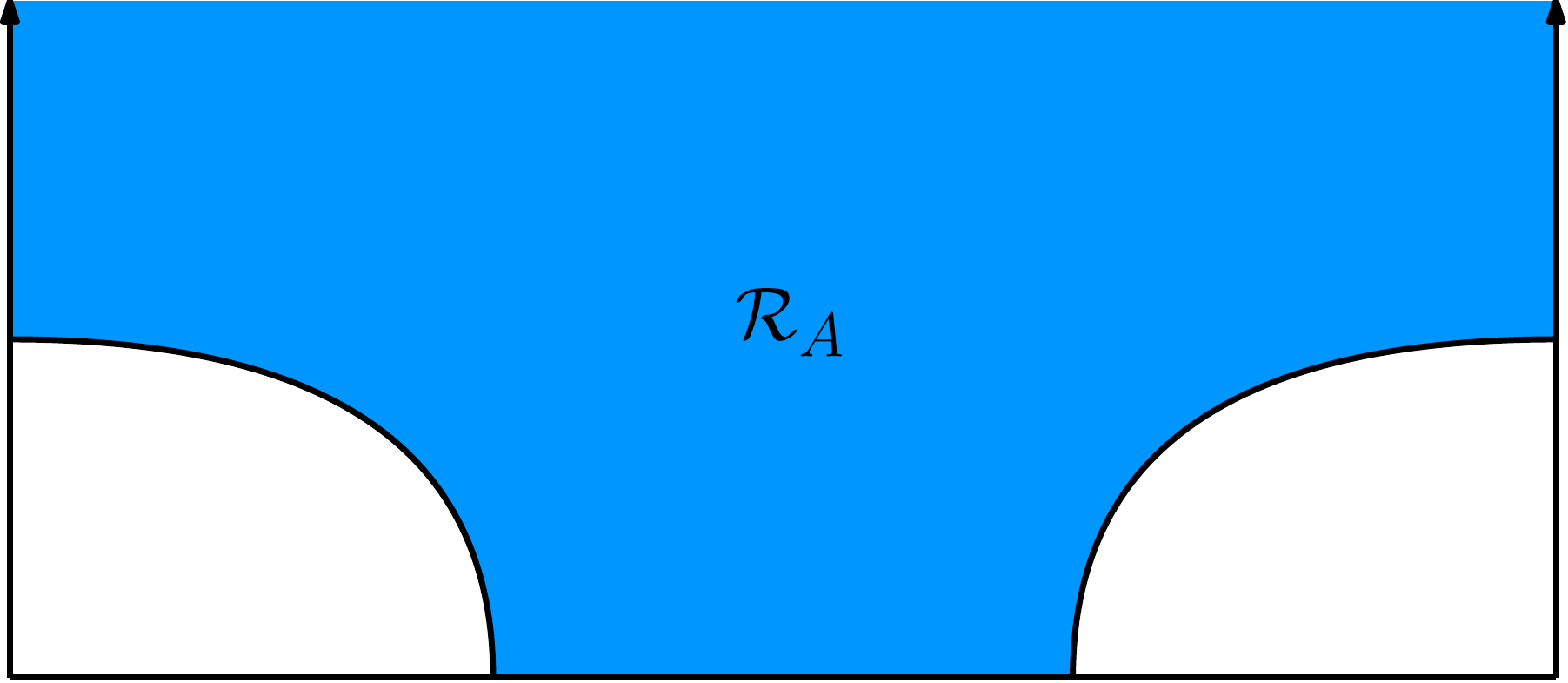} }
  \subfloat[Rainbow]{
    \includegraphics[width=.3\linewidth]{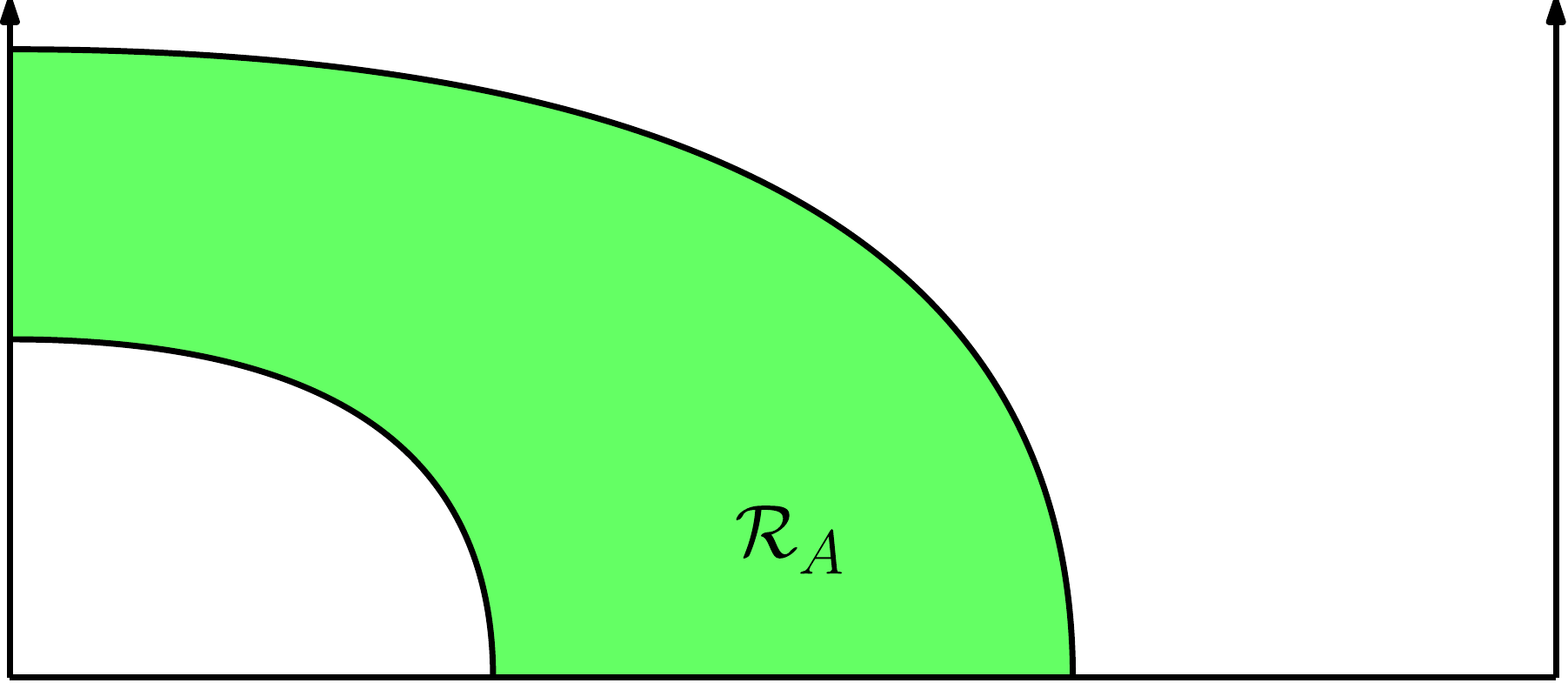} }
  \caption{The minimal surfaces in the pure AdS bulk spacetime. We fill the color red/blue/green in the entanglement wedge $\mathcal{R}_{ {A} }$ of the sunset/rainbow/sky phase.
  (a) $\mathcal{R}_{ {A} }$ is in the sunset phase when the entangled region $ {A} $ is relatively small.
  (b) $\mathcal{R}_{ {A} }$ is in the sky phase when the entangled region $ {A} $ is large enough.  (c) $\mathcal{R}_{ {A} }$ is in the rainbow phase when the entangled region $ {A} $ is close to the boundary.}
  \label{HEEshapes}
\end{figure}

The HEE corresponding to the different configurations of the minimal surfaces can be calculated as \cite{1805.06117},
\begin{align}
\text{sunset} &  :S_{ {A} }^{s}=S_{ {A} }(a),\label{ss} \\
\text{sky} &  :S_{ {A} }^{k}=\frac{1}{2}S_{ {A} }{(l-a+2|x|)+\frac
{1}{2}S_{ {A} }(l-a-2|x|)},\\
\text{rainbow} &  :S_{ {A} }^{r}=\frac{1}{2}S_{ {A} %
}{(l+a-2|x|)+\frac{1}{2}S_{ {A} }(l-a-2|x|)},
\end{align}
which are shown in Fig.\ref{HEEshapes}.

Although each of the HEE in the above three cases represents the local minimum, the global minimum is the smallest one of them. Depending on the size $a$ and the location $x$ of the entangled region $ {A} $, the HEE will transit among the three phases.

The phase diagram is an equilateral triangle with its bottom
equal to its height, which is set as $l$ in this work. Because any
entangled region $ {A} $ has to be inside of the conformal boundary
$\left[-l/2,l/2\right]  $, the width $a$ of the entangled region satisfy the condition: $a\leq l-2{\left\vert {x}\right\vert }$.

\begin{figure}
 \subfloat[Phase diagram]{
  \includegraphics[width=.4\linewidth]{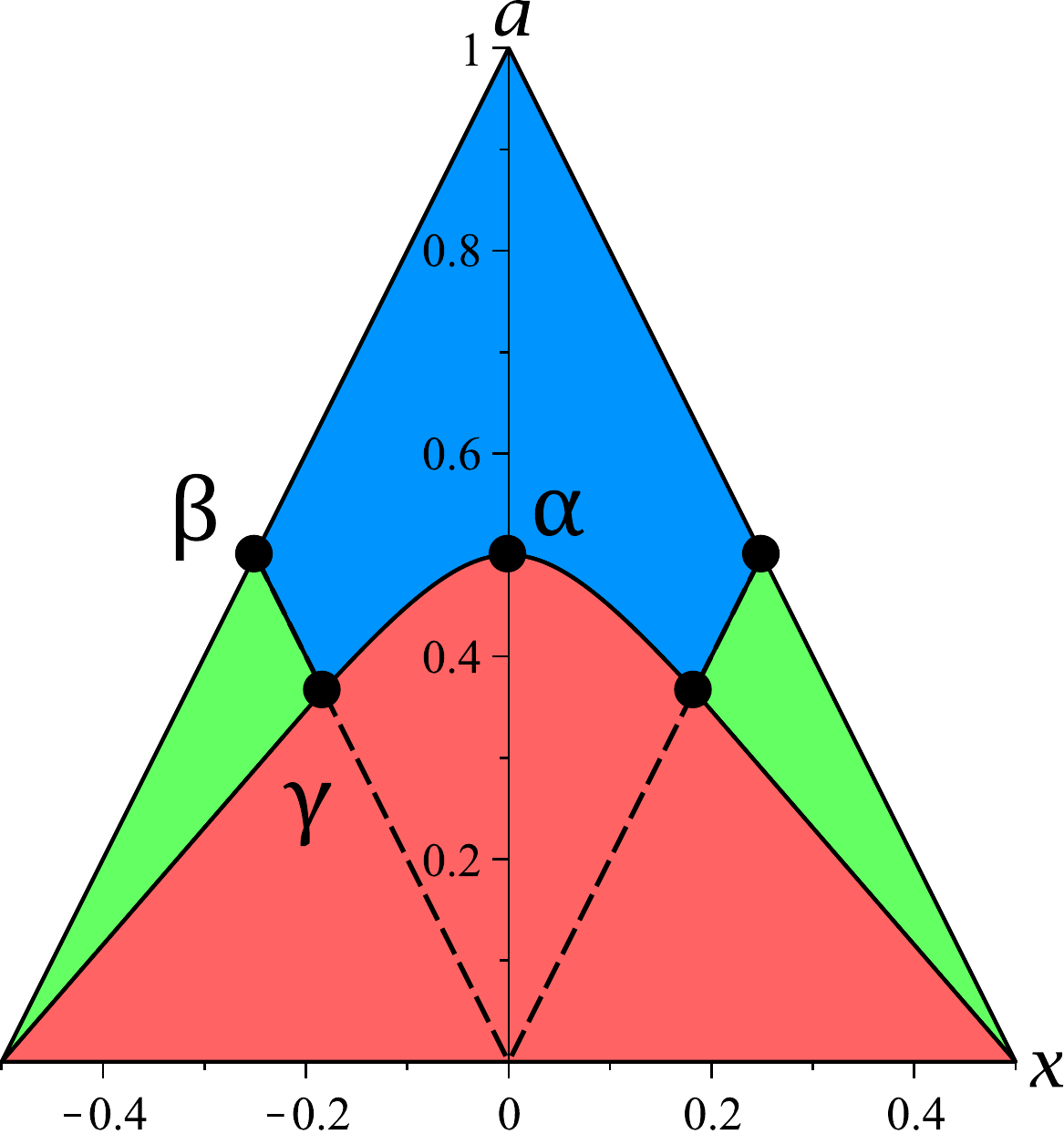}}
  \hspace{2cm}
 \subfloat[Character triangle]{
  \includegraphics[width=.4\linewidth]{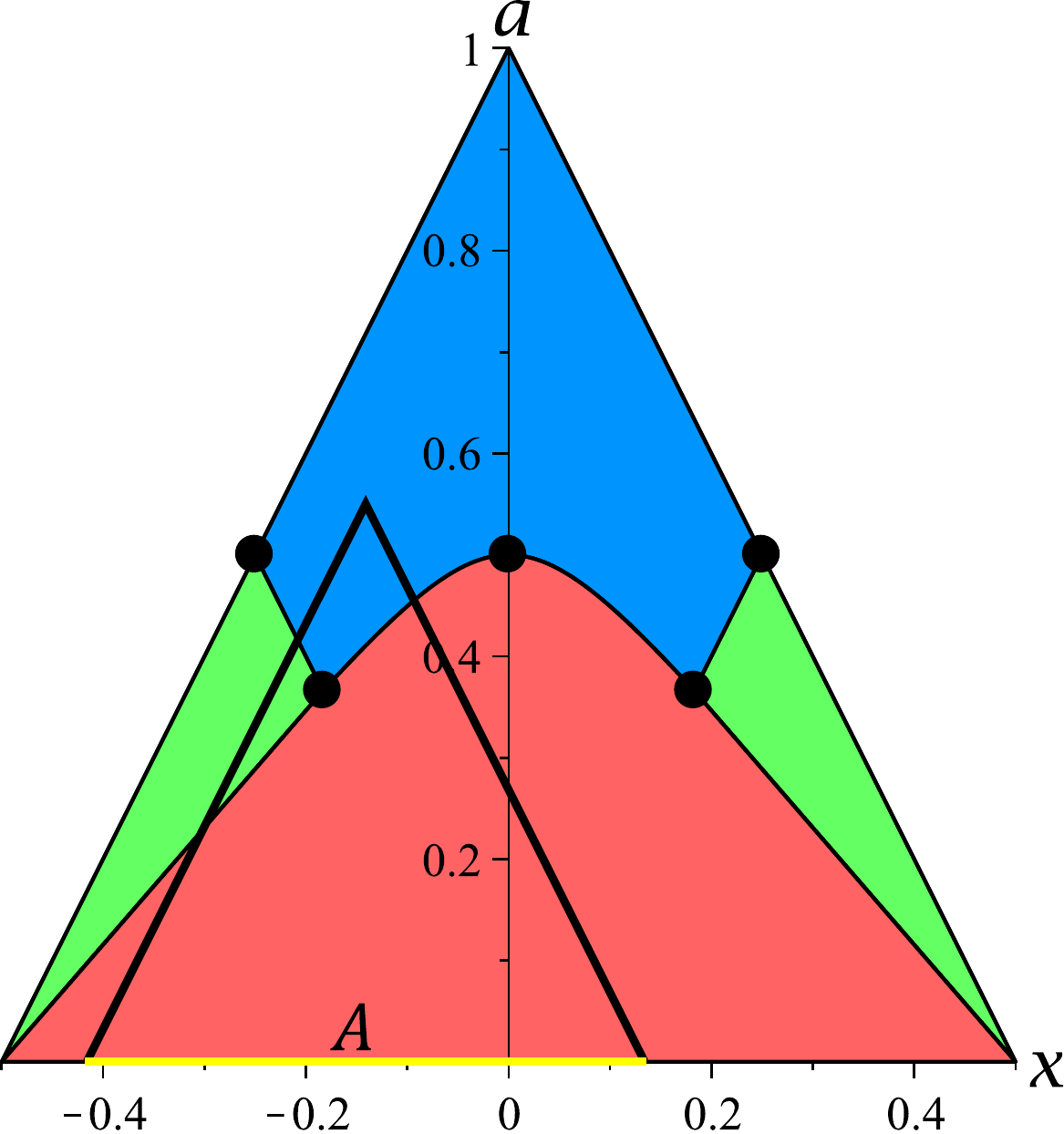}}
  \caption{(a) Phase Diagram of the holographic entanglement entropy in the pure AdS bulk spacetime. (b) Character triangle of ${A}$. The red/green/blue region represents the sunset/rainbow/sky phase and we will set $l = 1$ in the plots.}
  \label{PureAdsPD}
\end{figure}

The phase diagram is plotted in Fig.\ref{PureAdsPD}(a) with the red/blue/green region representing the sunset/sky/rainbow phase. The phase boundary between the sky and rainbow phases can be determined by $S_{ {A} }^{k}=S_{ {A} }^{r}$, which gives ${\left\vert
{x}\right\vert }=a/2$. We notice that these phase boundaries are parallel to the corresponding edges of the equilateral triangle, and their extensions intersect at the origin point $(0,0$) on the bottom of the phase triangle. This fact will be crucial for our later analysis.

Similarly, the phase boundaries between the sunset and
sky/rainbow phases can be determined by $S_{ {A} }^{s}=S_{ {A} }%
^{k}$ and $S_{ {A} }^{s}=S_{ {A} }^{r}$ respectively, which gives\begin{align}
\left(  \frac{1}{{l-a-2}\left\vert {x}\right\vert }\right)  ^{d-1}{+\left(
\frac{1}{{l-a-2}\left\vert {x}\right\vert }\right)  ^{d-1}} &  {=2\left(
\frac{1}{{a}}\right)  ^{d-1},}\label{sk line}\\
\left(  \frac{1}{{l+a-2}\left\vert {x}\right\vert }\right)  ^{d-1}{+\left(
\frac{1}{{l-a-2}\left\vert {x}\right\vert }\right)  ^{d-1}} &  {=2\left(
\frac{1}{{a}}\right)  ^{d-1}.}\label{sr line}%
\end{align}
Let us now pay attention to some special points in the phase diagram. First,
at $x_{1}=0$, Eq.(\ref{sk line}) leads to $a_{1}=l/2$ which labelled as point $\alpha:\left(x_{1},a_{1}\right)=(0,l/2)$ in the phase diagram. Second, the phase
boundaries between the sky and rainbow phases reach the two edges at the
point $\beta:\left(  x_{2},a_{2}\right)  =\left(  \pm l/4,l/2\right)$. In addition, the three phase boundaries meet at a triple point $\gamma:\left(  x_{3},a_{3}\right)  =\left(  a_{3}/2,a_{3}\right)  $ with
\begin{equation}
\left(  \frac{1}{{l}}\right)  ^{d-1}{+\left(  \frac{1}{{l-}a_{3}}\right)
^{d-1}=2\left(  \frac{1}{a_{3}}\right)  ^{d-1}.}%
\end{equation}
In the pure AdS case, it is easy to see that, an entangled region $ {A} $ and its complementary $ {A} ^{c}$ share the same RT surfaces so that $\mathcal{E}_{ {A} }=\mathcal{E}_{ {A} ^{c}}$.

Once we have the phase diagram, we can determine the phase for any
entangled region by marking the entangled region $ {A} $ on the bottom of the phase diagram and drawing an equilateral triangle on $ {A} $ with the height $a$. Then the location of the top vertex of the triangle indicates the corresponding phase for $ {A} $, which is shown in Fig.\ref{PureAdsPD}(b) as an example of the HEE in the sky phase. We will call this equilateral triangle for the region $ {A} $ the character triangle of $ {A} $.

\subsection{Bipartite system}
\begin{figure}
  \includegraphics[width=.5\linewidth]{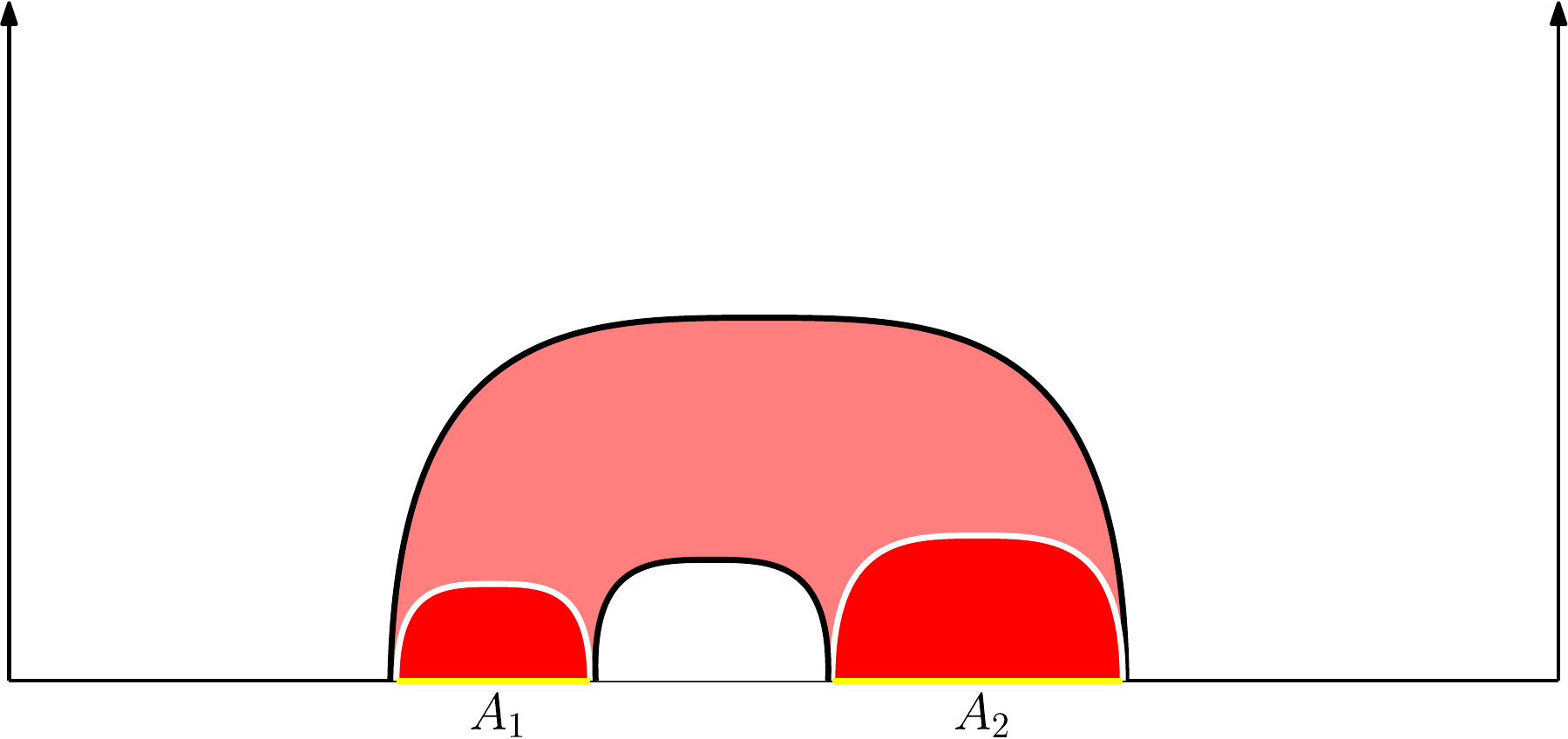}
  \caption{Two entangled regions in the connected and disconnected configurations.}
  \label{bipartite}
\end{figure}

We have studies the HEE for a single entangled region and its phase diagram.
In this section, we will consider the bipartite system, i.e. the HEE for two
disjointed entangled regions\footnote{The two entangled regions could be
connected side by side or even overlap, but their structure will reduce to
certain special cases. To be general, we consider the two entangled regions are disjointed in this paper.}. Although the two entangled regions we consider here are disjointed, their entanglement wedge could be connected as shown in Fig.\ref{bipartite}. The two solid curves represent the RT surface in the connected configuration, and the two dashed curves  represent the RT surface in the disconnected configuration. Whether the entanglement wedge is connected or not is determined by their HEE. The situation now is a little bit more complicated than the single region case. In addition to the connected and disconnected phase transition for the bipartite system, we also have the phase transitions among the sunset, sky and rainbow phases that we discussed in the last section.

To elaborate, let us define our notations for the multipartite system. Each entangled region $ {A} _{i}$ is described by its location $x_i$ and size $a_i$, respectively. Without loss of generality, we label the entangled regions $ {A} _{i}$'s by the order from left to right, i.e. $x_{i}<x_{j}$ for $i<j$.

Any pair of two entangled regions $ {A} _{i}$ and $ {A} _{j}%
$ with $i<j$ must satisfy the following two conditions in the presence of the boundaries at $x=\pm l/2$:
\begin{itemize}
\item The disjointed condition:  $ {A} _{i}$ and $ {A} _{j}%
$ does not overlap:
\begin{equation}
  x_{i} + \frac{a_{i}}{2} \leq x_{j} - \frac{a_{j}}{2}.
\end{equation}
\item The enclosed condition:  $ {A} _{i}$ and $ {A} _{j}%
$ are confined in the domain $[-l/2,l/2]$:
\begin{equation}
  -\frac{l}{2} \leq x_{i} - \frac{a_{i}}{2} \quad \text{and} \quad x_{j} + \frac{a_{j}}{2} \leq \frac{l}{2}.
\end{equation}
\end{itemize}

\begin{figure}
  \includegraphics[width=0.9\linewidth]{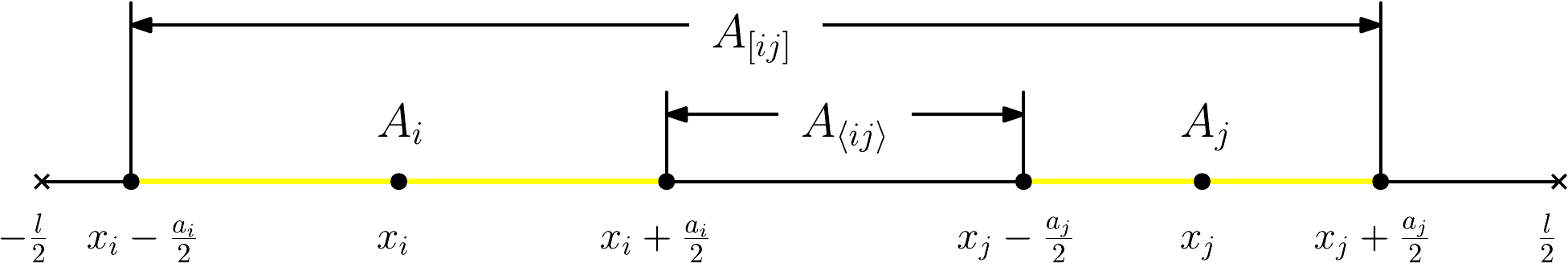}
  \caption{The definition of $A_{\langle ij \rangle}$ and $A_{[ ij ]}$. The entanglement regions of $A_i:(x_i,a_i)$ and $A_j:(x_j,a_j)$ are labelled with yellow color.}
  \label{Entanglement-Region}
\end{figure}

As usual, we define $ {A} _{ij}\equiv {A} _{i}\cup {A} _{j}$ to be the union of the two regions. In addition, as shown in Fig.\ref{Entanglement-Region}, we define $ {A} _{\left\langle
ij\right\rangle }$ to be the part between $ {A} _{i}$ and $ {A} _{j}$ with
\begin{equation}
\left(  x_{\left\langle ij\right\rangle },a_{\left\langle ij\right\rangle
}\right)  =\left(  \frac{x_{i}+x_{j}}{2}+\frac{a_{i}-a_{j}}{4},x_{j}%
-x_{i}-\frac{a_{i}+a_{j}}{2}\right)  ,\label{small region}%
\end{equation}
and $ {A} _{\left[  ij\right]  }\equiv
 {A} _{i}\cup {A} _{\left\langle ij\right\rangle }\cup
 {A} _{j}$ to be the combined region with
\begin{equation}
\left(  x_{\left[  ij\right]  },a_{\left[  ij\right]  }\right)  =\left(
\frac{x_{i}+x_{j}}{2}-\frac{a_{i}-a_{j}}{4},x_{j}-x_{i}+\frac{a_{i}+a_{j}}%
{2}\right)  .\label{large region}%
\end{equation}

Now, let us focus on the bipartite system. As we mentioned, there are two
configurations, with disconnected or connected entanglement wedges. Based on the RT prescription, the actual HEE for the union region $ {A} _{12}$ is the minimum value of the HEEs for the two configurations
\begin{equation}
S_{12}=\min\left(  S_{12}^{d},S_{12}^{c}\right), \label{S12}
\end{equation}
where the HEE for the disconnected configuration is $S_{12}^{d}=S_{1}+S_{2}$ and the HEE
for the connected configuration is $S_{12}^{c}=S_{\left[  12\right]
}+S_{\left\langle 12\right\rangle }$.

\begin{figure}
  \subfloat[sss]{
    \includegraphics[  width=.18\linewidth]{image/bipartite/bipartite-sss.pdf} }
  \subfloat[rss]{
    \includegraphics[  width=.18\linewidth]{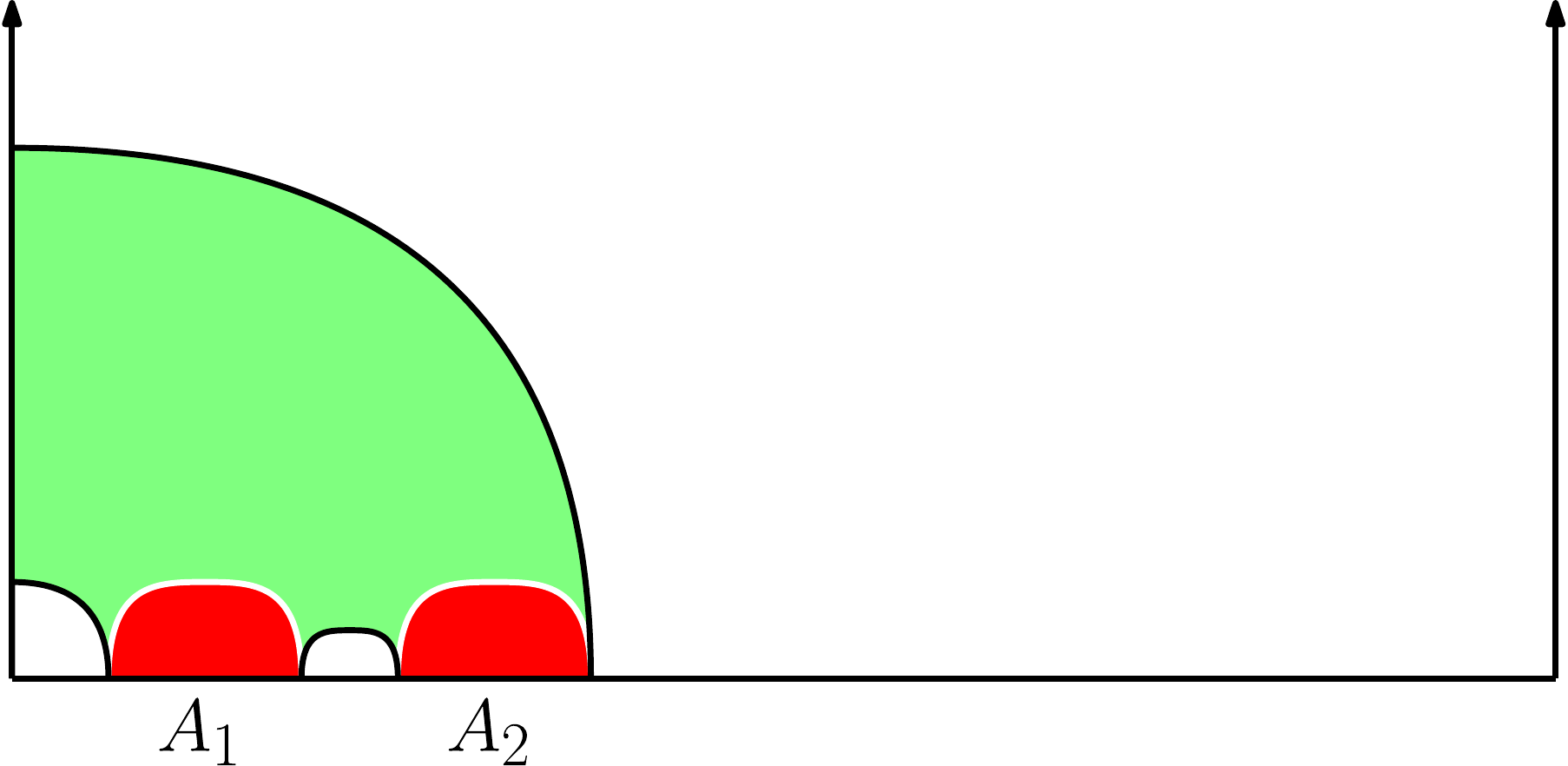} }
  \subfloat[rsr]{
    \includegraphics[  width=.18\linewidth]{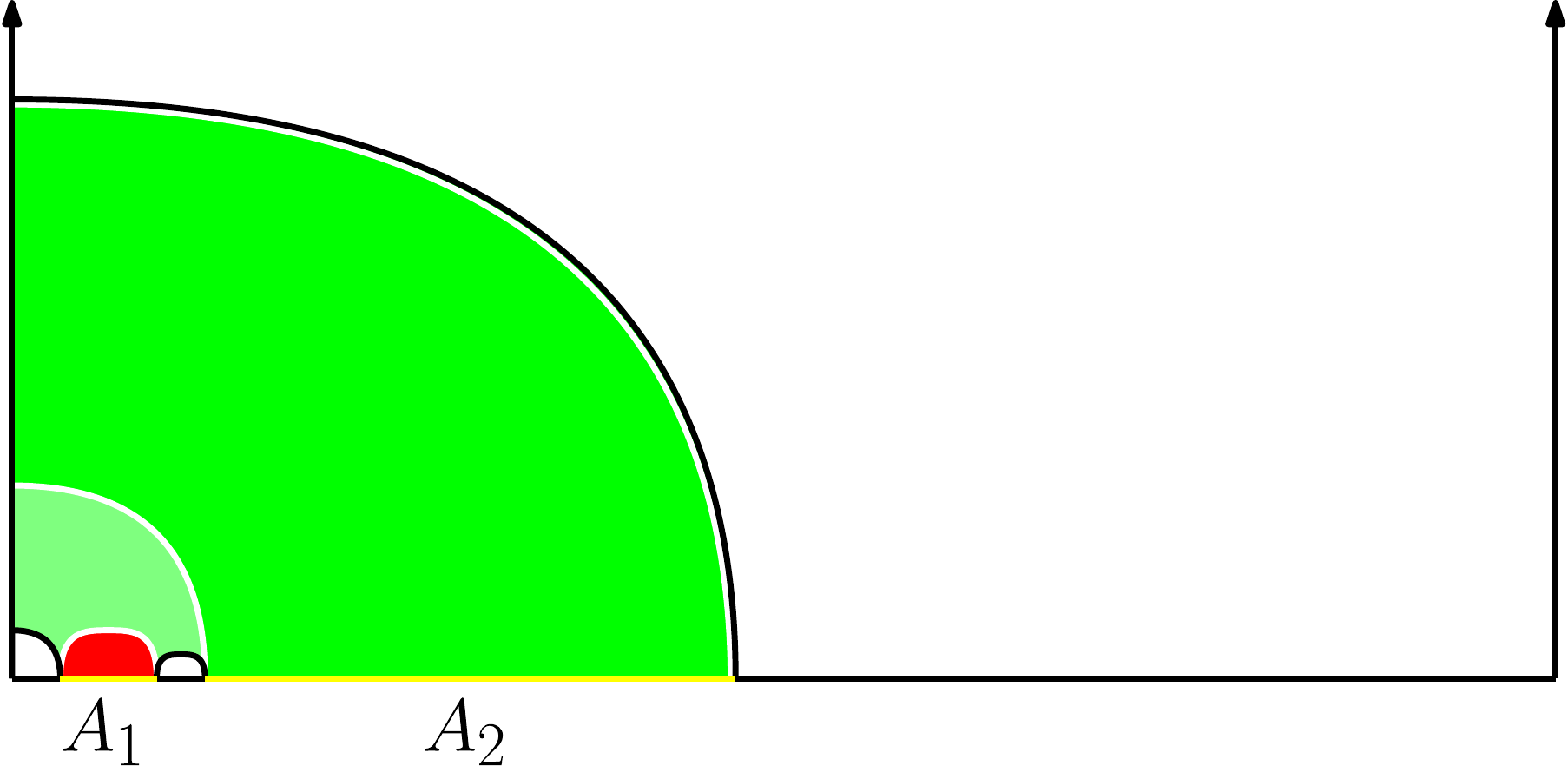} }
  \subfloat[rrs]{
    \includegraphics[  width=.18\linewidth]{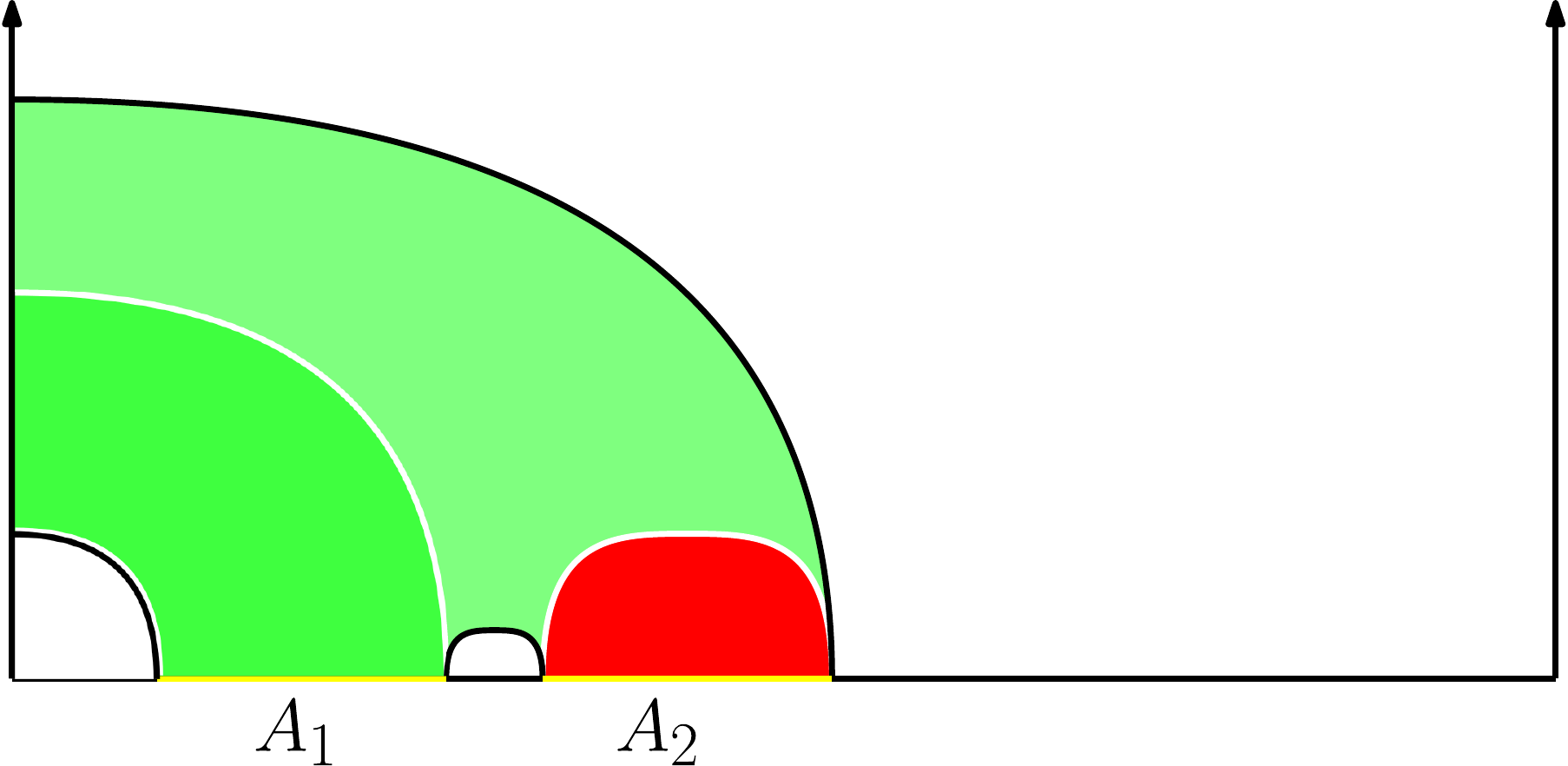} }
  \subfloat[rrr]{
    \includegraphics[  width=.18\linewidth]{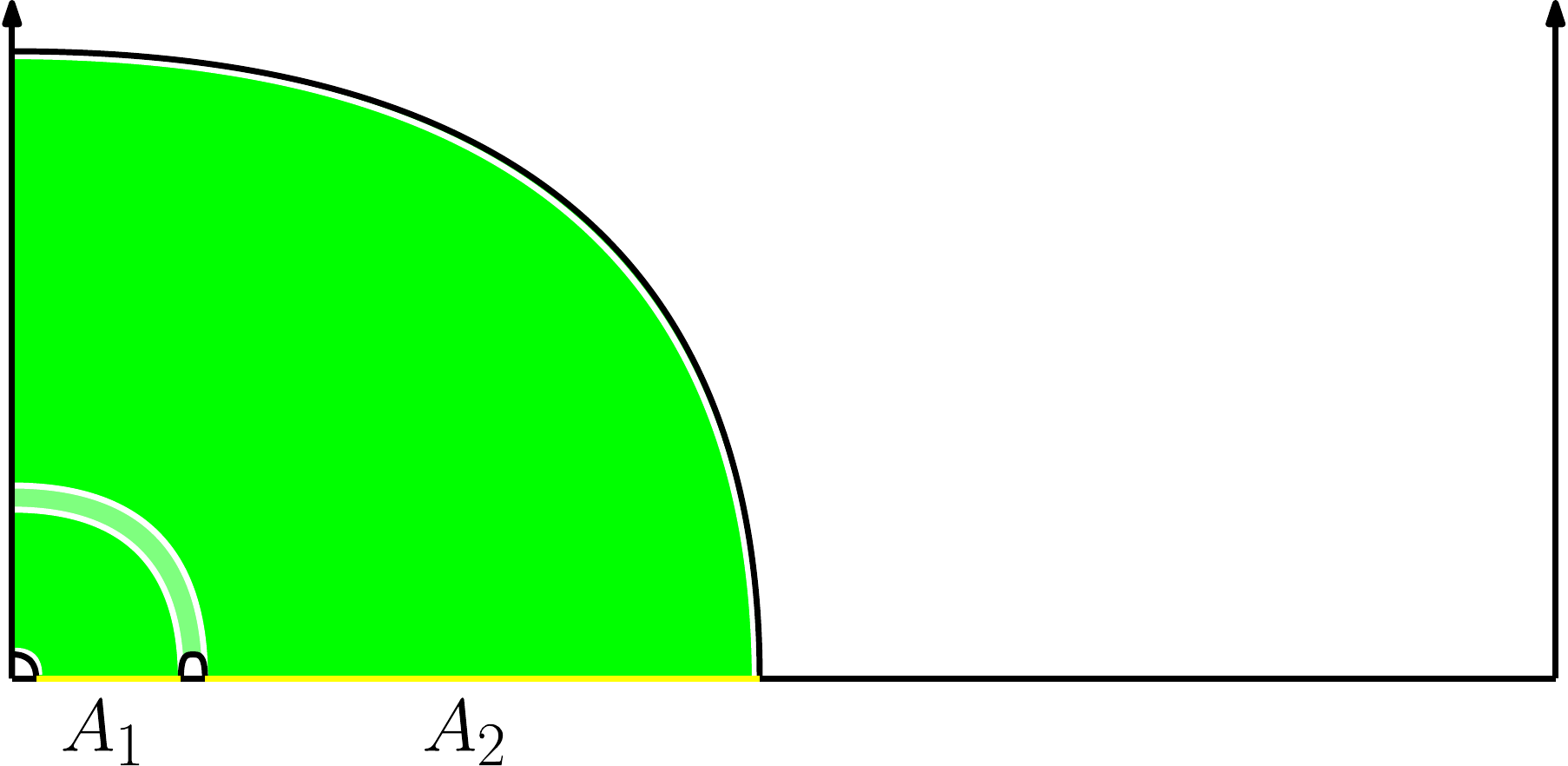} }\\
  \subfloat[kss]{
    \includegraphics[  width=.18\linewidth]{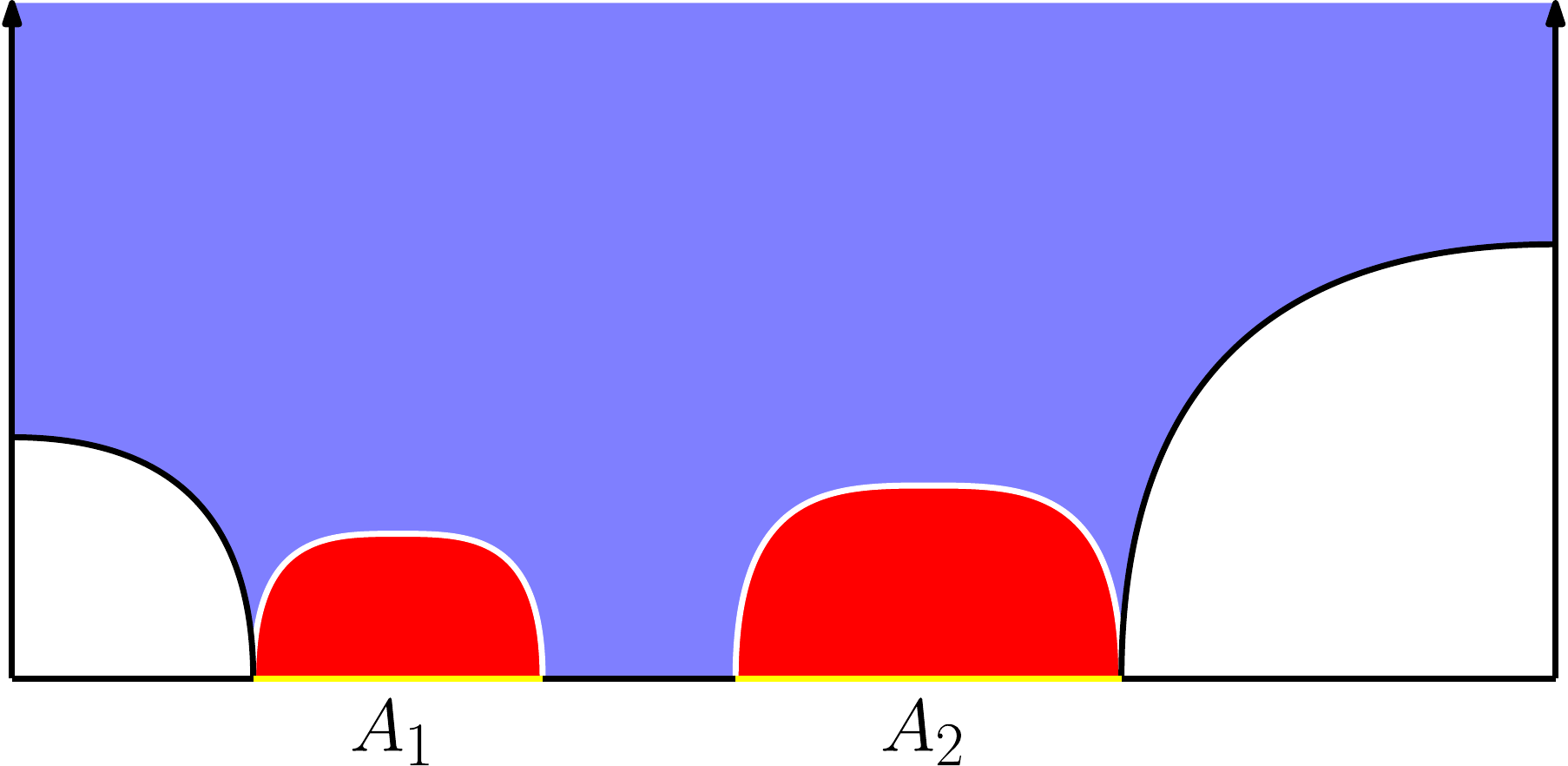} }
  \subfloat[ksk]{
    \includegraphics[  width=.18\linewidth]{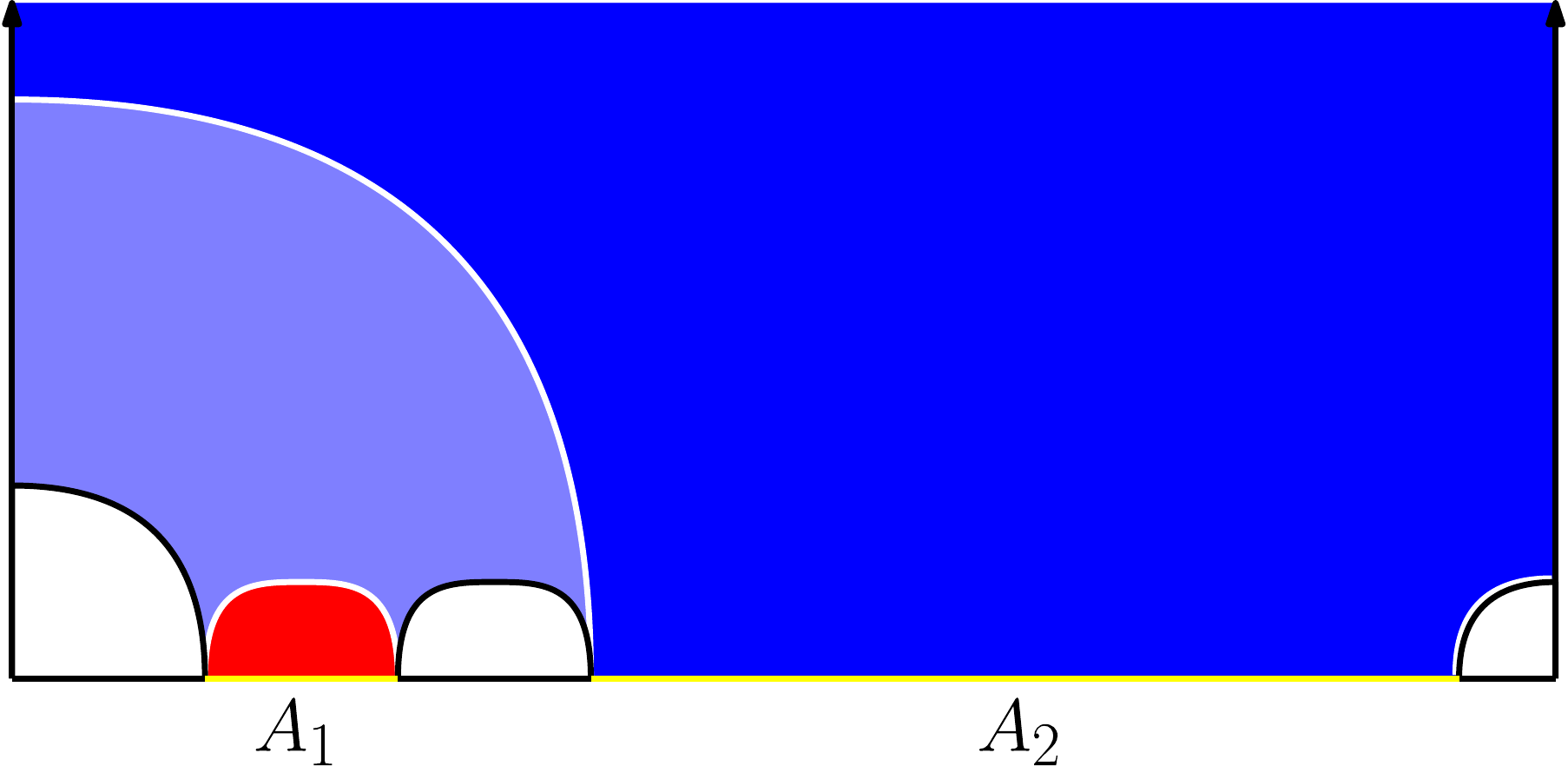} }
  \subfloat[krs]{
    \includegraphics[  width=.18\linewidth]{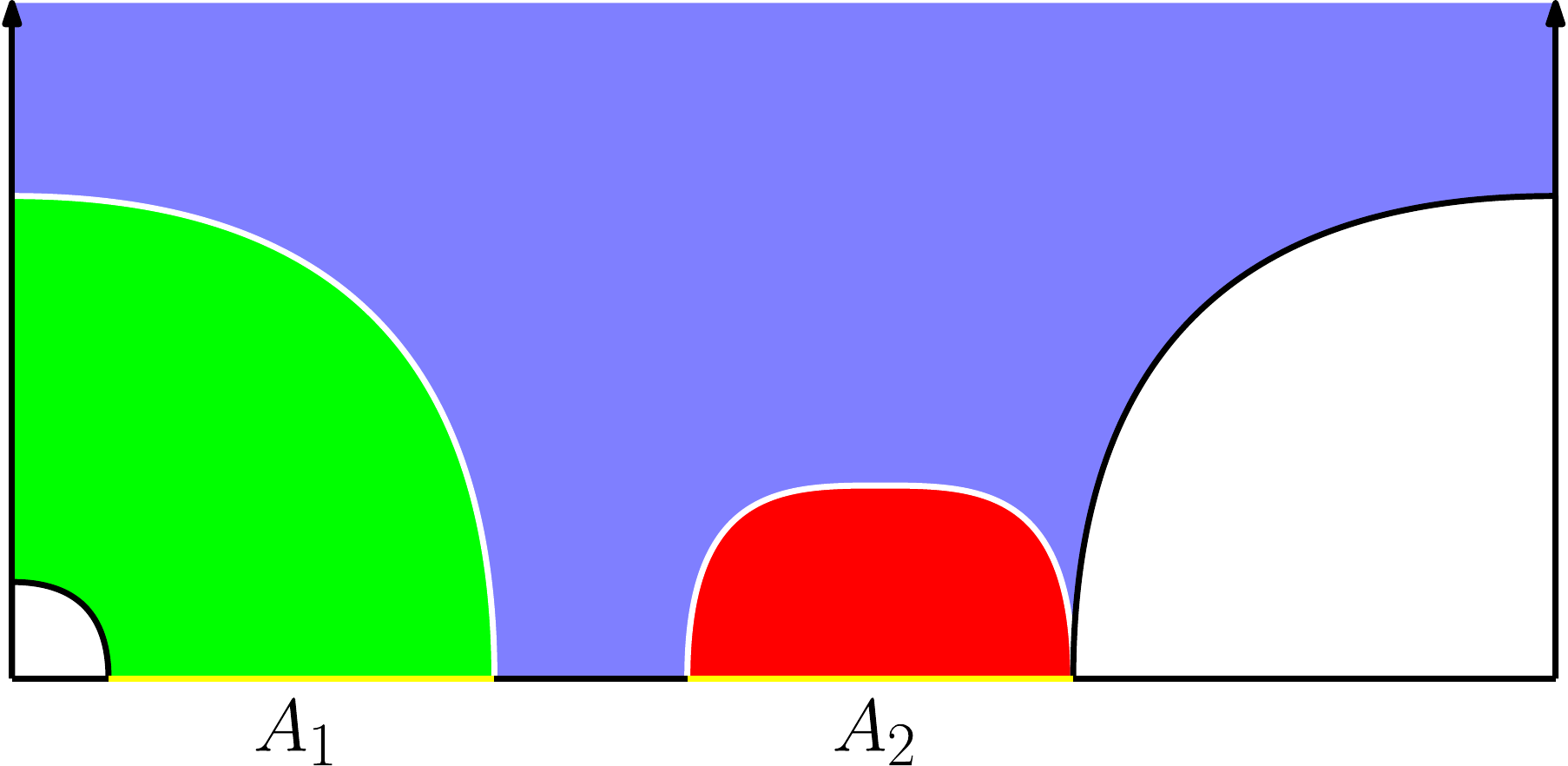} }
  \subfloat[krr]{
    \includegraphics[  width=.18\linewidth]{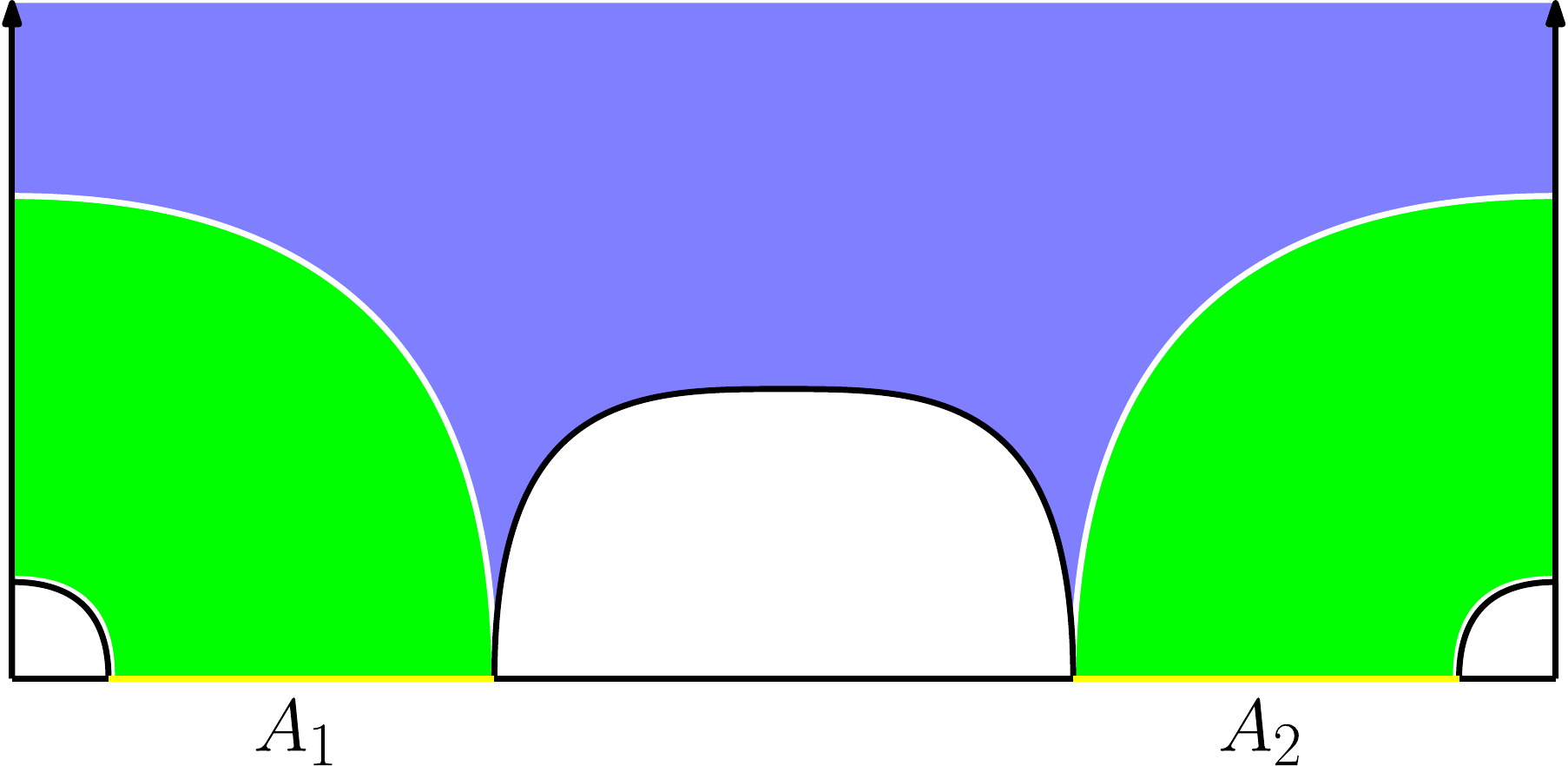} }
  \subfloat[krk]{
    \includegraphics[  width=.18\linewidth]{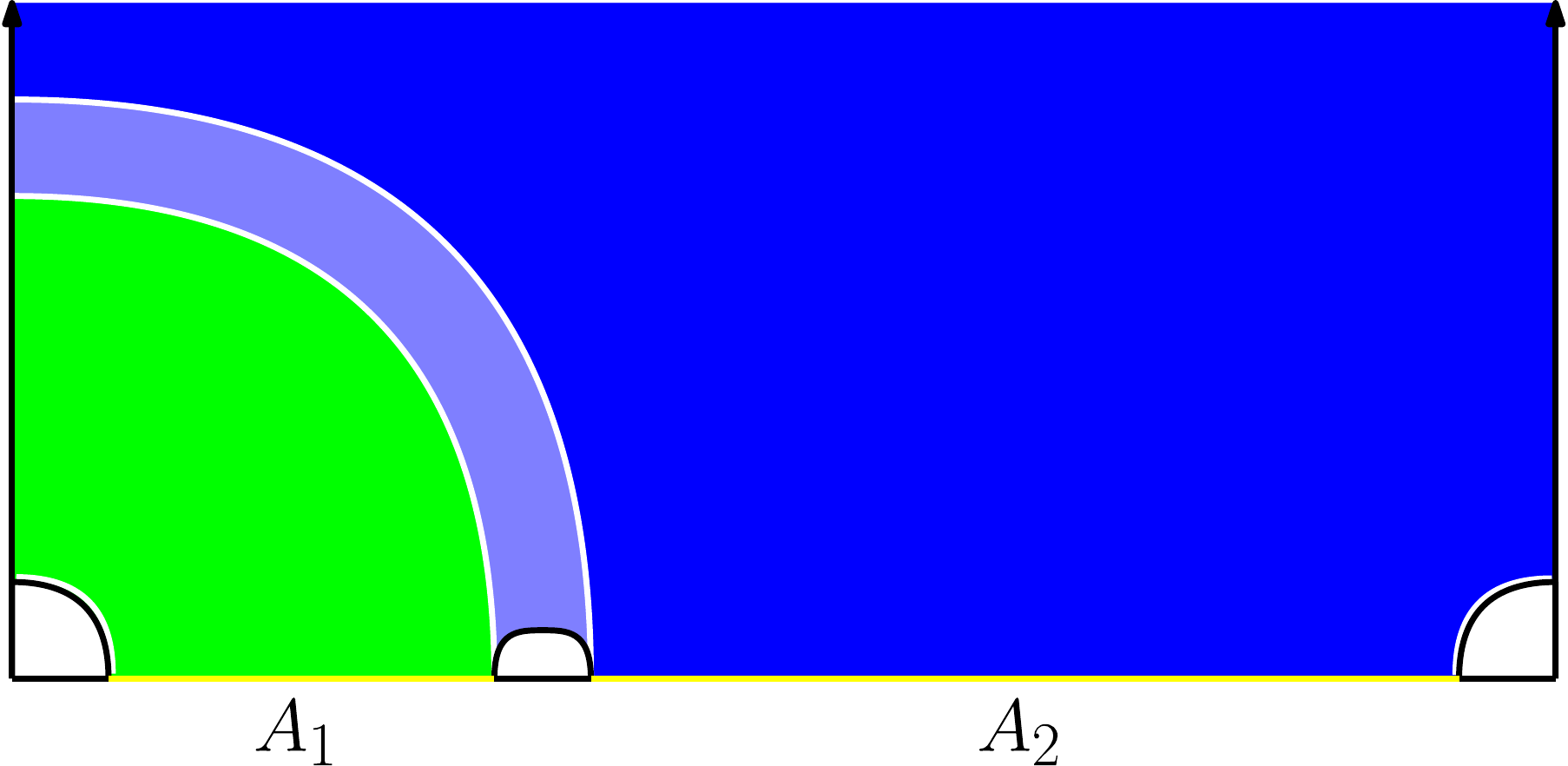} }
  \caption{The ten allowed cases for a bipartite system.}
  \label{fig_bipartite}
\end{figure}

The further complicated issue is that, for each HEE $S_{1}$, $S_{2}$,
$S_{\left[  12\right]  }$ and $S_{\left\langle 12\right\rangle }$, there could
be three different phases: sunset, sky and rainbow. Naively, we have to
consider all the $3^{4}=81$ combinations and choose the minimum one to be the true HEE. Fortunately, a simple observation helps to simplify the situation a lot: $S_{\left\langle 12\right\rangle }$ must be in the sunset phase to
ensure the entanglement wedge of $ {A} _{12}$ is connected. So we only
need to consider the phases of the other three regions, $S_{1}$, $S_{2}$ and
$S_{\left[  12\right]  }$ with $3^{3}=27$ choices. We can further
reduce the number of the choices by showing that most of the choices are
not consistent with the phase diagram of HEE we obtained in the last section.
For example, let's consider the case $S_{1}\sim$ rainbow, $S_{2}\sim$ sunset,
$S_{\left[  12\right]  }\sim$ sunset and $S_{\left\langle 12\right\rangle
}\sim$ sunset. By plotting the character triangle of $ {A} _{1}$ in the phase diagram, we find that it is impossible for the top vertex of the
character triangle of $ {A} _{\left[  12\right]  }$ to locate in the
sunset phase, so we can rule out this choice. After ruling out all the
impossible ones, there are only ten independent choices left to be
considered\footnote{Since the phase diagram is symmetric between left and
right, we only consider one side of the equivalent choices.}. We list all ten independent choices in Table \ref{SAtable}. The abbreviations labelled the different choices are based on the phases of $S_{1}$, $S_{2}$ and $S_{\left[  12\right]  }$ since the region $S_{\left\langle 12\right\rangle }$ is always in the sunset phase to ensure that the union region $S_{12}$ is connected. The corresponding entanglement wedges and phase diagrams are plotted in Fig.\ref{fig_bipartite}.

\begin{table}[htbp]
	\centering
	\caption{The ten allowed choices for a bipartite system.}
	\label{SAtable}
\setlength{\tabcolsep}{7mm}{
\begin{tabular}[c]
{|c|c|c|c|c|}\hline
choice & $ {S} _{\left\langle 12\right\rangle }$ & $ {S} %
_{\left[  12\right]  }$ & $ {S} _{1}$ & $ {S} _{2}$\\\hline
sss & sunset & sunset & sunset & sunset\\\hline
rss & sunset & rainbow & sunset & sunset\\\hline
rsr & sunset & rainbow & sunset & rainbow\\\hline
rrs & sunset & rainbow & rainbow & sunset\\\hline
rrr & sunset & rainbow & rainbow & rainbow\\\hline
kss & sunset & sky & sunset & sunset\\\hline
ksk & sunset & sky & sunset & sky\\\hline
krs & sunset & sky & rainbow & sunset\\\hline
krr & sunset & sky & rainbow & rainbow\\\hline
krk & sunset & sky & rainbow & sky\\\hline
\end{tabular}}
\end{table}

To obtain the complete list of the allowed choices, we use the following rules:
\begin{enumerate}
\item If $S_{\left[  ij\right]  }$ is in the sunset phase, then both $S_i$ and $S_j$ must be in the sunset phase.
\item If $S_{\left[  ij\right]  }$ is in the rainbow phase, then both $S_i$ and $S_j$  must not be in the sky phase.
\item If $S_{\left[  ij\right]  }$ is in the sky phase, then any $S_{k\ne i,j}$ must not be in the sky phase.
\item If $S_i$ is in the sky phase, then any $S_{k\ne i}$ must not be in the sky phase.
\end{enumerate}
The above rules can be easily checked by using the phase diagram for the pure AdS case. For the black hole case we will discuss in the next section, some of them will change.

The subadditivity (SA) inequality for a bipartite system is
\begin{equation}
S_{12}\leq S_{1}+S_{2},
\end{equation}
which is trivially satisfied by the definition of $S_{12}$ in Eq.(\ref{S12}). The equality is saturated if the entanglement wedge of $ {A}_{12}$ is disconnected. Equivalently, we can define the non-negativity of the mutual information,
\begin{equation}
I_{2}=S_{1}+S_{2}-S_{12}\geq0.
\end{equation}

\subsection{Tripartite System and Inequality of HEE}
In this section, we consider three disjointed entangled regions. The phase structure can be
obtained similarly as for the bipartite system, but with more complicated.
One of the most interesting issues for the multipartite
systems is the inequalities of HEE. The inequalities of the entanglement entropy are important in both classical and quantum information theories. Some of the inequalities are rather difficult to prove
in the frame of field theory. Recently, it has been shown that many of them
are reasonably easier by using holographic correspondence.

Before focusing on some particular inequalities, let us introduce a useful
property of the inequalities of HEE. In a multipartite inequality, each term represents a HEE for an union of several disjointed entangled regions $ {A} _{ijk\cdots}$. This
term is defined as the minimum of the HEE for all possible configurations with connected, disconnected or partially connected entanglement wedges. Thus, before verifying the inequality, we have
to calculate the HEE for all possible configurations to find out the actual
configuration for each term. However, at least for the tripartite system, it can be shown that if an inequality of HEE satisfies all its terms being in the connected configuration, then other inequalities would also follow from other configurations. Therefore, we only need to consider the inequalities of HEE with all the terms being in
the connected configurations.

For example. For a union of three regions $ {A} _{ijk}\equiv {A} _{i}\cup {A} _{j}%
\cup {A} _{k}$ with $i<j<k$, the HEE of $ {A} _{ijk}$
corresponding to the connected entanglement wedge is
\begin{equation}
S_{ijk}^{c}=S_{\left[  ik\right]  }+S_{\left\langle ij\right\rangle
}+S_{\left\langle jk\right\rangle },
\end{equation}
where the notations have been defined in Eqs.(\ref{small region}) and (\ref{large region}). Remember $S_{\left\langle
ij\right\rangle }$ and $S_{\left\langle jk\right\rangle }$ must be in the sunset phase to ensure the entanglement wedge of $ {A} _{ijk}$ is connected. While $S_{\left[  ik\right]  }$ can be in any one of the three phases.

\subsubsection{Strong subadditivity}
\begin{figure}
    \includegraphics[width=.4\linewidth]{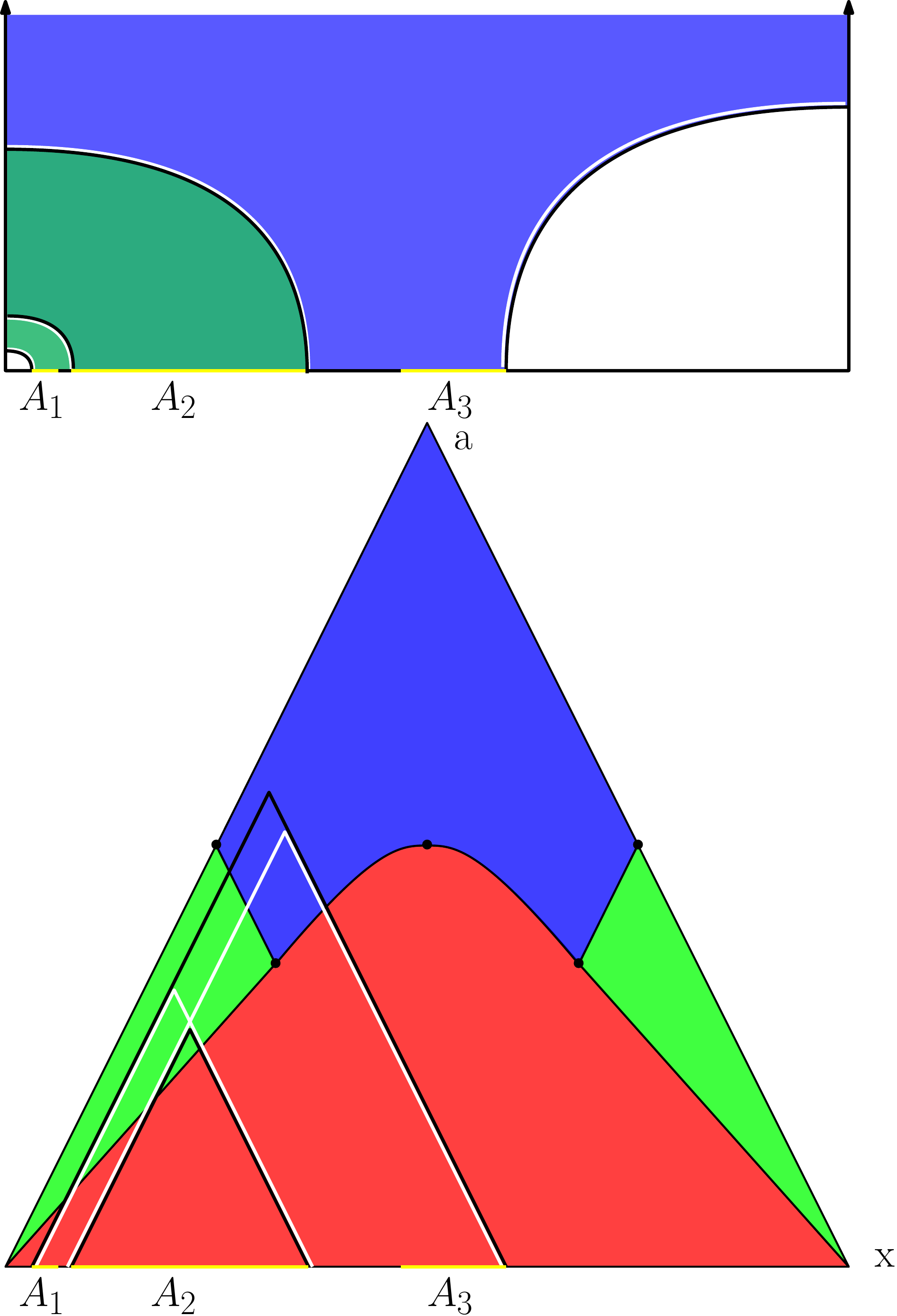}
  \caption{The example that $S_{2}$ is not in the sunset phase.}
  \label{S2-example}
\end{figure}

The strong subadditivity (SSA) in a tripartite system reads,
\begin{align}
S_{12}+S_{23} & \ge S_{123}+S_{2}. \label{SSA}
\end{align}
We first consider the case that $S_{12}$, $S_{23}$ and $S_{123}$ are all in the connected configuration. In this case, we need to prove the following inequality,
\begin{align}
S_{12}^{c}+S_{23}^{c} & \ge S_{123}^{c}+S_{2}, \label{SSAc}
\end{align}
where the HEE's with the connected entanglement wedges are
\begin{align}
S_{12}^{c} &  =S_{\left[  12\right]  }+S_{\left\langle 12\right\rangle },\label{S12c}\\
S_{23}^{c} &  =S_{\left[  23\right]  }+S_{\left\langle 23\right\rangle }, \label{S23c}\\
S_{123}^{c} &  =S_{\left[  13\right]  }+S_{\left\langle 12\right\rangle
}+S_{\left\langle 23\right\rangle }.\label{S123c}
\end{align}
Plugging Eqs.(\ref{S12c} - \ref{S123c}) into Eq.(\ref{SSAc}), the SSA becomes
\begin{align}
S_{\left[  12\right]  }+S_{\left[  23\right]  } & \ge S_{\left[  13\right]  }+S_{2}. \label{SSAf}
\end{align}
It is easy to show that $S_2$ must be in the sunset phase, otherwise the SSA reduces to a trivial equality. Actually, if $S_2$ is not in the sunset phase, all other regions, $S_{\left[  12\right]  }$, $S_{\left[  23\right]  }$ and $S_{\left[  13\right]}$ could not be in the sunset phase either, and the both sides of HEE are exactly the same. Fig.\ref{S2-example} is an example of $S_2$ being in the rainbow phase.

\begin{table}[htbp]
	\centering
	\caption{The ten allowed cases for strong subadditivity inequality.}
	\label{SSAtable}
\setlength{\tabcolsep}{7mm}{
\begin{tabular}[c]
{|c|c|c|c|c|}\hline
choice & $S_2$ & $S_{[12]}$ & $S_{[23]}$ & $S_{[13] }$ %
\\\hline
sss & sunset & sunset & sunset & sunset\\\hline
ssr & sunset & sunset & sunset & rainbow\\\hline
srr & sunset & sunset & rainbow & rainbow\\\hline
rsr & sunset & rainbow & sunset & rainbow\\\hline
rrr & sunset & rainbow & rainbow & rainbow\\\hline
ssk & sunset & sunset & sunset & sky\\\hline
rsk & sunset & rainbow & sunset & sky\\\hline
ksk & sunset & sky & sunset & sky\\\hline
rkk & sunset & rainbow & sky & sky\\\hline
kkk & sunset & sky & sky & sky\\\hline
\end{tabular}}
\end{table}

Given that $S_2$ is in the sunset phase, the other three regions, $S_{\left[  12\right]  }$, $S_{\left[  23\right]  }$ and $S_{\left[  13\right]}$ could be any of the three phases that leads to $3^{3}=27$ cases. However, based on the rules we list in the last section, only ten of them are allowed as shown in table \ref{SSAtable}, in which the three letters represent the phases of $S_{\left[  12\right]  }$, $S_{\left[  23\right]  }$ and $S_{\left[  13\right]}$ since the phase of $S_2$ is always in the sunset phase.

\begin{figure}[h]
    \includegraphics[width=.4\linewidth]{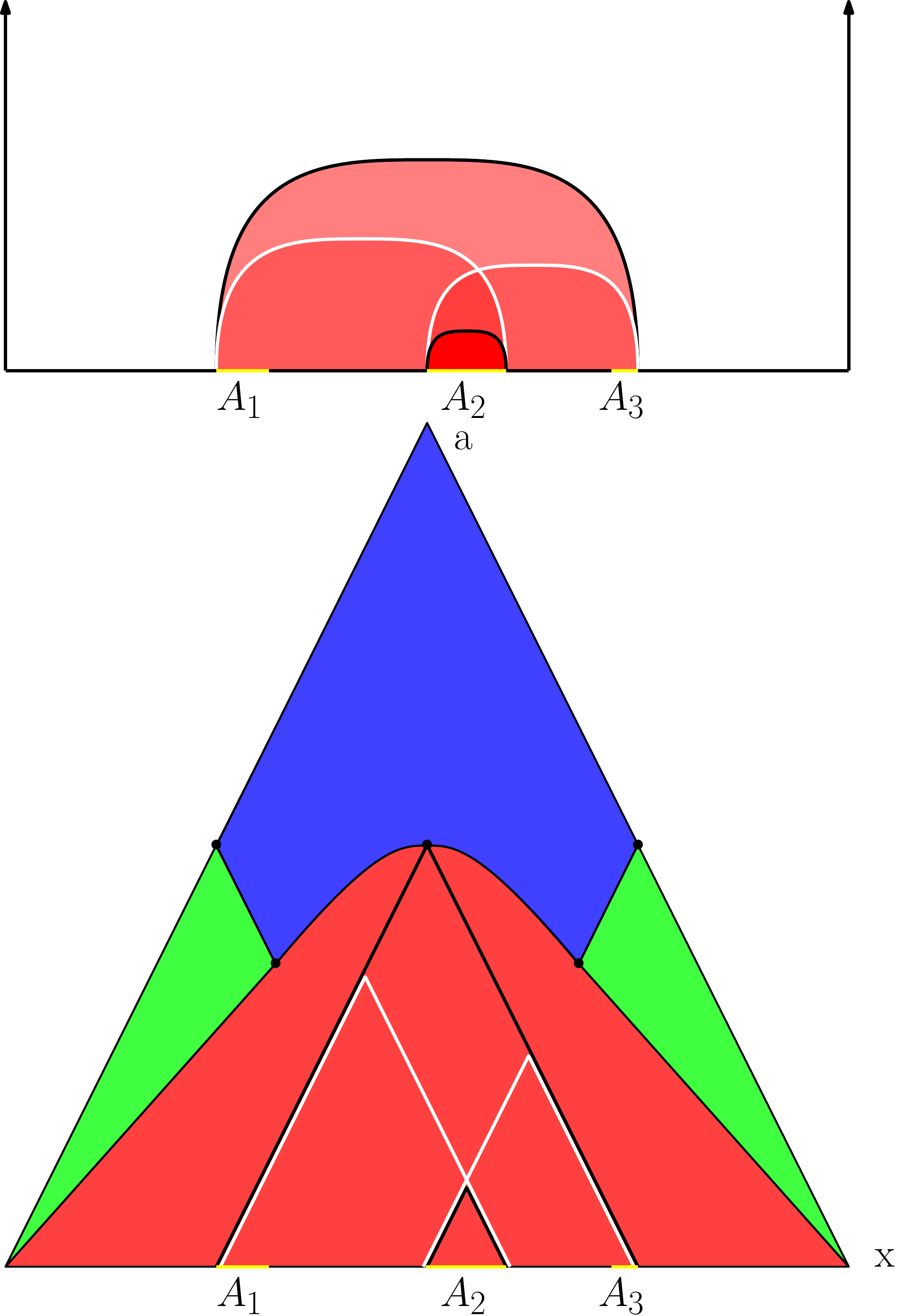}
  \caption{The case of $S_{\left[  12\right]  }$, $S_{\left[  23\right]  }$ and $S_{\left[  13\right]}$ being in the sunset phase, i.e sss.}
  \label{fig_sss}
\end{figure}

Among the ten cases, the simplest one is that $S_{\left[  12\right]  }$, $S_{\left[  23\right]  }$ and $S_{\left[  13\right]}$ are all in the sunset phase, namely sss. The corresponding entanglement wedge and the phase diagram are plotted in in Fig.\ref{fig_sss}. We plot the RT surfaces anchored on the boundaries of the entangled region in the upper part, and the associated critical triangles in the phase diagram in the lower part. The RT surfaces and the associated critical triangles of $S_{[12]}$ and  $S_{[23]}$ are labelled with white color, and that of $S_{[13]}$ and $S_2$ are labelled with black color. The entanglement wedges are filled with the same color of their corresponding phases. For example, the case of all $S_{\left[  12\right]  }$, $S_{\left[  23\right]  }$ and $S_{\left[  13\right]}$ being in the sunset phase, i.e sss, is plotted in Fig.\ref{fig_sss}.

To prove the SSA in the form of Eq.(\ref{SSAf}), we need to show that the sum of the two white curves is larger than the sum of the two black curves in the upper part of Fig.\ref{fig_sss}. To do so, we cut the two white curves at their intersection, and rejoin them to two joint curves. One of the joint curves is in the same homology with the entanglement wedge $A_2$, and the other is in the same homology with the entanglement wedge $A_{[13]}$. According to the RT perspective, it is straightforward to see that the joint curve in homology with $A_2$ must be larger than the HEE $S_2$ which is the minimal surface of the entangled region $A_2$ by definition. In this simple case, it is also straightforward to see that the other joint curve is larger than $S_{\left[  13\right]}$ by definition. We thus proved the inequality Eq.(\ref{SSAc}) in the sss case. Actually, this simple case is the same as the HEE in CFT without boundary. The presence of the boundary induces two new phases - sky and rainbow, which apparently modifies the above discussion.

\begin{figure}[h]
  \subfloat[srr]{
    \includegraphics[width=.25\linewidth]{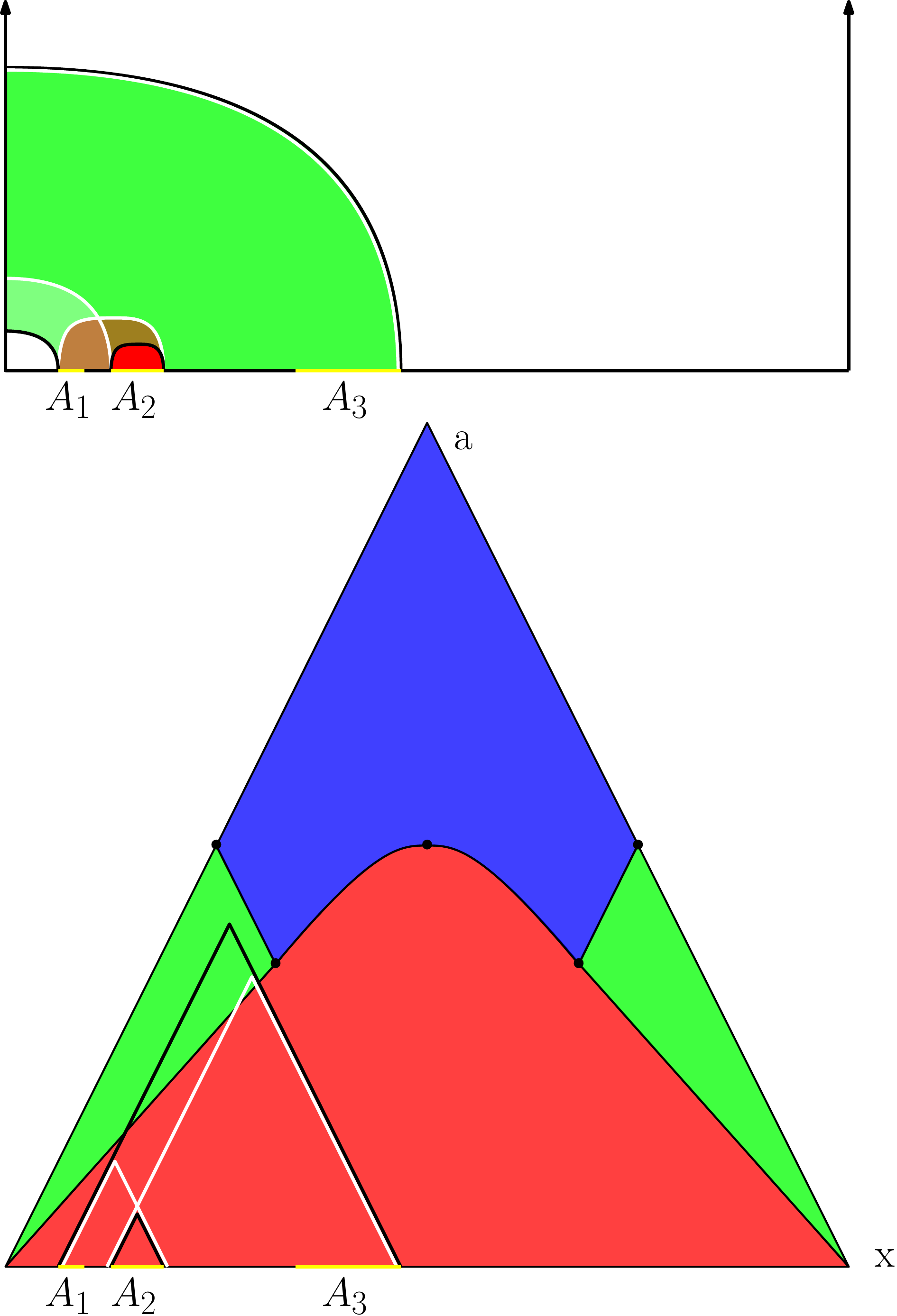} }\hspace{0.5cm}
  \subfloat[rsr]{
    \includegraphics[width=.25\linewidth]{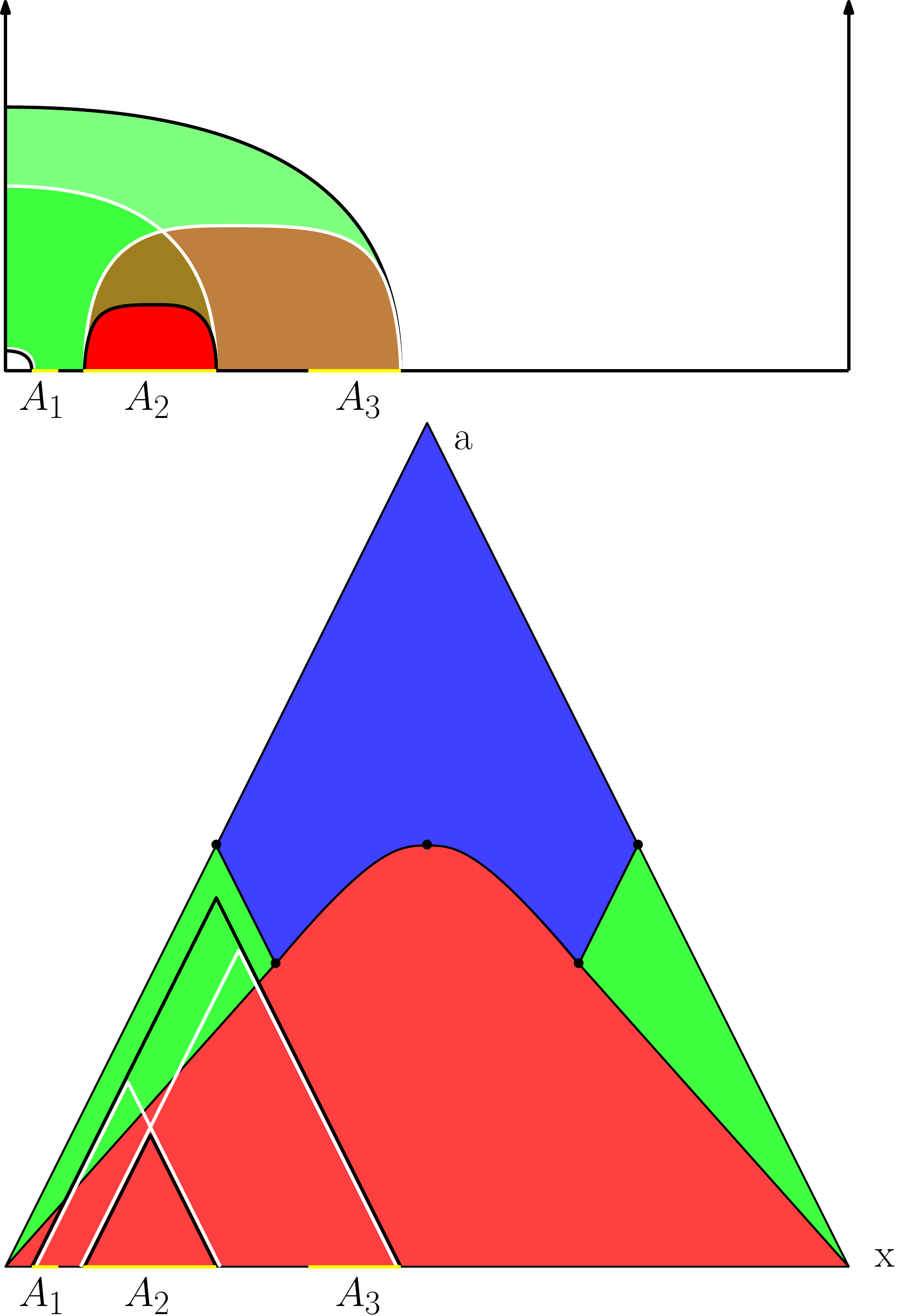} }\hspace{0.5cm}
  \subfloat[ksk]{
    \includegraphics[width=.25\linewidth]{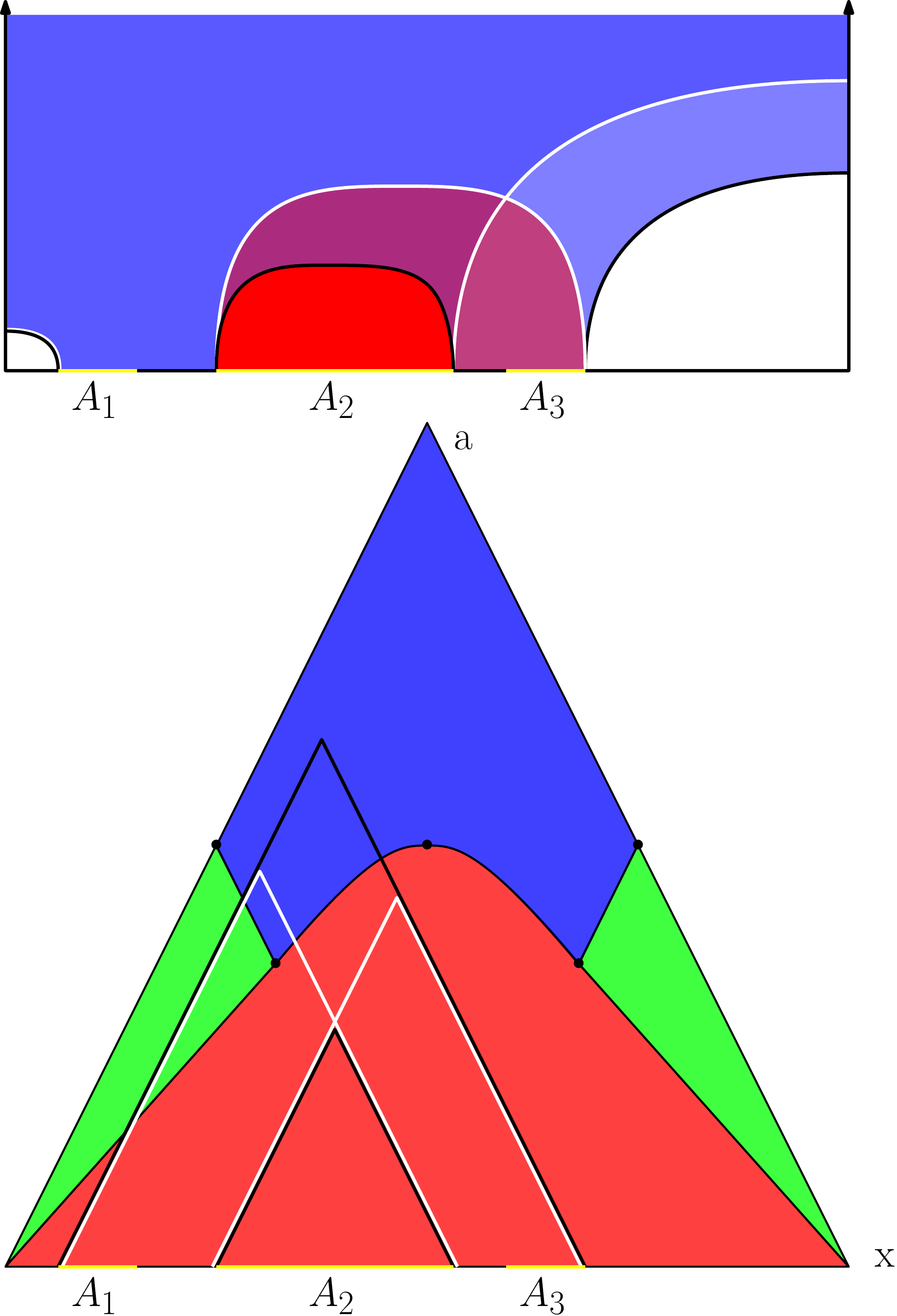} }\hspace{0.5cm}
  \caption{The cases of srr, rsr and ksk.}
  \label{fig_srr}
\end{figure}

The cases of srr, rsr and ksk are plotted in Fig.\ref {fig_srr}. There are three black curves and three white curves in each case, and we need to prove that the sum of the three white curves is larger than the sum of the three black ones. However, there is a black curve that overlaps with a white one because their corresponding entangled regions share a part of the RT surface. For example, in the srr case, both $S_{[23]}$ and  $S_{[13]}$ are in the rainbow phase, and they share a quarter-circle-shaped curve anchored on the right boundary of the entangled region $A_3$ as a part of their RT surface.  Since we want to show that the sum of the white curves is larger than the sum of the black curves, the overlapped black and white curves will cancel out each other and can be ignored in the discussion.

Ignoring the pair of the overlapped black and white curves, we cut the other two white curves at their intersection, and rejoin them to two joint curves. Similarly, the sss case, one of the joint curves is in the same homology with $A_2$, and it must be larger than $S_2$ by definition. In addition, it is easy to see that the other joint curve is larger than the quarter-circle-shaped curve anchored on the left boundary of $A_1$ in the srr case, or on the right boundary of $A_3$ in the rsr and ksk cases. On the other hand, the sum of this quarter-circle-shaped curve and the curve that we have ignored is just $S_{[13]}$, We thus prove the inequality Eq.(\ref{SSAc}) in the cases of srr, rsr and ksk.

\begin{figure}[h]
  \subfloat[ssr]{
    \includegraphics[width=.25\linewidth]{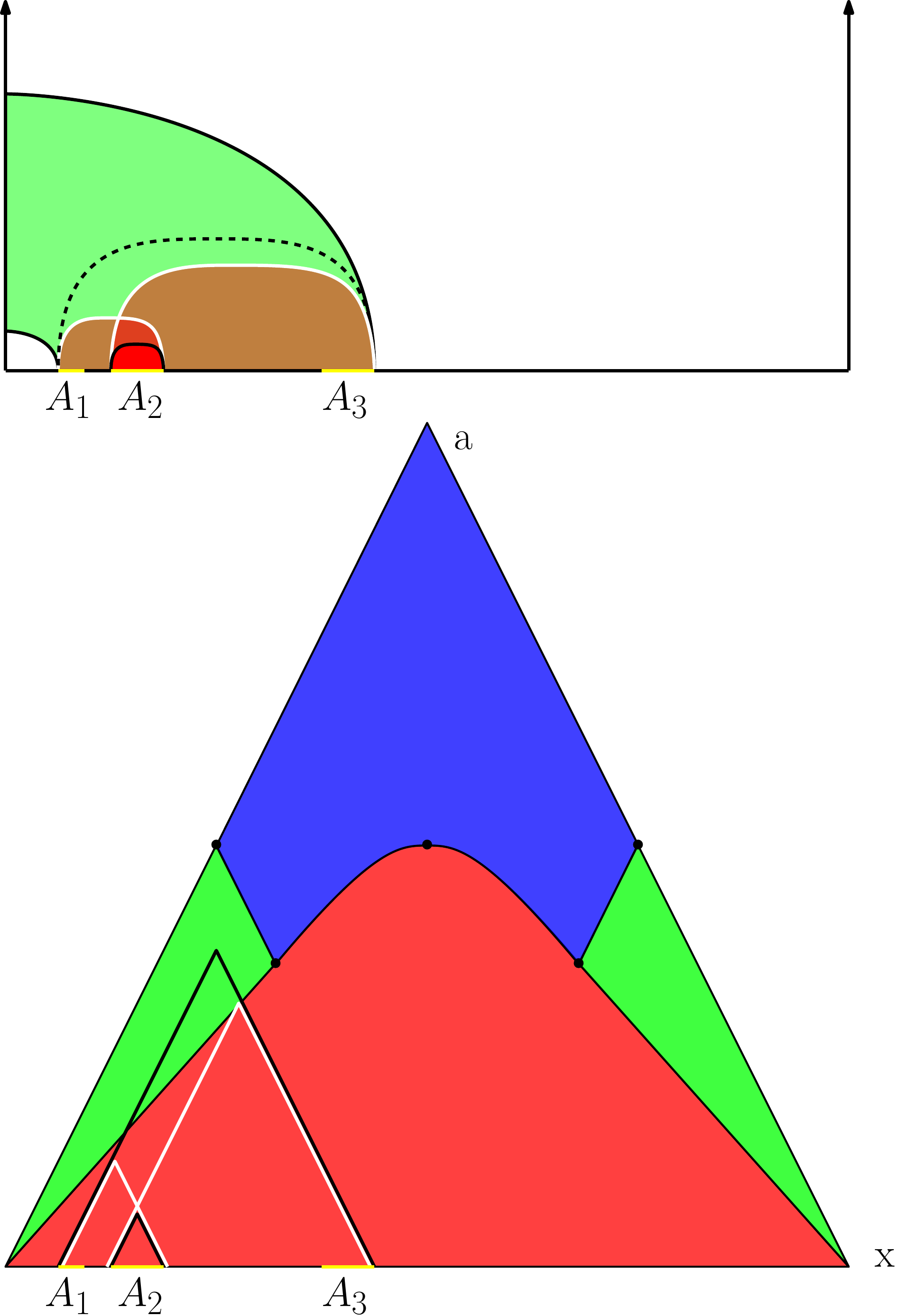} }\hspace{0.5cm}
  \subfloat[ssk]{
    \includegraphics[width=.25\linewidth]{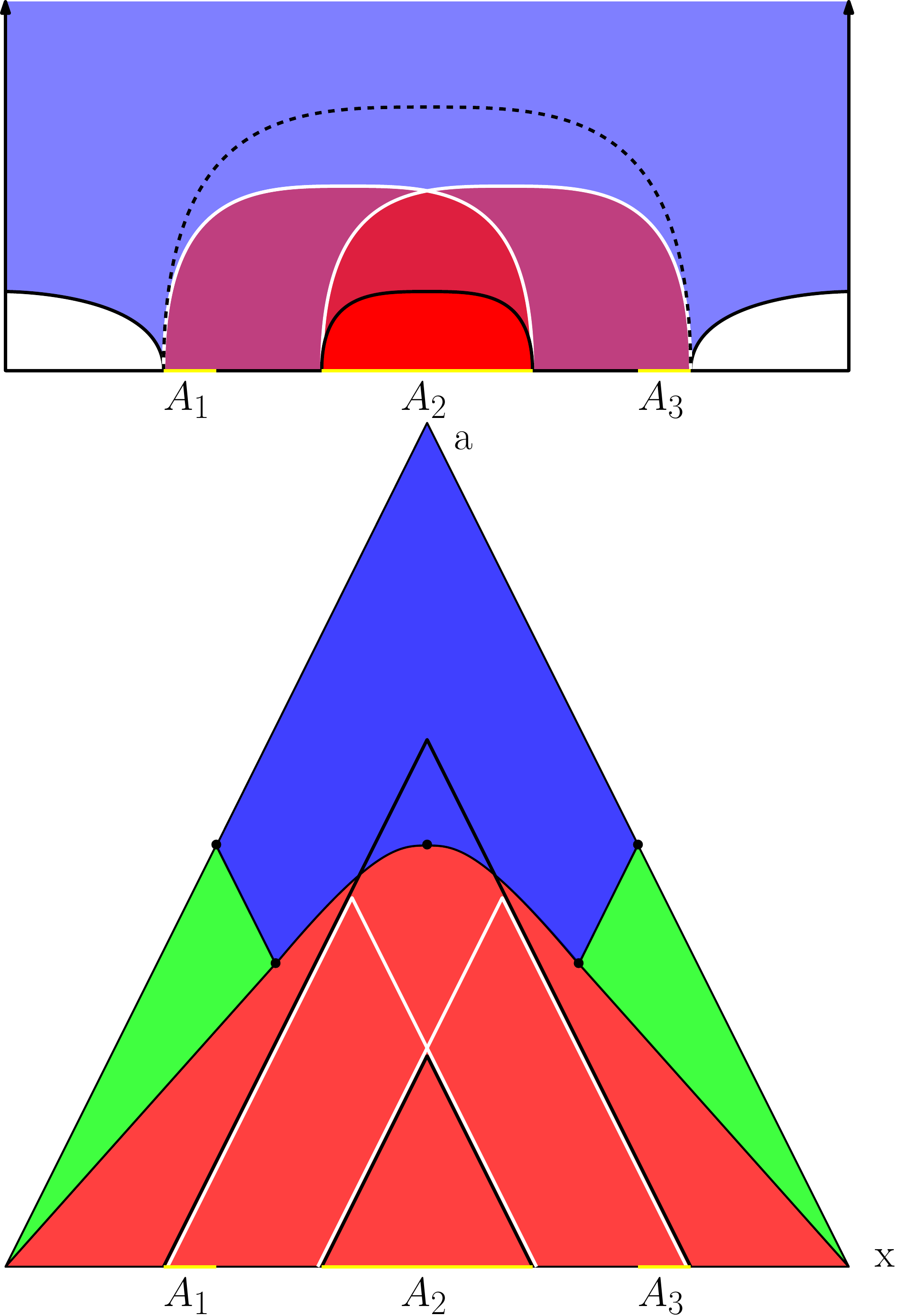} }\hspace{0.5cm}
  \subfloat[rsk]{
    \includegraphics[width=.25\linewidth]{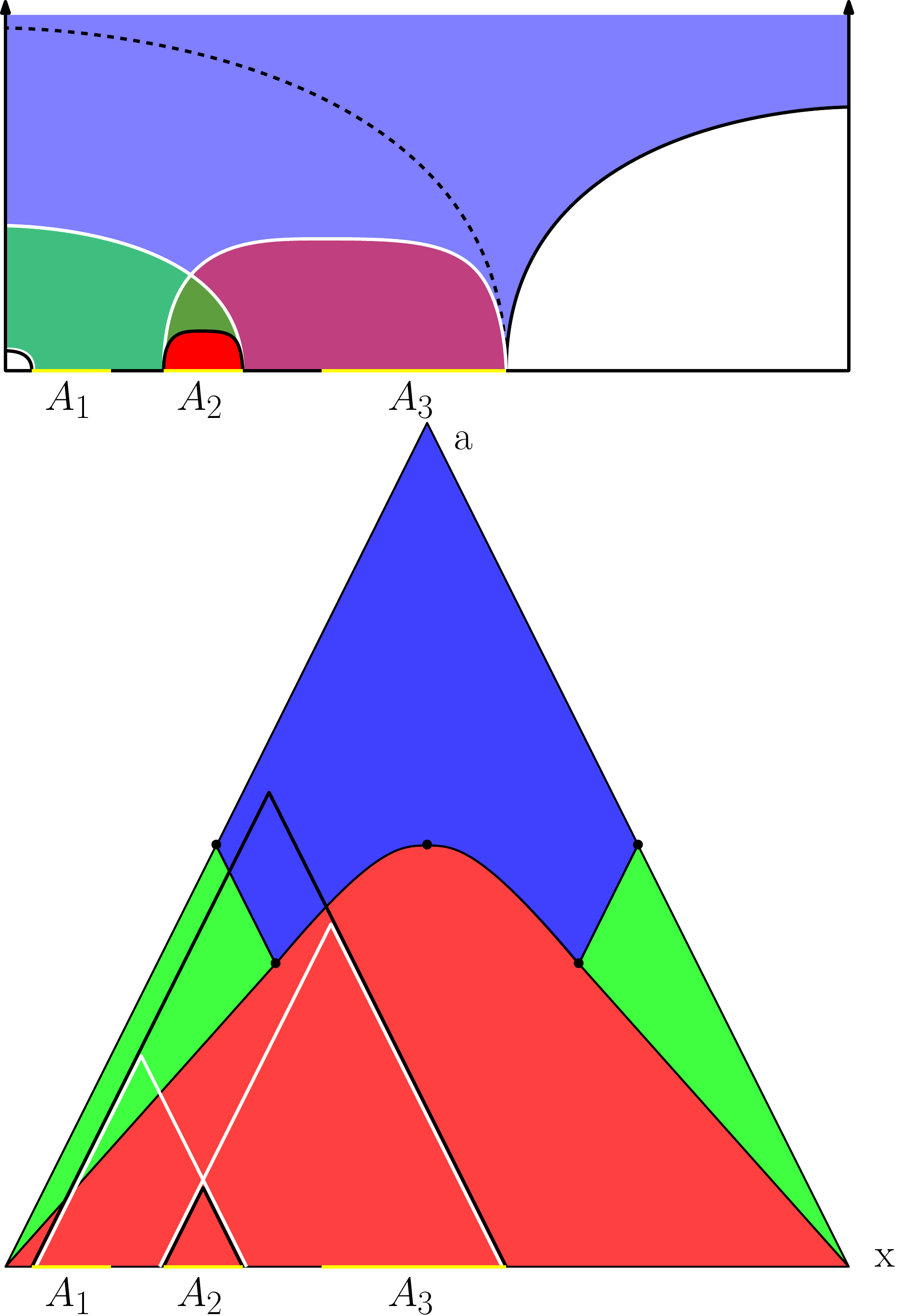} }\hspace{0.5cm}
  \caption{The cases of ssr, ssk and rsk. The dashed black curves are auxiliary.}
  \label{fig_ssr}
\end{figure}

Next, we consider the cases of ssr, ssk and rsk as shown in Fig.\ref{fig_ssr}. As before, we cut the other two white curves at their intersections\footnote{We have ignored the overlapped curves in ssr and rsk cases for the same reason that we argued before.}, and rejoin them to two joint curves.  Similarly, the joint curve in the same homology with $A_2$ is larger than $S_2$ by definition. On the other hand, to deal with the other joint curve, we need to add an auxiliary curve (the dashed black curve) in each case as shown in Fig.\ref{fig_ssr}. It is easy to see that the other joint curve is larger than the auxiliary curve. Furthermore, in the ssr and ssk cases, the auxiliary curve is just the $S_{\left[  13\right]}$ in the sunset phase, which is larger than the $S_{\left[  13\right]}$ in the rainbow or sky phase, as we assumed in these cases. In the rsk case, the sum of the auxiliary curve and the curve, which is ignored, is  the $S_{\left[  13\right]}$ in the rainbow phase. This sum is larger than $S_{\left[  13\right]}$ in the sky phase, as we assumed in this case. We thus prove the inequality Eq.(\ref{SSAc}) in the cases of ssr, ssk and rsk.

\begin{figure}
  \subfloat[rrr]{
    \includegraphics[width=.25\linewidth]{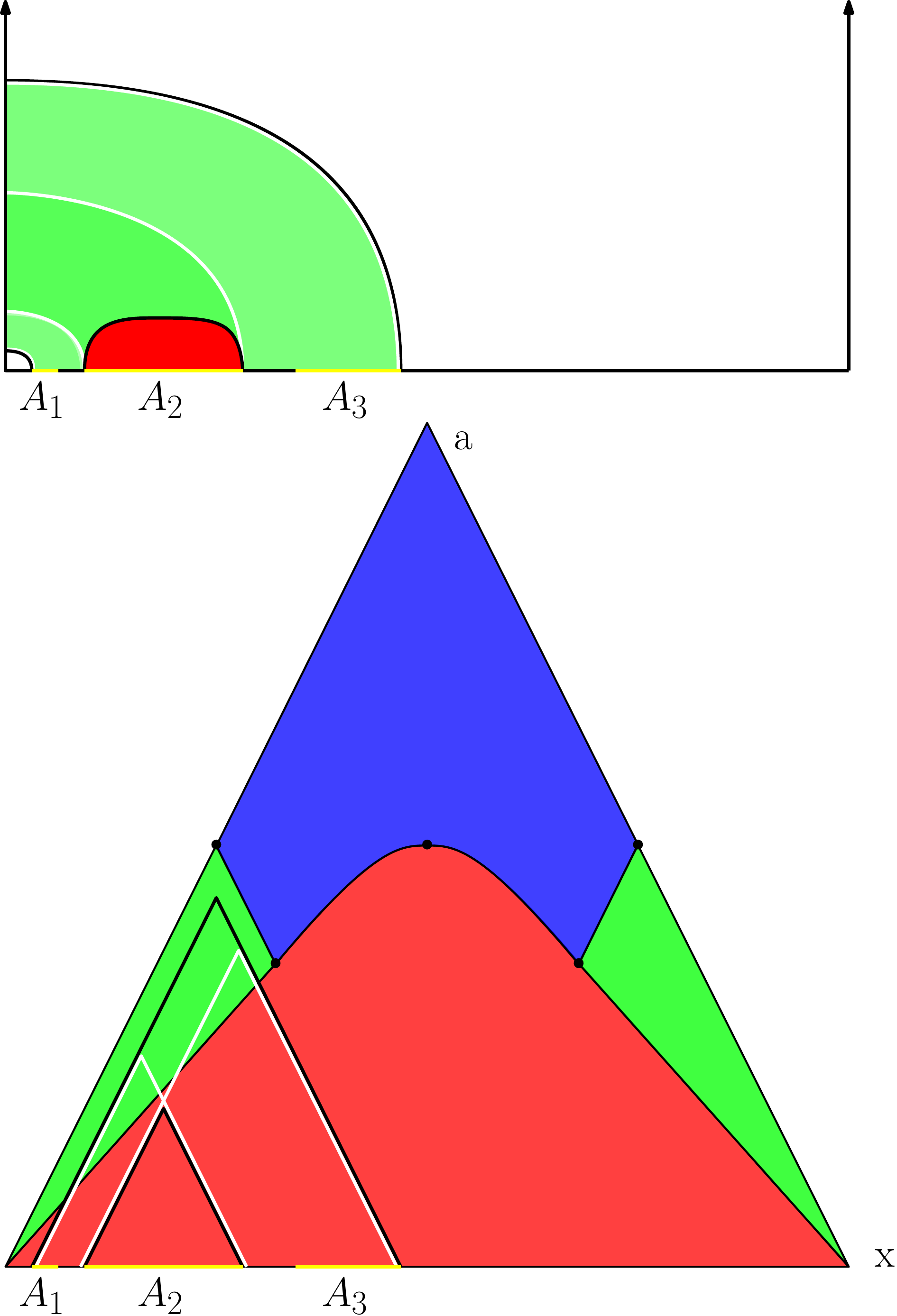} }\hspace{0.5cm}
  \subfloat[rkk]{
    \includegraphics[width=.25\linewidth]{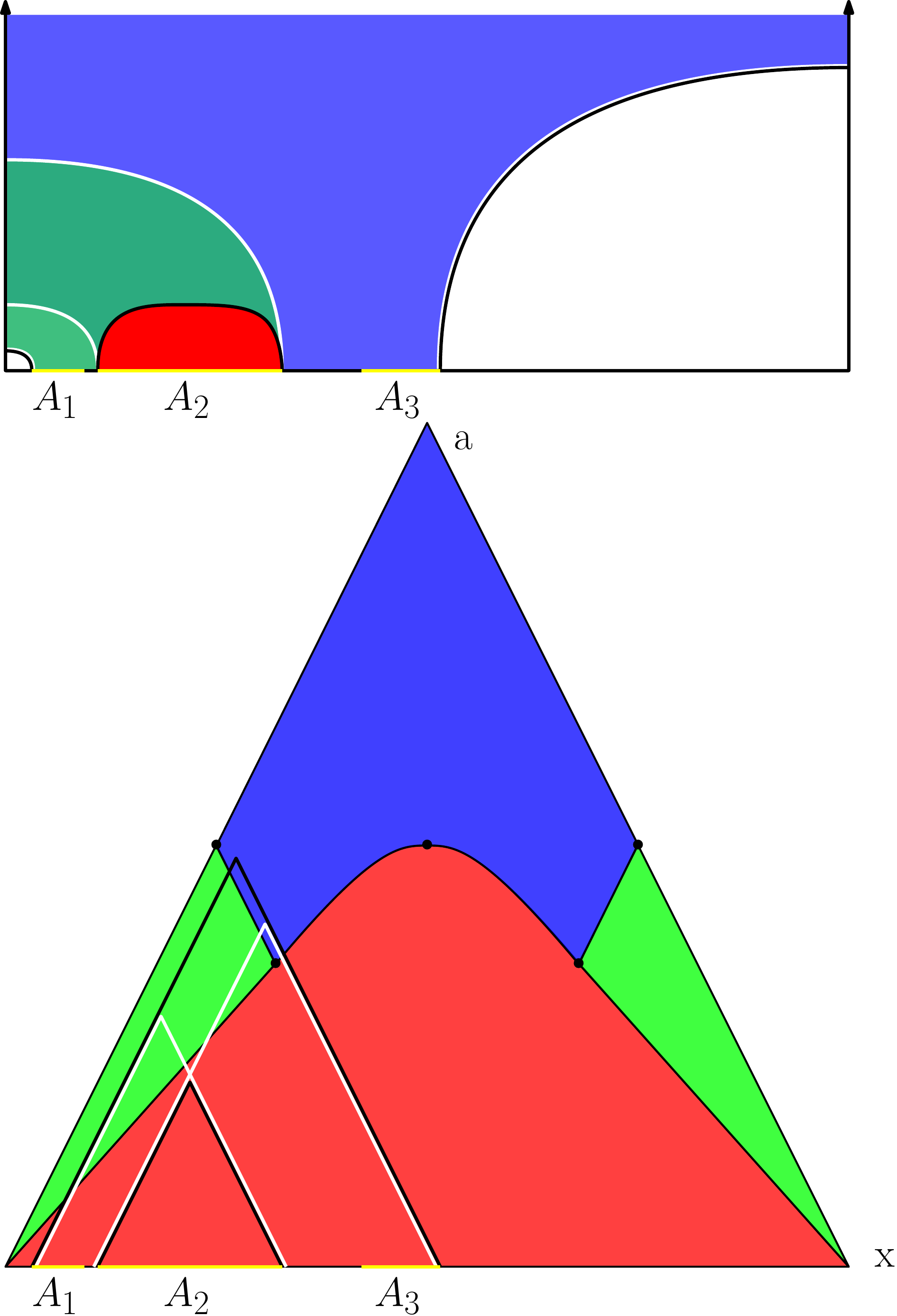} }\hspace{0.5cm}
  \subfloat[kkk]{
    \includegraphics[width=.25\linewidth]{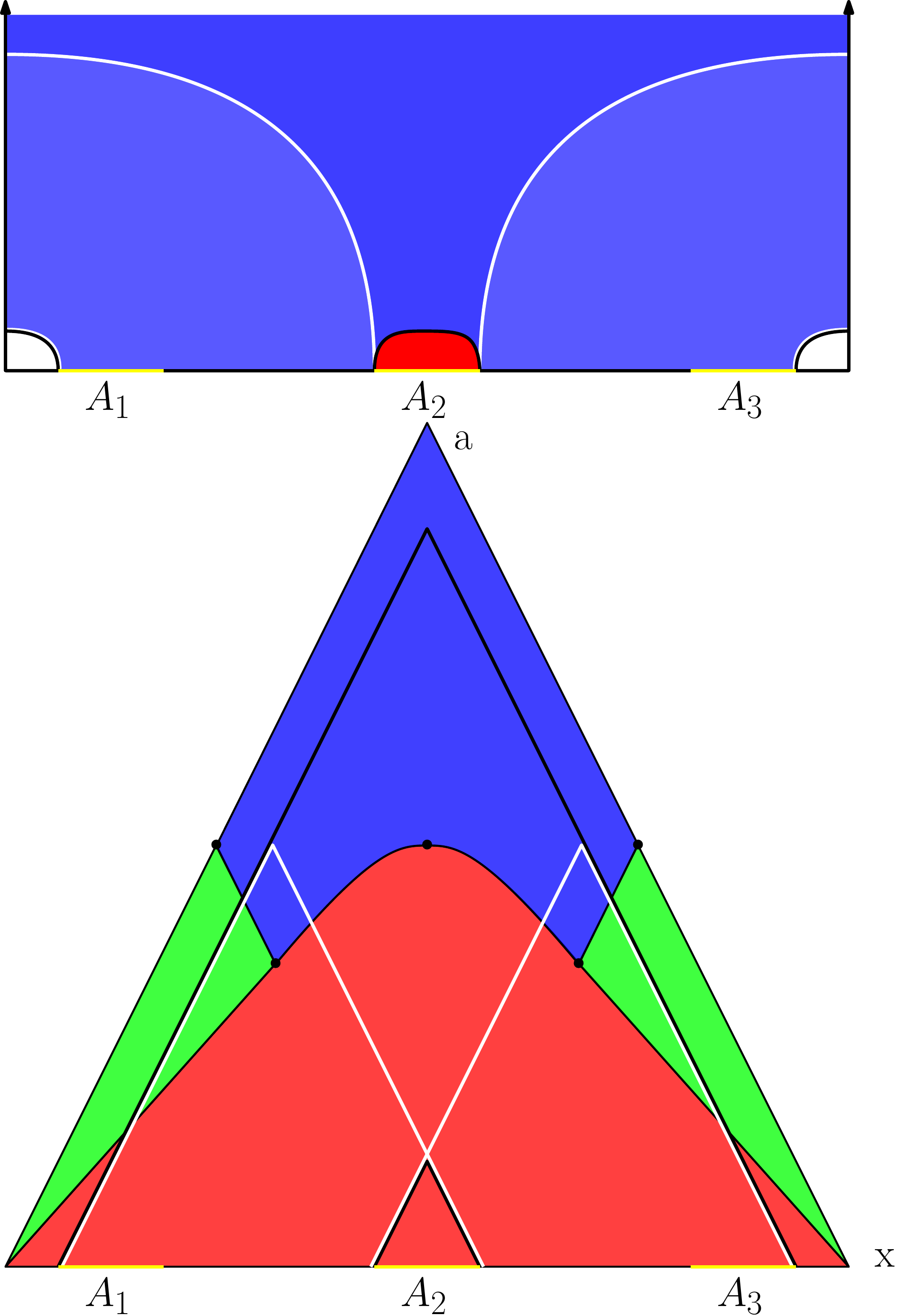} }\hspace{0.5cm}
  \caption{The cases of rrr, rkk and kkk.}
  \label{fig_rrr}
\end{figure}

Finally, the cases of rrr, rkk and kkk are shown in Fig.\ref{fig_rrr}. In these cases, the two pairs of the black and white quarter-circle-shaped curves, anchored on the left boundary of $A_1$ and the right boundary of $A_3$, overlap and cancel out each other. In addition, the other two white curves do not intersect, and the sum of them is just $S_2$ in the rainbow (in the rrr and rkk cases) or the sky phase (in the kkk case), which is larger than $S_2$ in the sunset phase as we assumed. We thus prove the inequality Eq.(\ref{SSAc}) in the cases of rrr, rkk and kkk.

We have proved the SSA with the assumption that $S_{12}$, $S_{23}$ and $S_{123}$ being all connected. Now let us consider the cases that not all of them are connected.

If $S_{12}$ is not in the connected configuration, i.e. $S_{12}=S_1+S_2$, then the SSA in Eq.(\ref{SSA}) reduces to
\begin{align}
S_{1}+S_{23} & \ge S_{123},
\end{align}
which can be proved using the SA by treating $ {A} _{23}$ as a single union region. Similar argument is also true for $S_{23}$. On the other hand, if both $S_{12}$ and $S_{23}$ are in the connected configuration, since we have proved the SSA in the connected case Eq.(\ref{SSAc}) , using Eqs.(\ref{S12c}) and (\ref{S23c}) as well as SA, we can show that
\begin{align}
S_{123}^{c} &\le S_{1}+S_{23}^{c} ,\\
S_{123}^{c} &\le S_{12}^{c}+S_{3},\\
S_{123}^{c} &\le S_{1}+S_{3}+S_{3},
\end{align}
which justifies that $S_{123}$ must be in the connected configuration. We thus complete the proof of the SSA in Eq.(\ref{SSA}) for BCFT.

\subsubsection{Monogamy of mutual information}
Another important inequality of HEE is the monogamy of mutual information (MMI),
\begin{equation}
S_{12}+S_{23}+S_{13}\geq S_{1}+S_{2}+S_{3}+S_{123}, \label{MMI}
\end{equation}
Similar to the case of SSA, we first prove that $S_{12}$, $S_{23}$, $S_{13 }$ and $S_{123  }$ are all in the connected configurations,
\begin{equation}
S_{12}^{c}+S_{23}^{c}+S_{13}^{c}\geq S_{1}+S_{2}+S_{3}+S_{123}^{c}, \label{MMIc}
\end{equation}
which can be expressed as follows by using the notation in Eqs.(\ref{S12c} - \ref{S123c}),
\begin{equation}
S_{\left[  12\right]  }+S_{\left[  23\right]  }+S_{\left\langle
13\right\rangle }\geq S_{1}+S_{2}+S_{3}.  \label{MMIcc}
\end{equation}
Among the six terms in Eq.(\ref{MMIcc}), $S_{\left\langle
13\right\rangle }$ must be in the sunset phase because $S_{13}$ is connected. In addition, $S_{\left\langle 13\right\rangle }$ being in the sunset phase induces that $S_2$ must also be in the sunset phase, since $ {A} _2$ is enclosed in $ {A} _{\left\langle 13\right\rangle }$. While the other four terms, $S_{[12]}$, $S_{[23]}$, $S_{1}$ and $S_{3}$, and $S_{\left[  23\right]  }$, could be in any of the three phases. The naive number of total choices is $3^4=81$, however, only 25 of them are allowed by using the rules we discussed in the last section. Accordingly, we list all the 25 cases in table \ref{MMItable}, in which the four letters represent the phases of  $S_{[12]}$, $S_{[23]}$, $S_{1}$ and $S_{3}$ respectively.
\begin{table}[h!]
	\centering
	\caption{The 25 allowed choices for monogamy of mutual information.}
	\label{MMItable}
    \setlength{\tabcolsep}{7mm}{
	\begin{tabular}
[c]{|c|c|c|c|c|}\hline
choice & $S_{\left[  12\right]  }$ & $S_{\left[
23\right]  }$ & $S_{1}$ & $S_{3}$\\\hline
ssss & sunset & sunset & sunset & sunset\\\hline
srss & sunset & rainbow & sunset & sunset\\\hline
srsr & sunset & rainbow & sunset & rainbow\\\hline
skss & sunset & sky & sunset & sunset\\\hline
sksr & sunset & sky & sunset & rainbow\\\hline
sksk & sunset & sky & sunset & sky\\\hline
rsss & rainbow & sunset & sunset & sunset\\\hline
rsrs & rainbow & sunset & rainbow & sunset\\\hline
rrss & rainbow & rainbow & sunset & sunset\\\hline
rrsr & rainbow & rainbow & sunset & rainbow\\\hline
rrrs & rainbow & rainbow & rainbow & sunset\\\hline
rrrr & rainbow & rainbow & rainbow & rainbow\\\hline
rkss & rainbow & sky & sunset & sunset\\\hline
rksr & rainbow & sky & sunset & rainbow\\\hline
rksk & rainbow & sky & sunset & sky\\\hline
rkrs & rainbow & sky & rainbow & sunset\\\hline
rkrr & rainbow & sky & rainbow & rainbow\\\hline
rkrk & rainbow & sky & rainbow & sky\\\hline
ksss & sky & sunset & sunset & sunset\\\hline
ksrs & sky & sunset & rainbow & sunset\\\hline
ksks & sky & sunset & sky & sunset\\\hline
kkss & sky & sky & sunset & sunset\\\hline
kksr & sky & sky & sunset & rainbow\\\hline
kkrs & sky & sky & rainbow & sunset\\\hline
kkrr & sky & sky & rainbow & rainbow\\\hline
\end{tabular}}
\end{table}

\begin{figure}
  \subfloat[ssss]{
    \includegraphics[width=.25\linewidth]{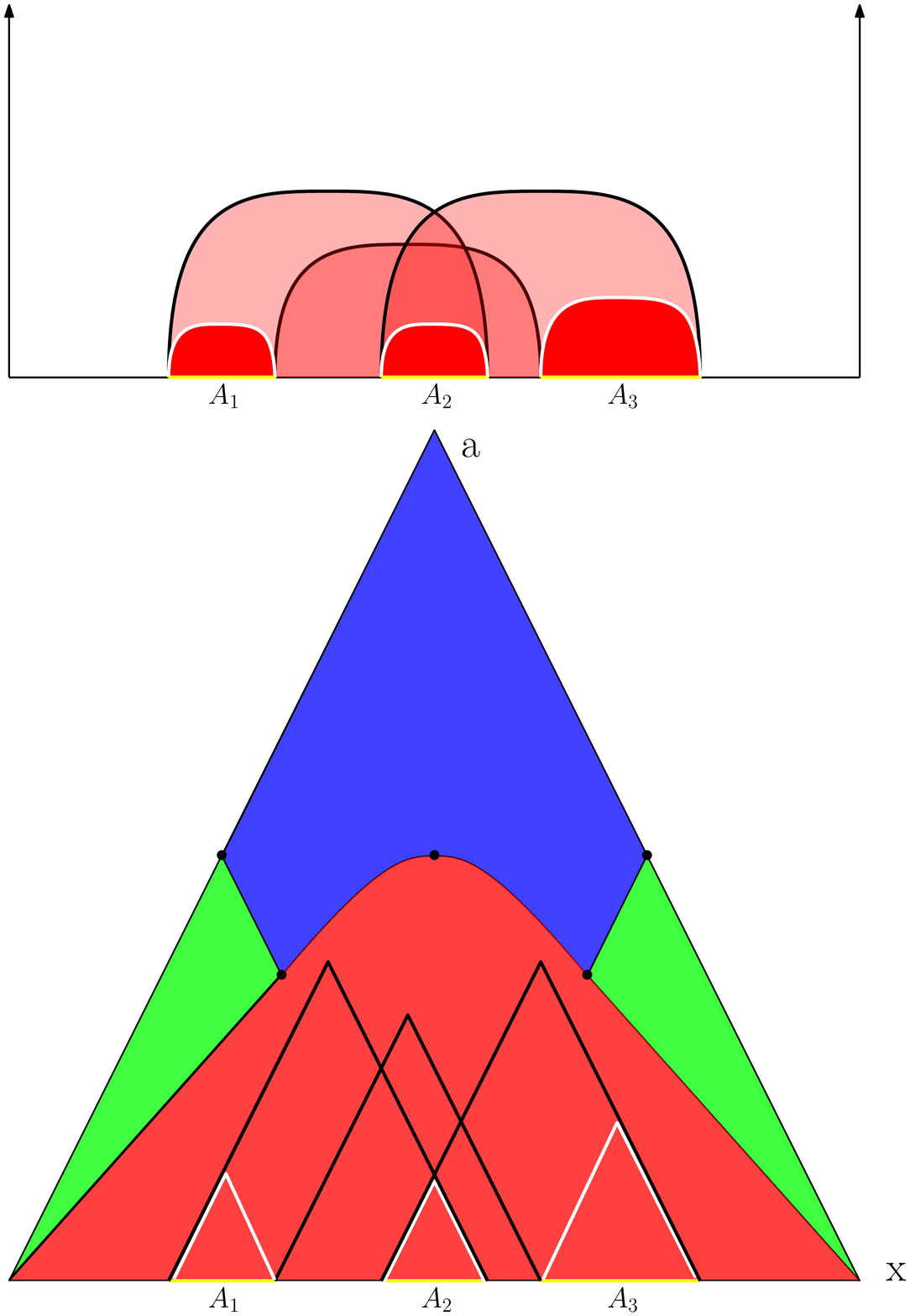} }\hspace{0.5cm}
  \subfloat[ksks]{
    \includegraphics[width=.25\linewidth]{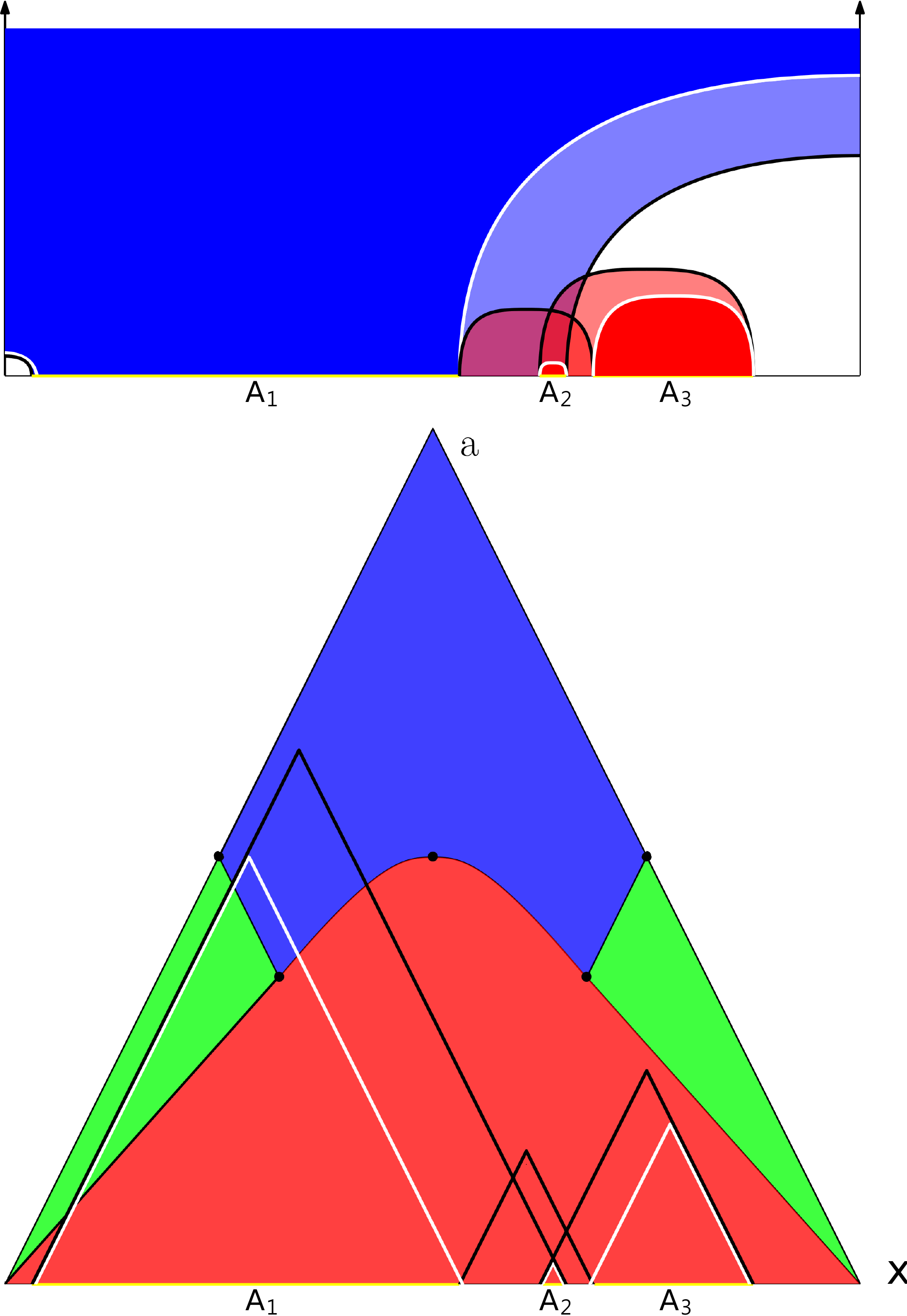} }\hspace{0.5cm}
  \subfloat[sksr]{
    \includegraphics[width=.25\linewidth]{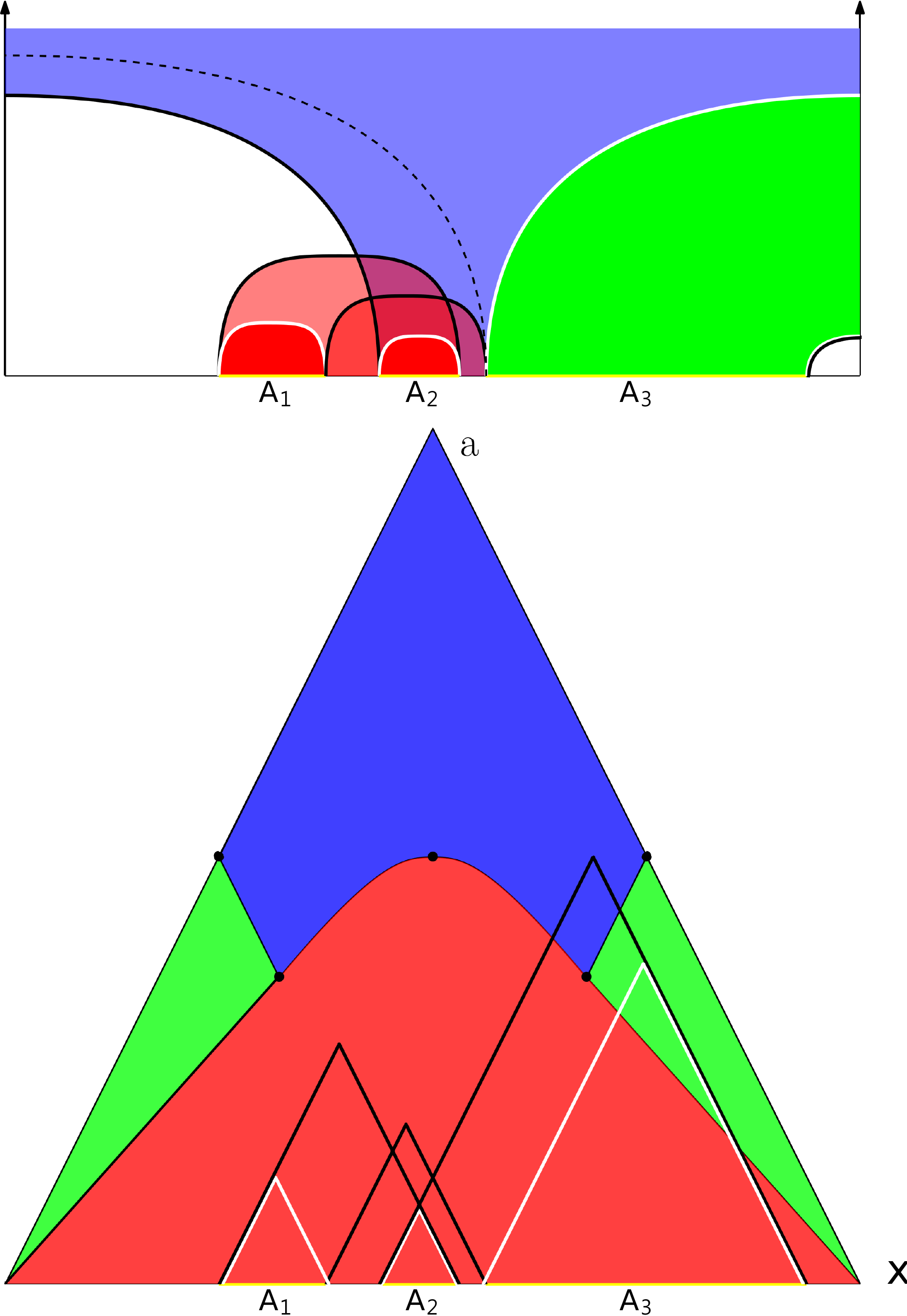} }\hspace{0.5cm}
  \caption{(a) The ssss case with $S_{\left[  12\right]  }$, $S_{\left[  23\right]  }$, $S_{1 }$ and $S_{3}$ in the sunset phase. (b) The ksks case with $S_{\left[  12\right]  }$ and $S_1$ are in the sunset phase, and $S_{\left[  23\right]  }$ and $S_{3}$ are in the sunset phase. (c) The sksr case with $S_{\left[  12\right]  }$ and $S_1$ are in the sunset phase, $S_{\left[  23\right]  }$ is in the sky phase and $S_{3}$ are in the rainbow phase.}
  \label{fig_ssss}
\end{figure}

The simplest case is that $S_{[12]}$, $S_{[23]}$, $S_{1}$ and $S_{3}$ are all in the sunset phase, i.e the ssss case. The phase diagram and the entanglement wedges corresponding to the ssss case are plotted in Fig.\ref{fig_ssss}(a). To prove the inequality Eq.(\ref{MMIcc}) in this case, we need to show that the sum of the three black curves is larger than the sum of the three white curves in the upper part of Fig.\ref{fig_ssss}(a). As in the proof of SSA, we cut the three black curves at their intersections, and rejoin them to three joint curves. The three joint curves are in the homology with the entangled regions $A_1$, $A_2$ and $A_3$ respectively. According to the RT perspective, it is straightforward to see that the joint curve in homology with $A_i$ must be larger than the HEE $S_i$ which is the minimal surface of the entangled region $A_i$ for $i=1, 2, 3$. We thus proved the inequality Eq.(\ref{MMIc}) in the ssss case. This simple case is the same as the MMI in CFT without boundary.

\begin{figure}
  \subfloat[srsr]{
    \includegraphics[width=.25\linewidth]{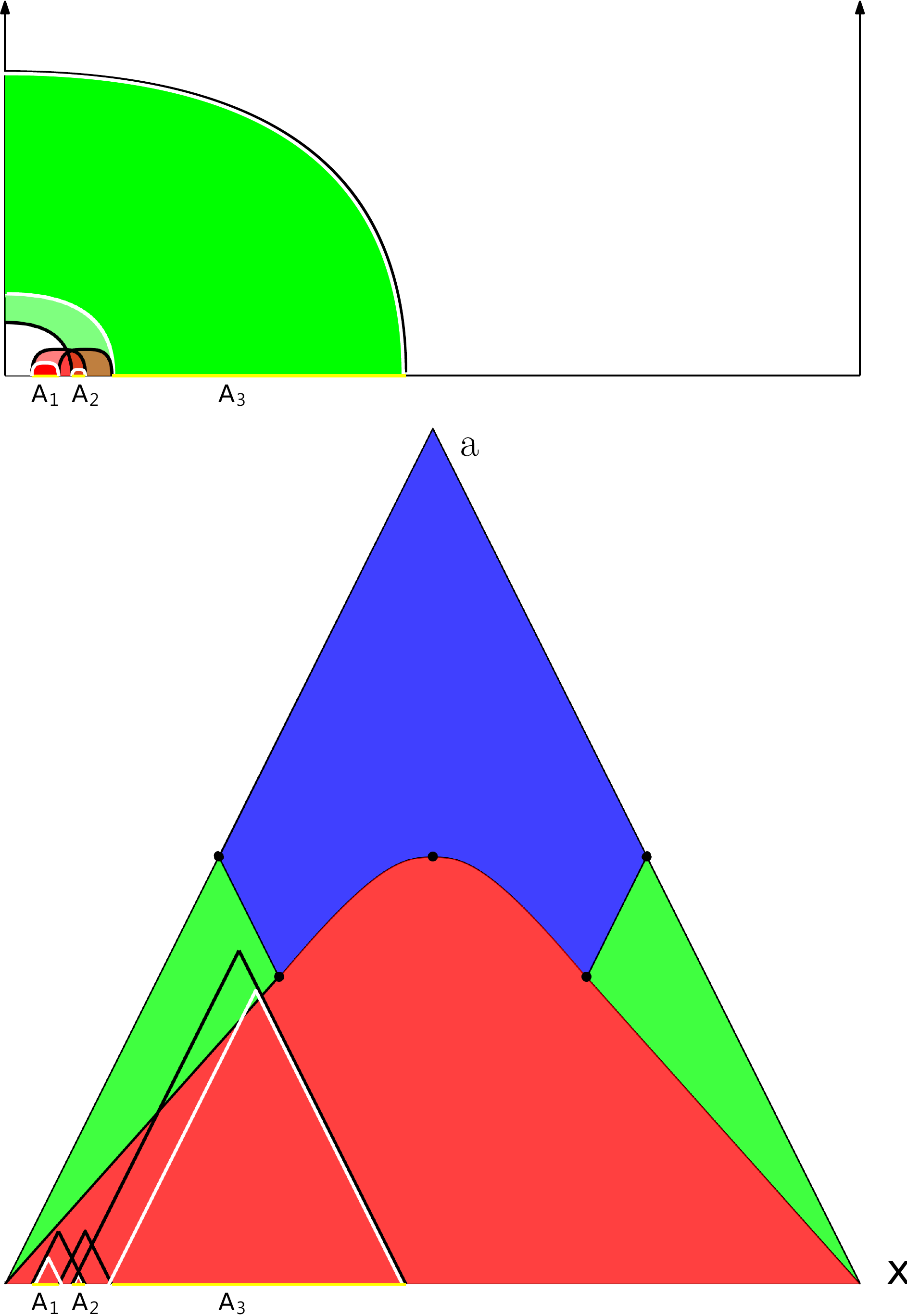} }\hspace{0.5cm}
  \subfloat[sksk]{
    \includegraphics[width=.25\linewidth]{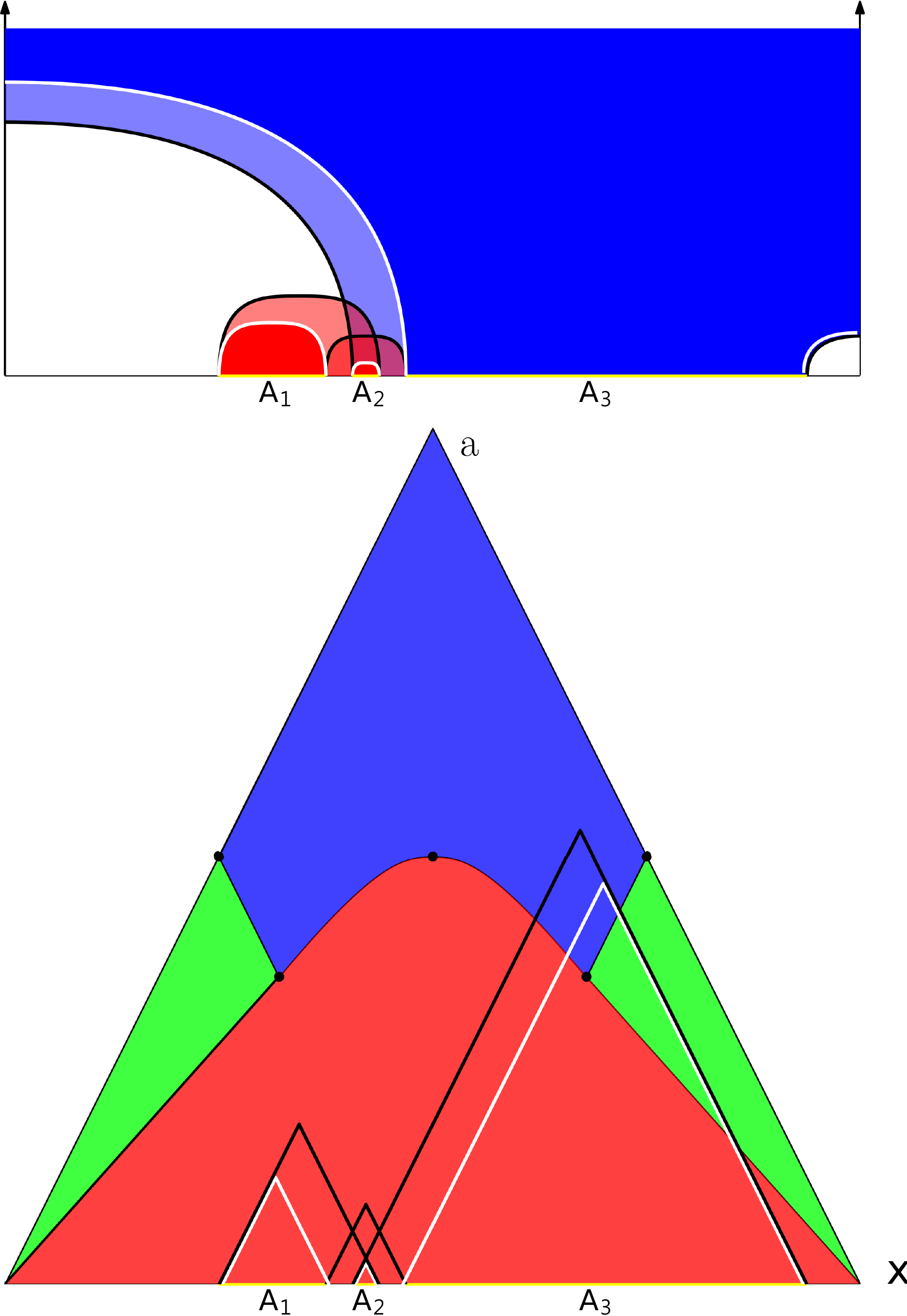} }\hspace{0.5cm}
  \subfloat[rsrs]{
    \includegraphics[width=.25\linewidth]{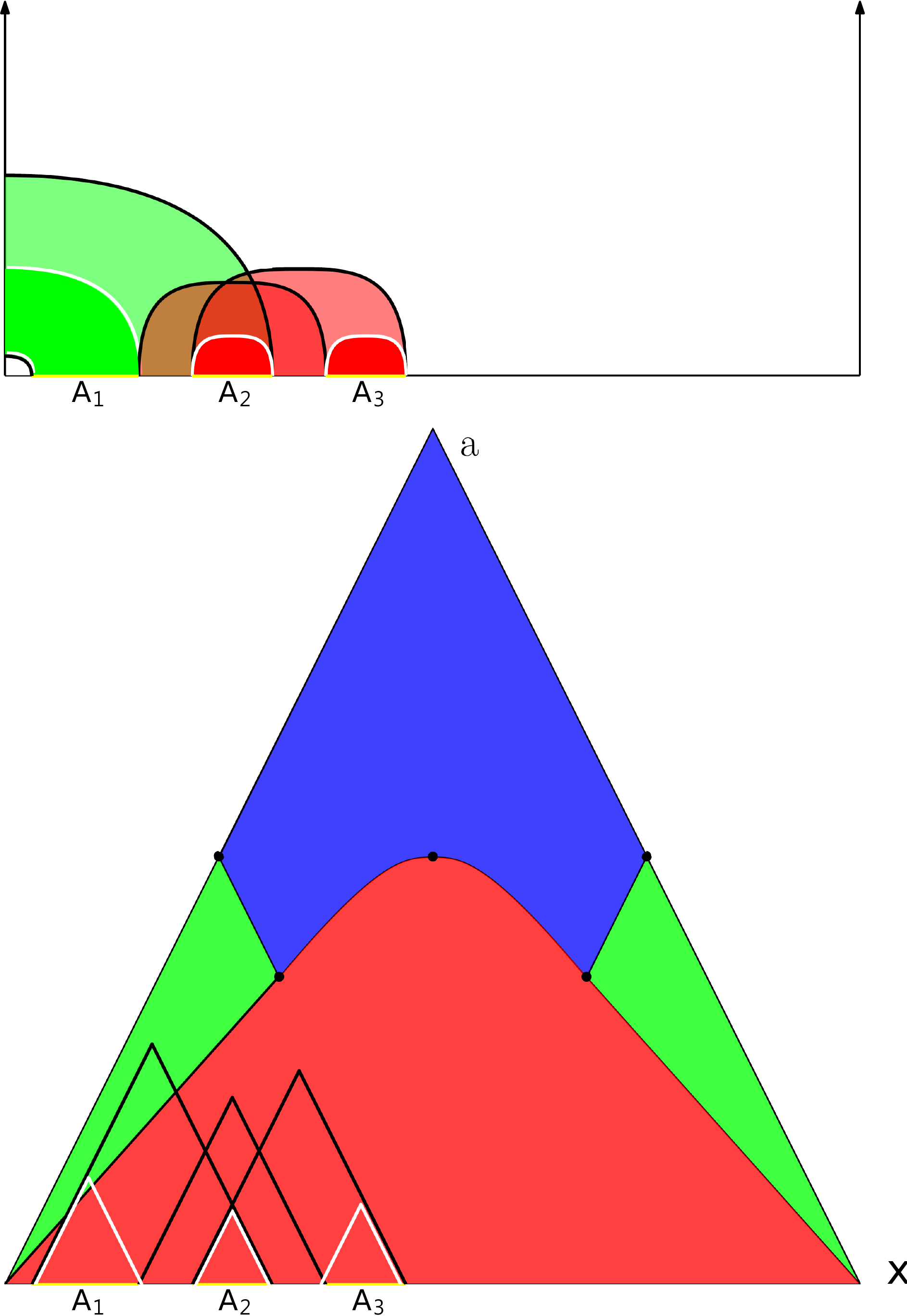} }\\
  \subfloat[rrrr]{
    \includegraphics[width=.25\linewidth]{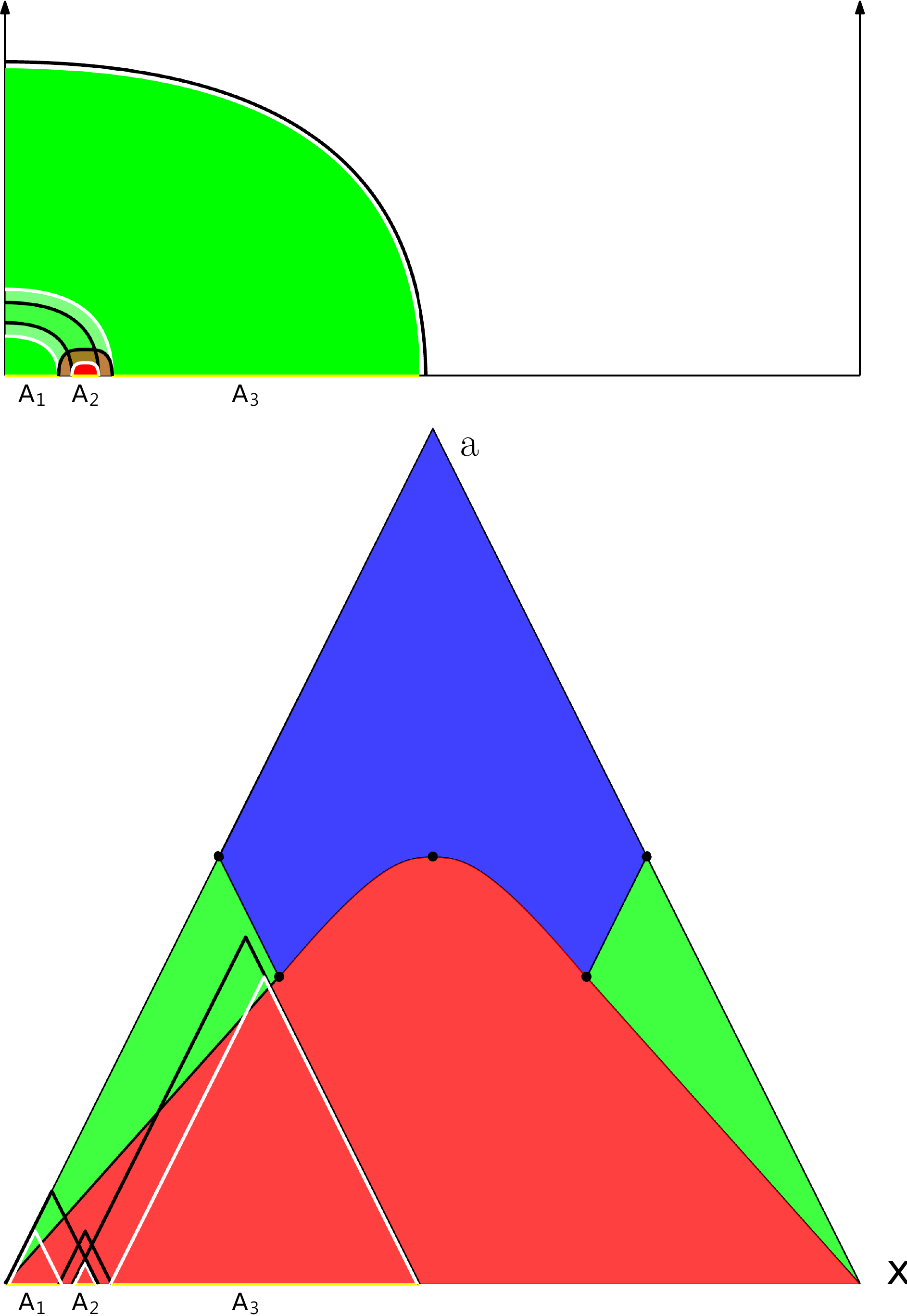} }\hspace{0.5cm}
  \subfloat[rkrk]{
    \includegraphics[width=.25\linewidth]{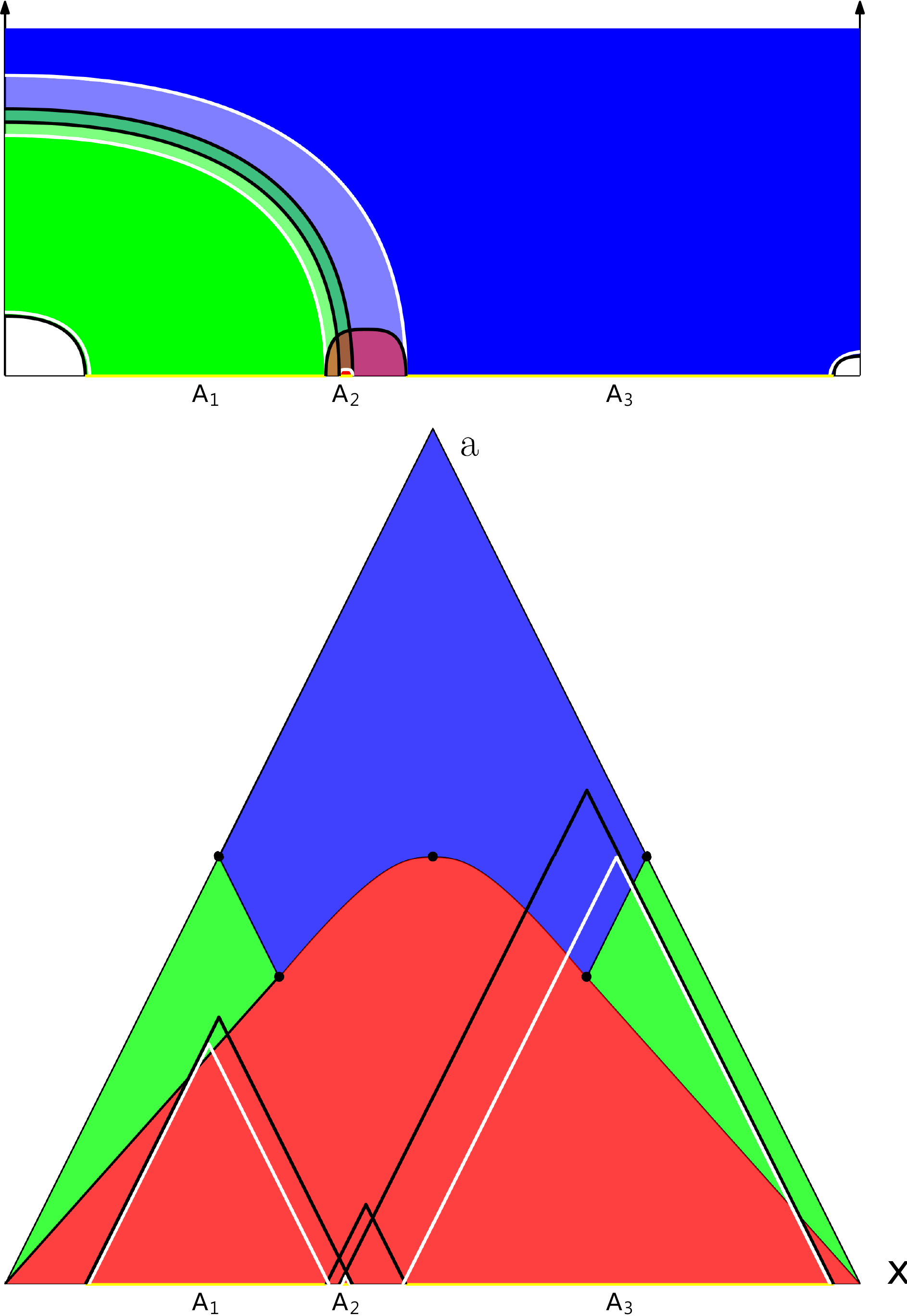} }\hspace{0.5cm}
  \subfloat[kkrr]{
    \includegraphics[width=.25\linewidth]{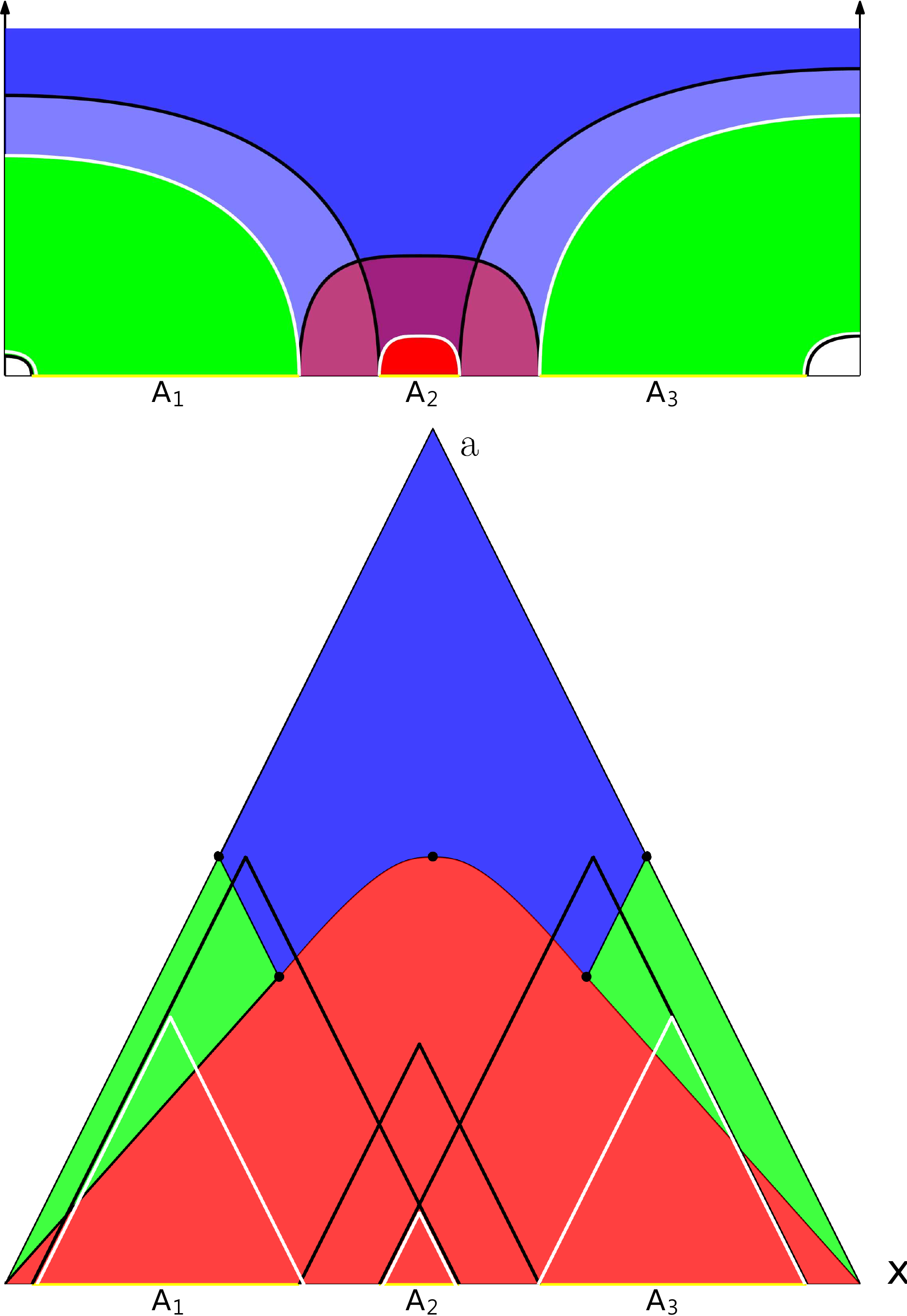} }
  \caption{The cases of srsr, sksk, rsrs, rrrr, rkrk, kkrr.}
  \label{fig_srsr}
\end{figure}

In the presence of boundaries, $S_{[12]}$, $S_{[23]}$, $S_{1}$ and $S_{3}$ could be in the rainbow or sky phase instead of the sunset phase. For example, the ksks case is plotted in Fig.\ref{fig_ssss}(b). According to the RT perspective, it is easy to see that the two joint curves in homology with $A_1$ and $A_3$ must be larger than their corresponding HEE's represented by the white curves. In addition, the third joint curve must larger than the white quarter-circle-shaped curve anchored on the right boundary of the entangled region $A_1$. Finally, because the two black and white quarter-circle-shaped curves anchored on the right boundary of the entangled region $A_3$ cancel out each other, we thus showed the sum of the black curves is larger than the sum of the white curves, which proves the inequality Eq.(\ref{MMIc}) in the ksks case. The similar cases also include srsr, sksk, rsrs rrrr, rkrk and kkrr, which are shown in Fig.\ref{fig_srsr}.

\begin{figure}
  \subfloat[srss]{
    \includegraphics[width=.2\linewidth]{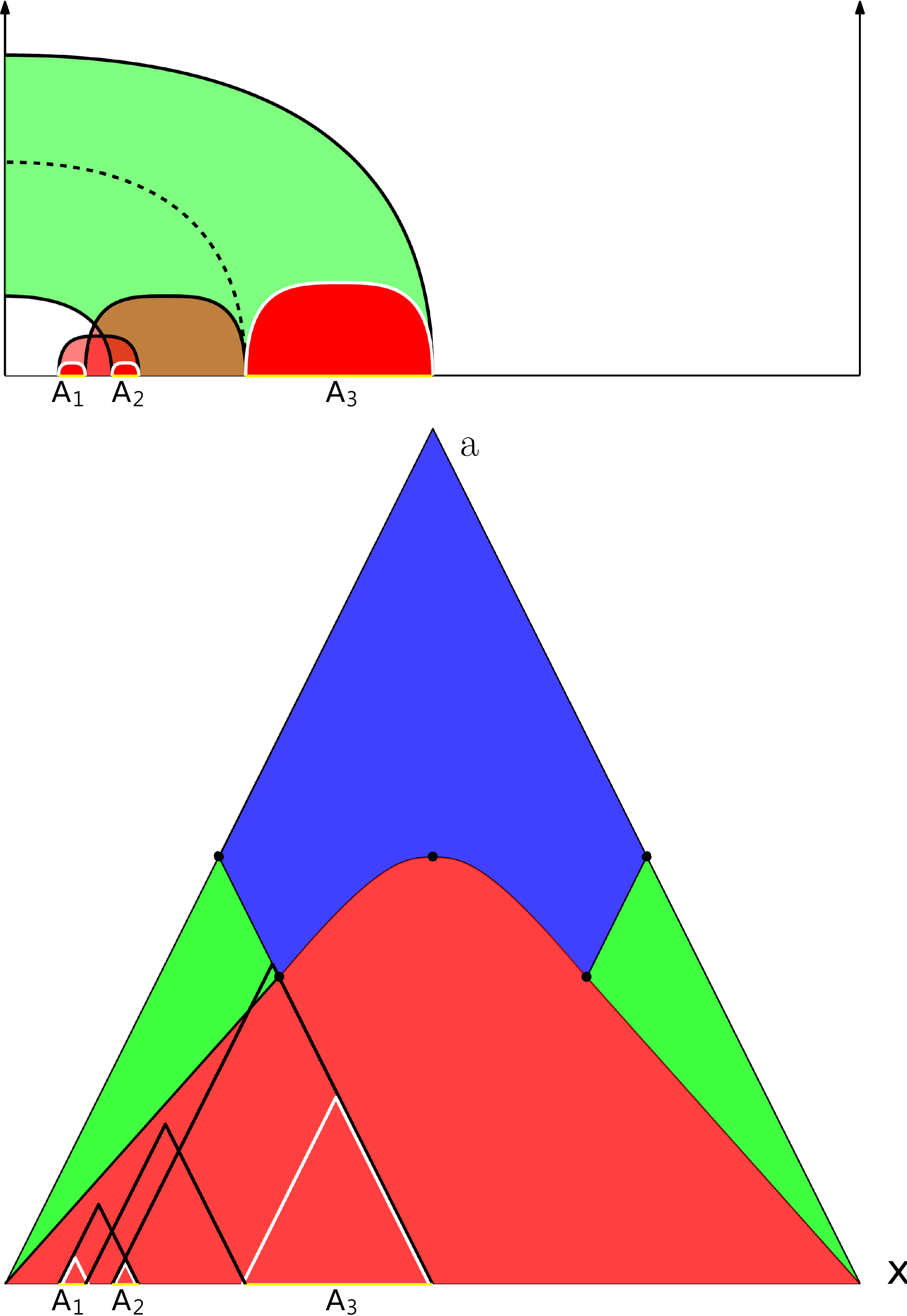} } \hspace{0.5cm}
  \subfloat[skss]{
    \includegraphics[width=.2\linewidth]{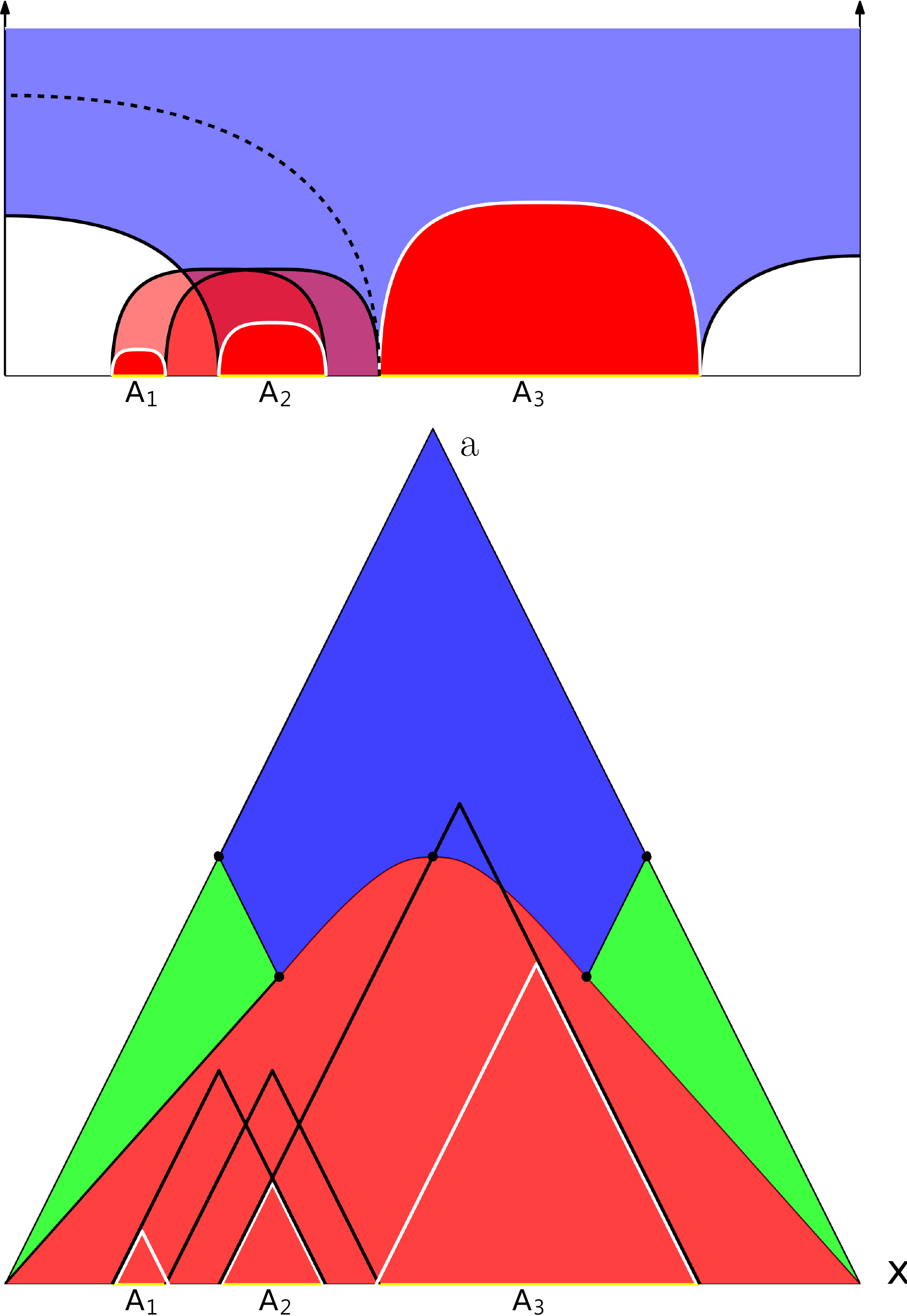} }\hspace{0.5cm}
  \subfloat[rsss]{
    \includegraphics[width=.2\linewidth]{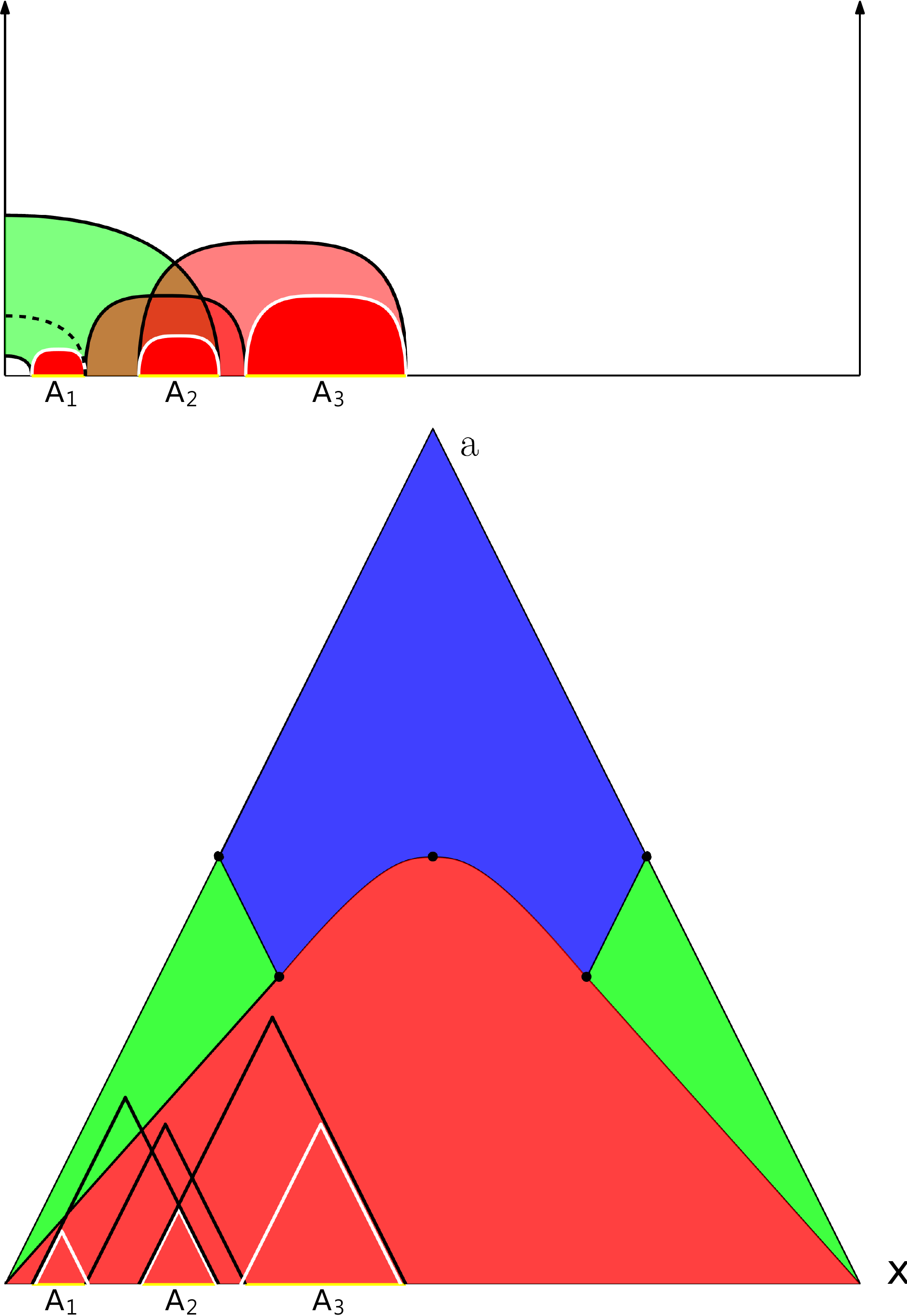} }\hspace{0.5cm}
  \subfloat[rrss]{
    \includegraphics[width=.2\linewidth]{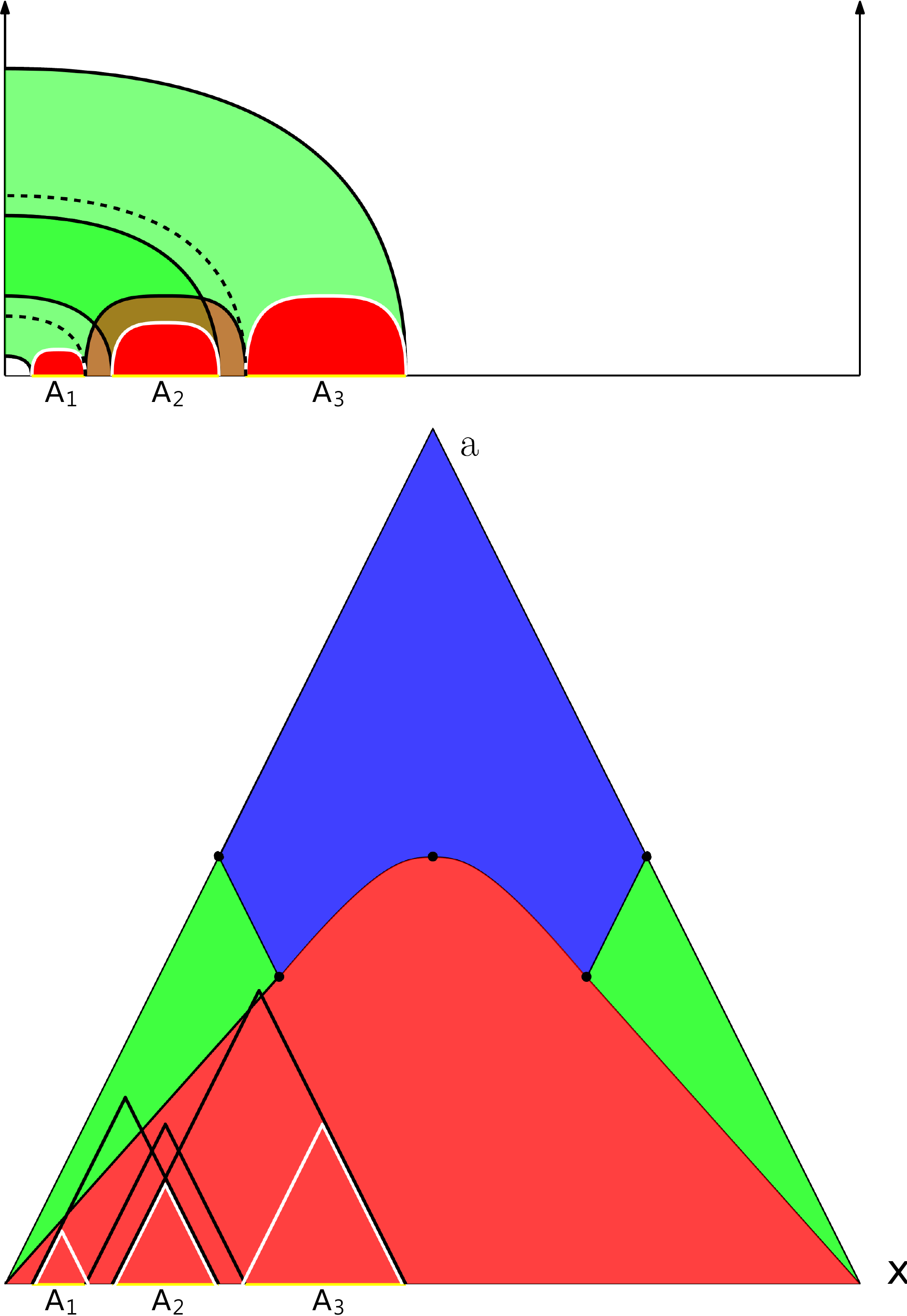} }\\
  \subfloat[rrsr]{
    \includegraphics[width=.2\linewidth]{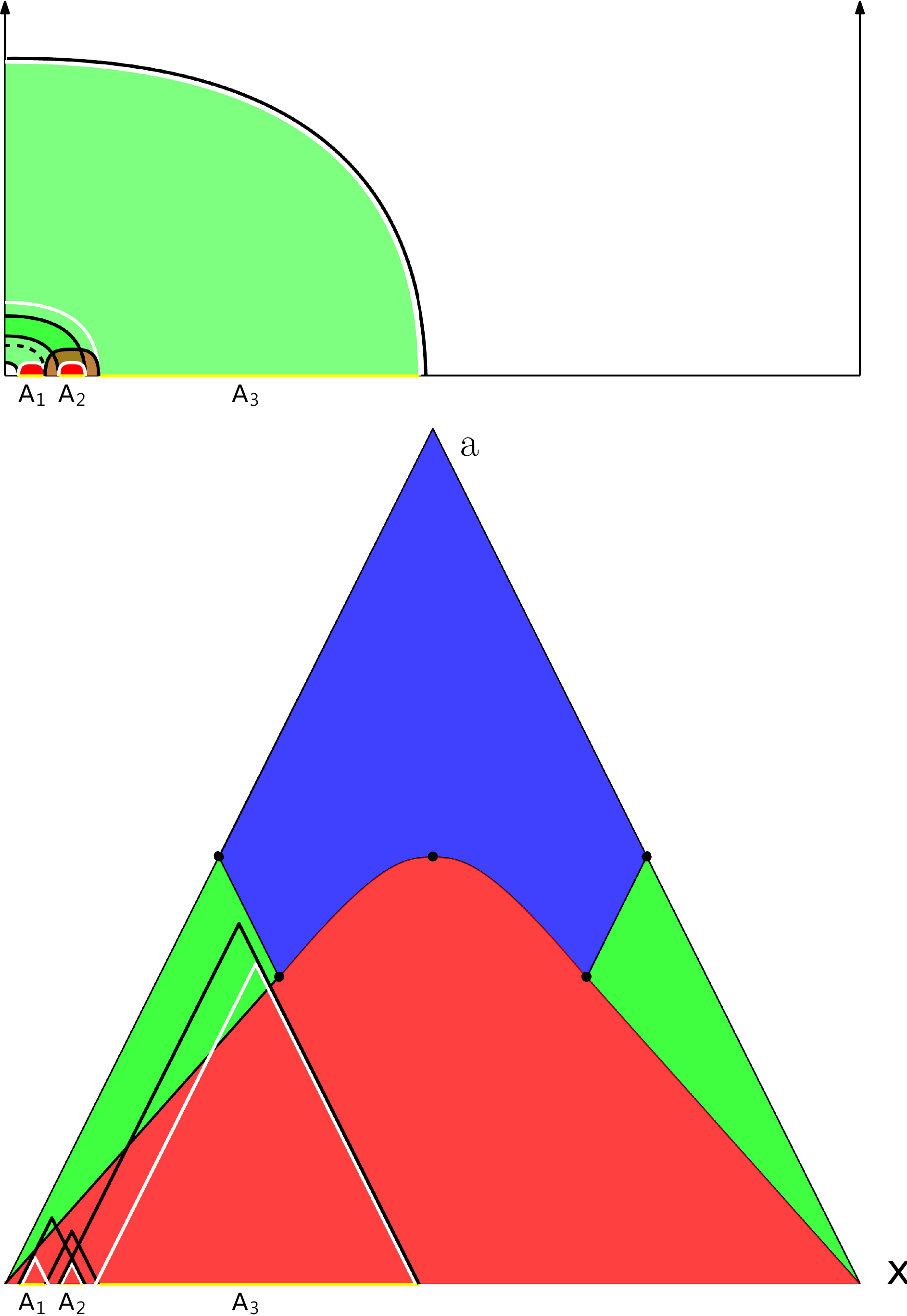} }\hspace{0.5cm}
     \subfloat[rrrs]{
    \includegraphics[width=.2\linewidth]{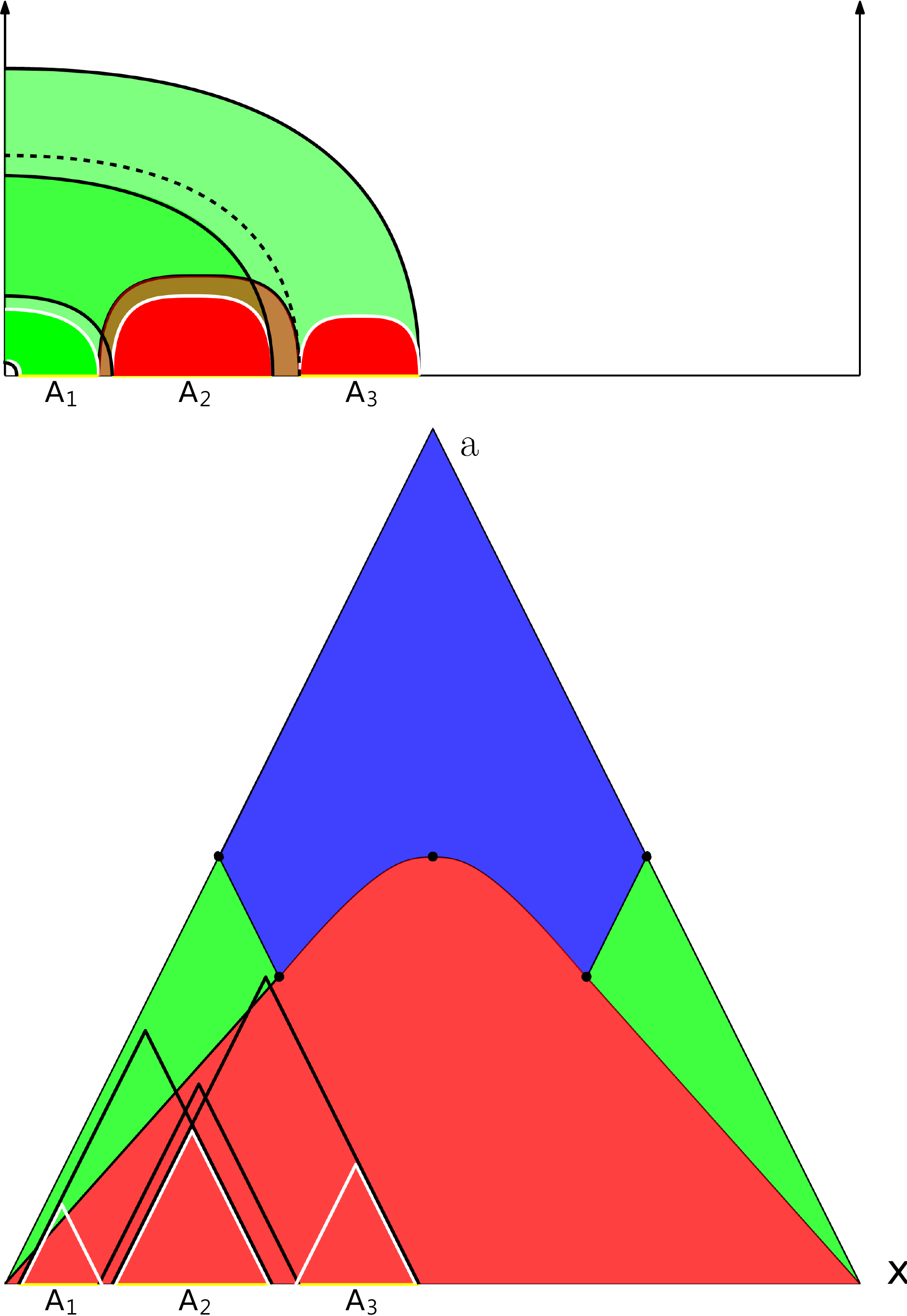} }\hspace{0.5cm}
  \subfloat[rkss]{
    \includegraphics[width=.2\linewidth]{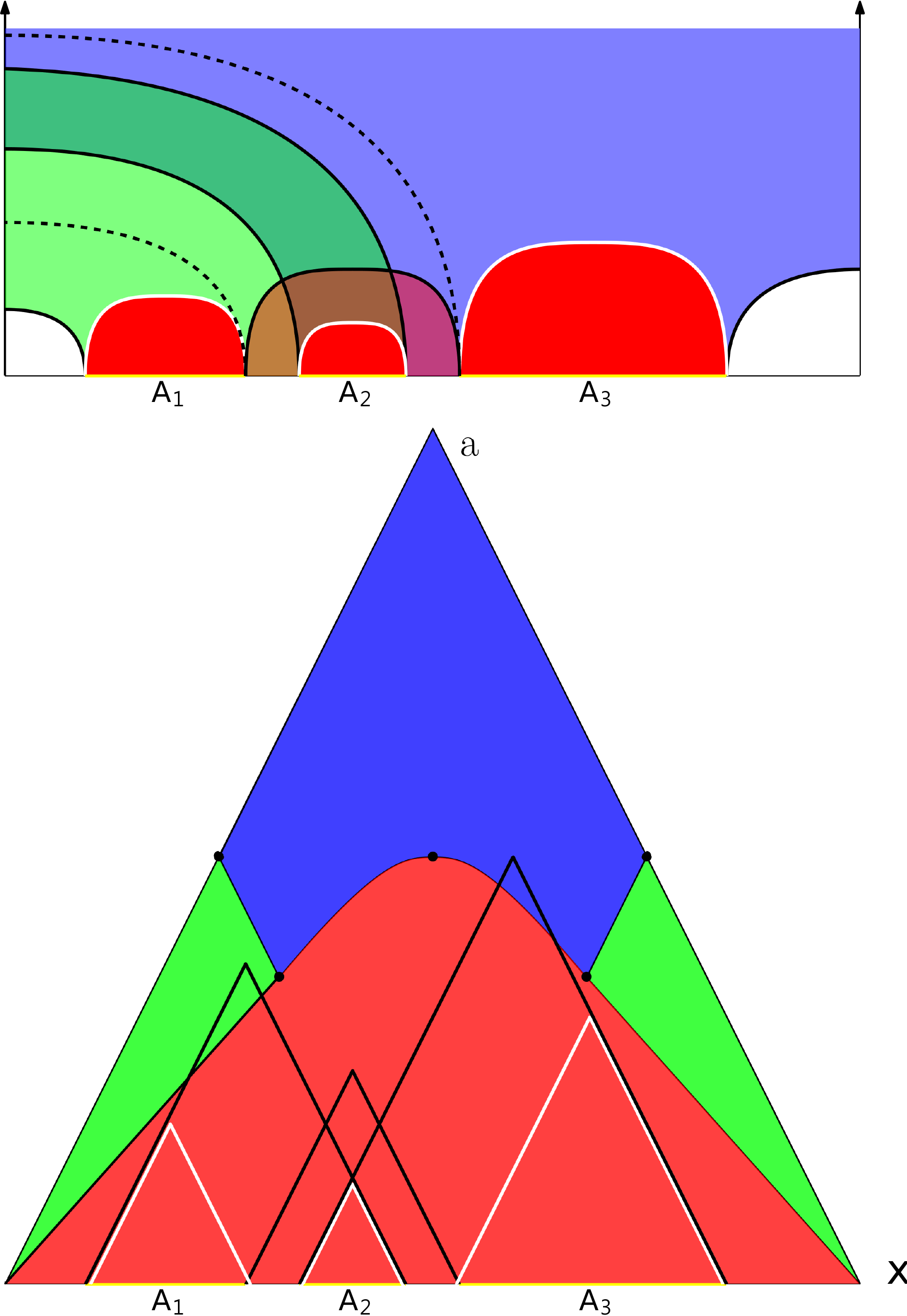} }\hspace{0.5cm}
  \subfloat[rksr]{
    \includegraphics[width=.2\linewidth]{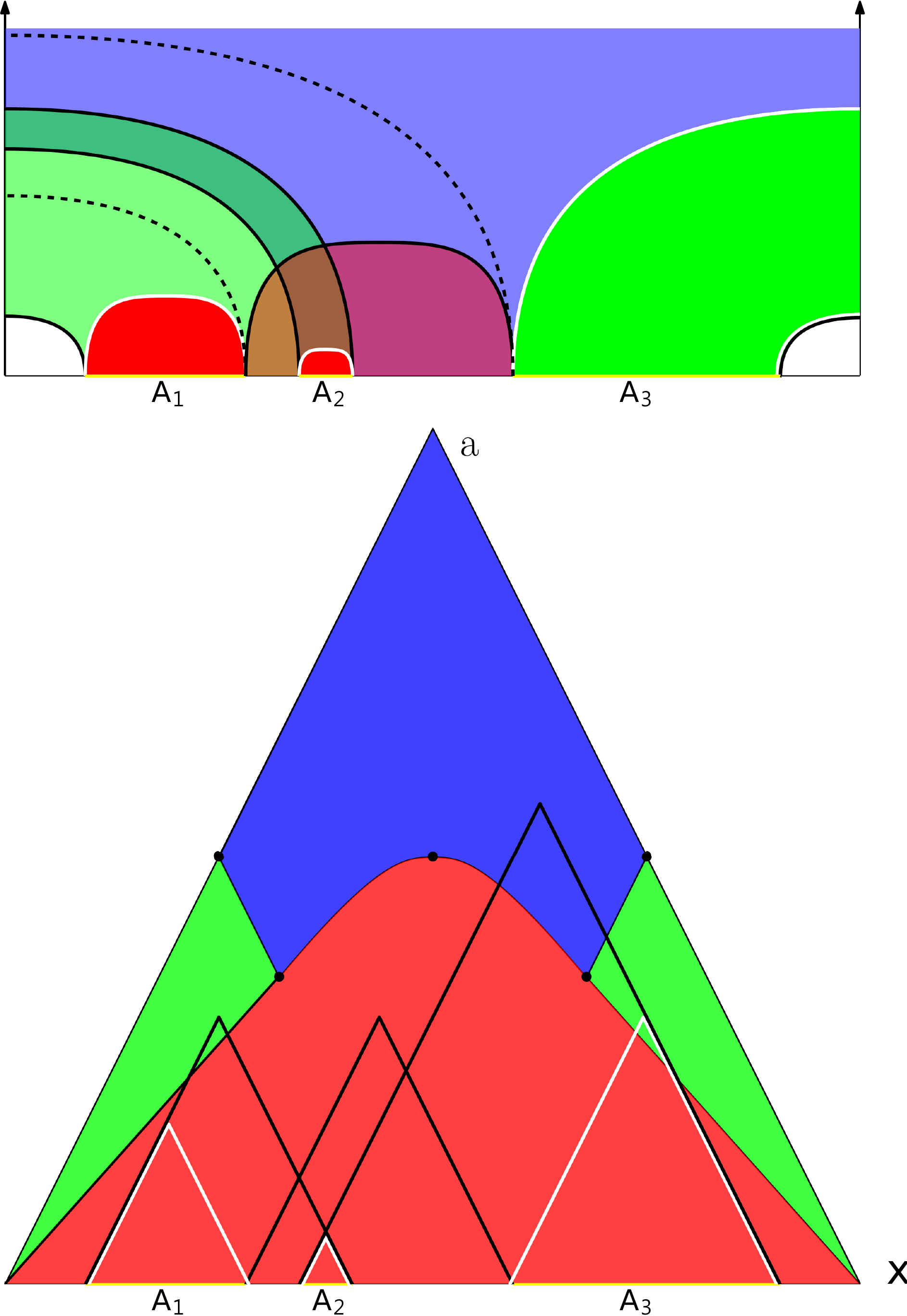} }\\
  \subfloat[rksk]{
    \includegraphics[width=.2\linewidth]{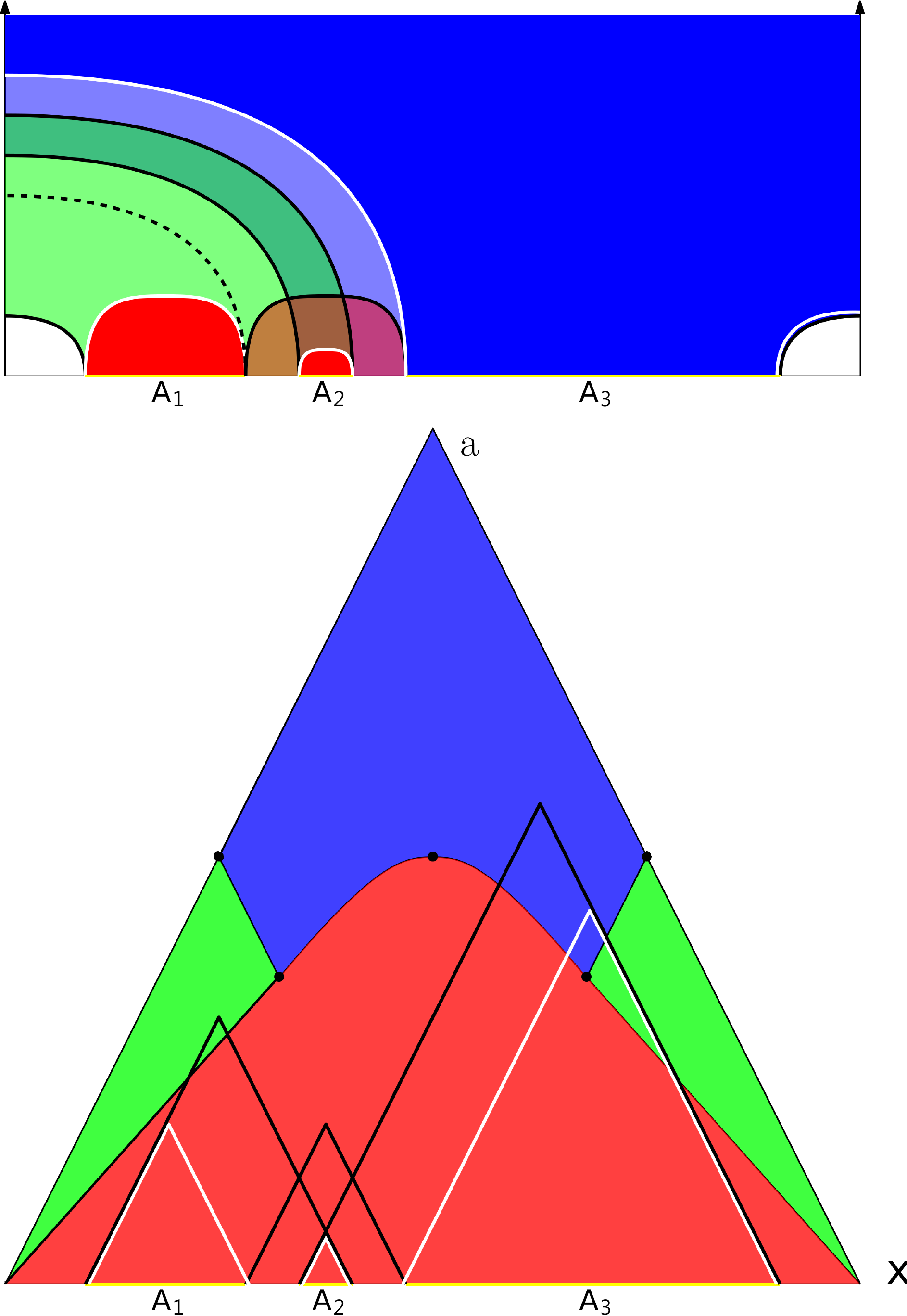} }\hspace{0.5cm}
  \subfloat[rkrs]{
    \includegraphics[width=.2\linewidth]{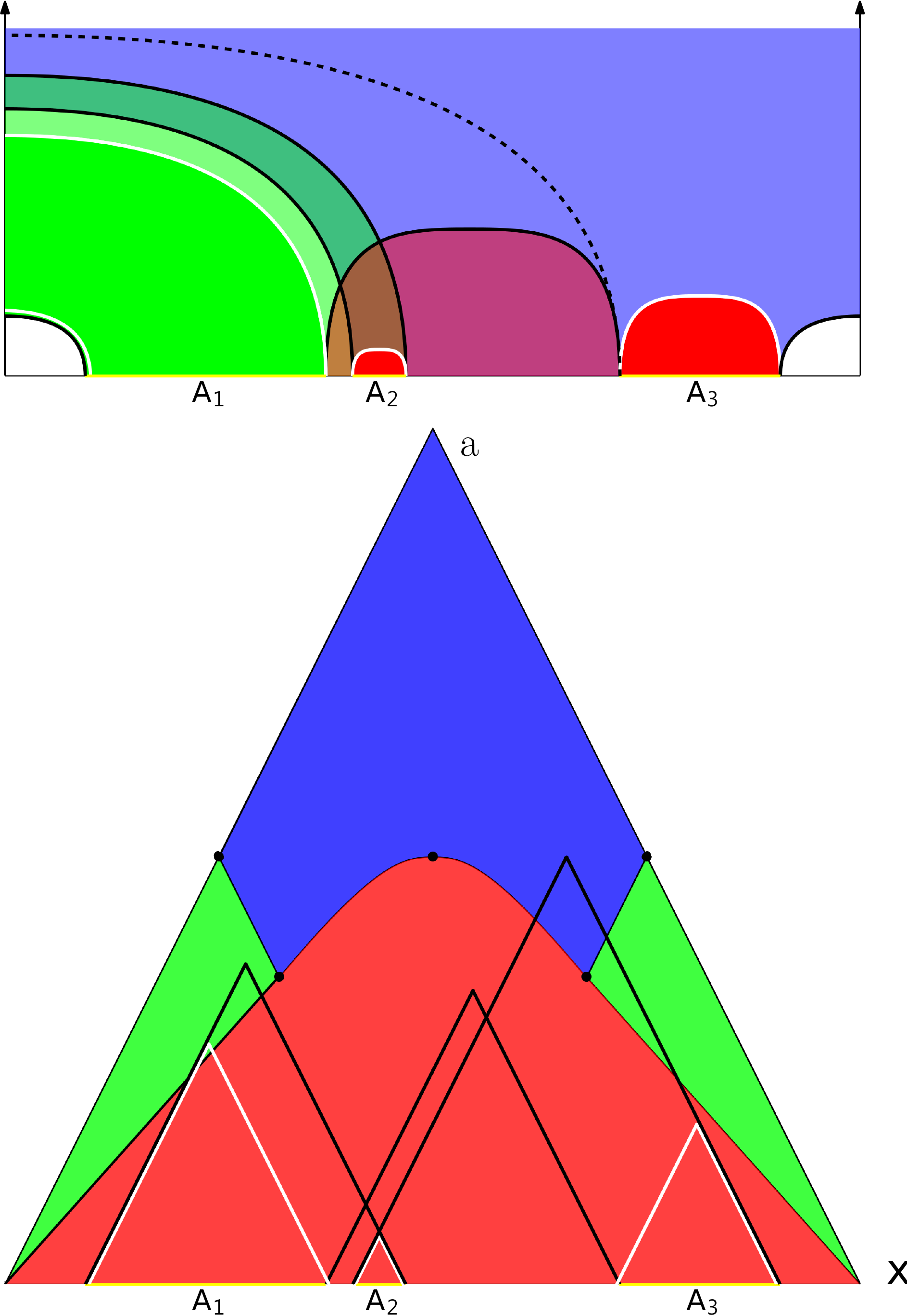} }\hspace{0.5cm}
  \subfloat[rkrr]{
    \includegraphics[width=.2\linewidth]{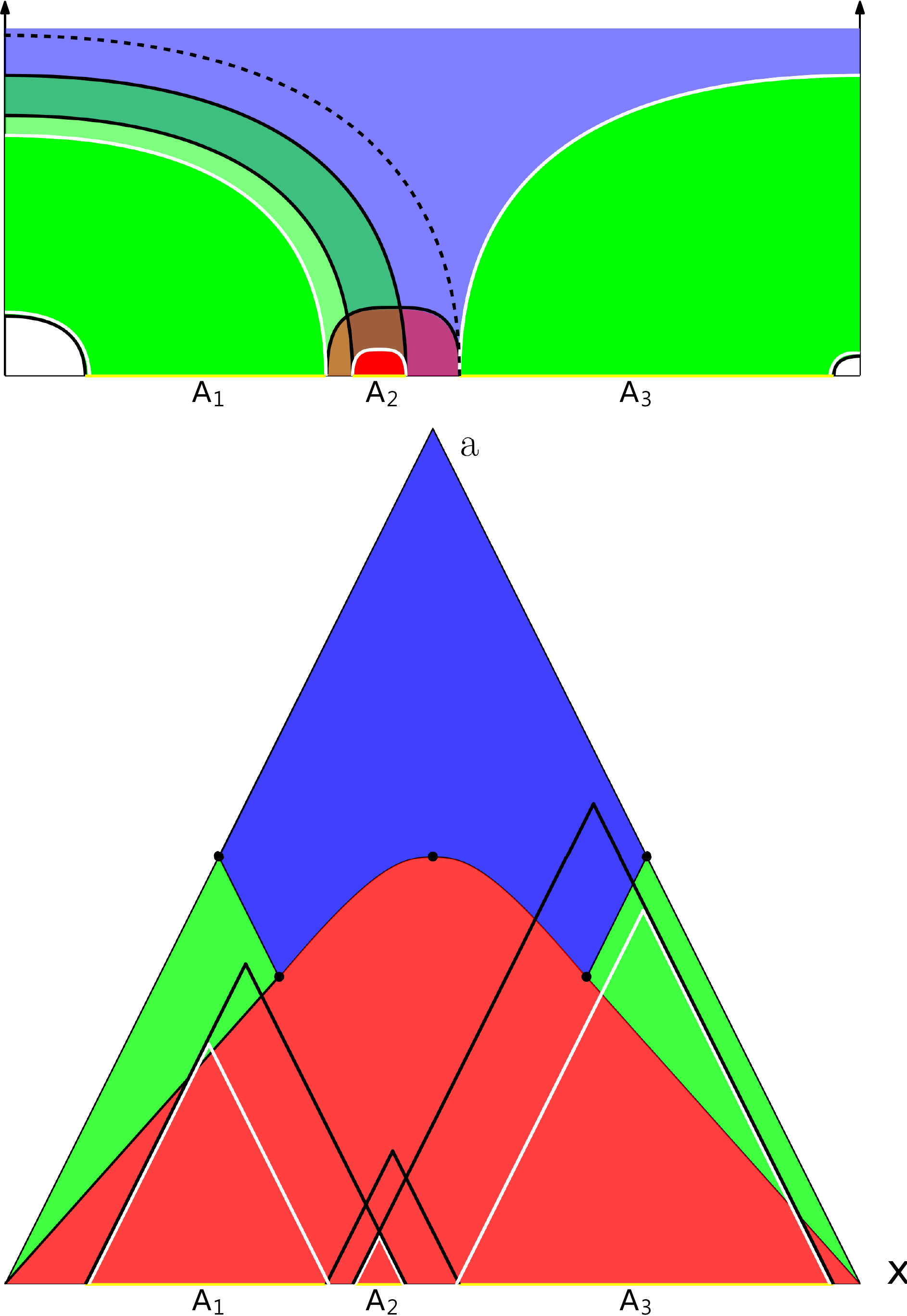} } \hspace{0.5cm}
 \subfloat[ksss]{
    \includegraphics[width=.2\linewidth]{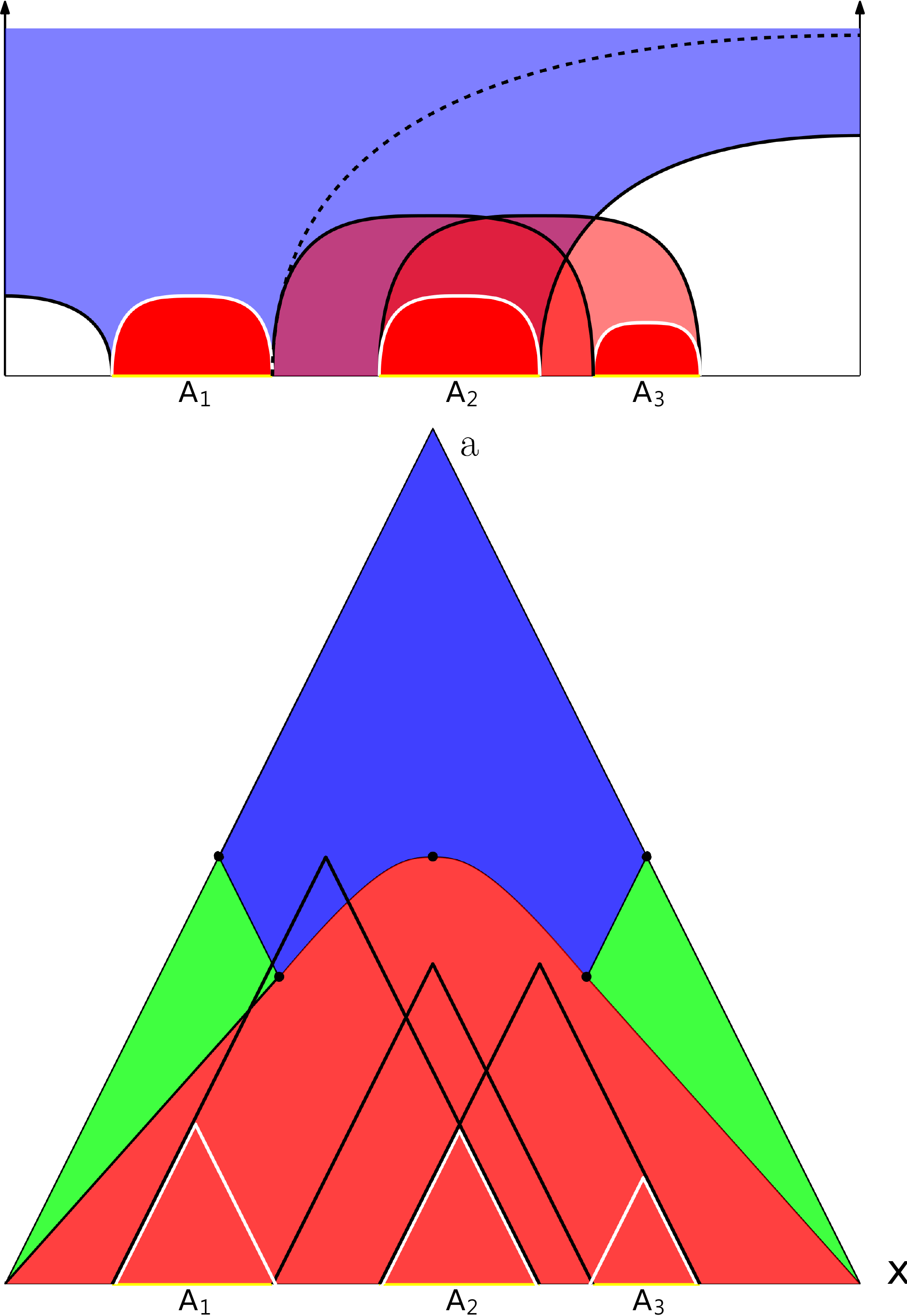} }\\
  \subfloat[ksrs]{
    \includegraphics[width=.2\linewidth]{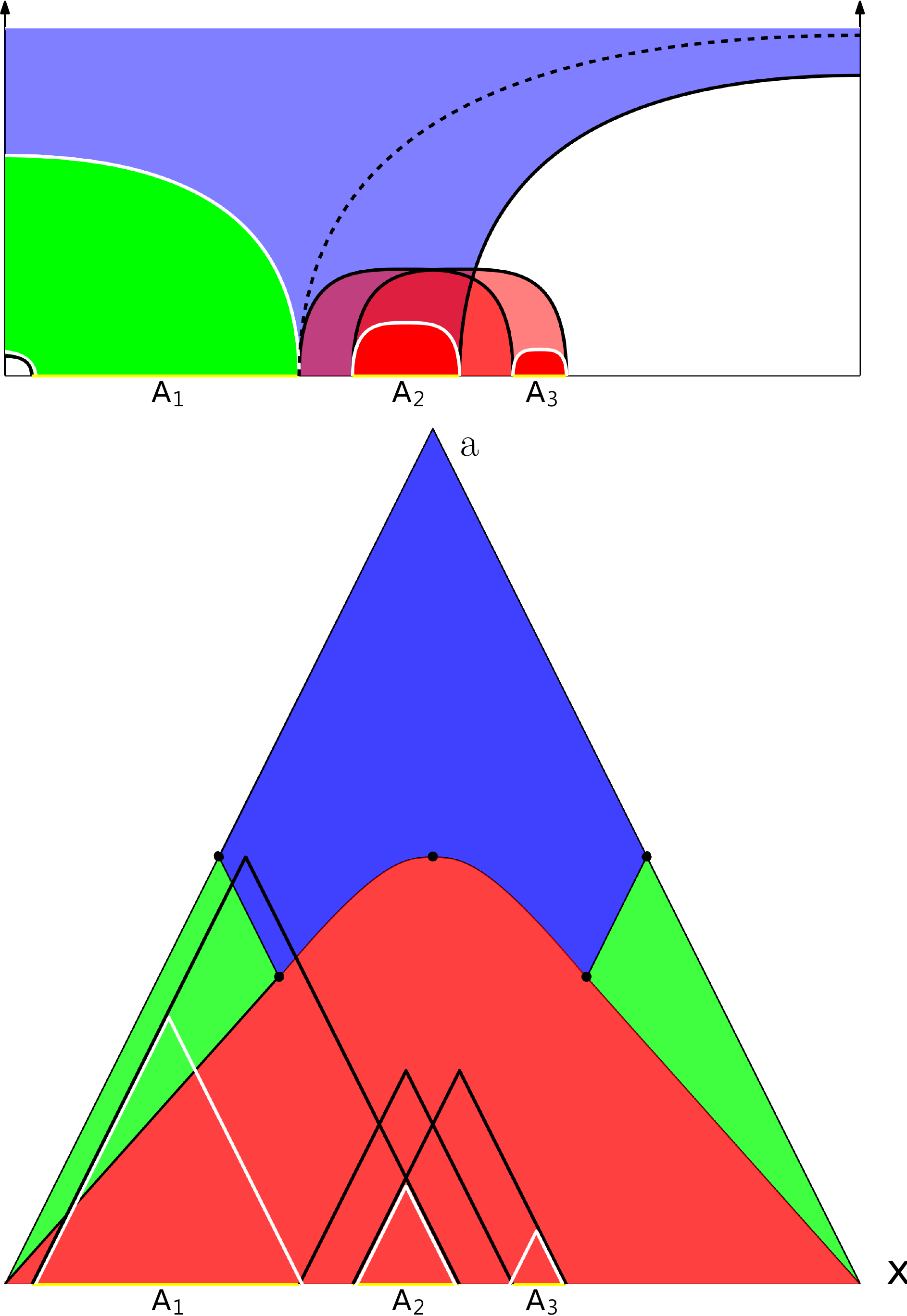} }\hspace{0.5cm}
  \subfloat[kkss]{
    \includegraphics[width=.2\linewidth]{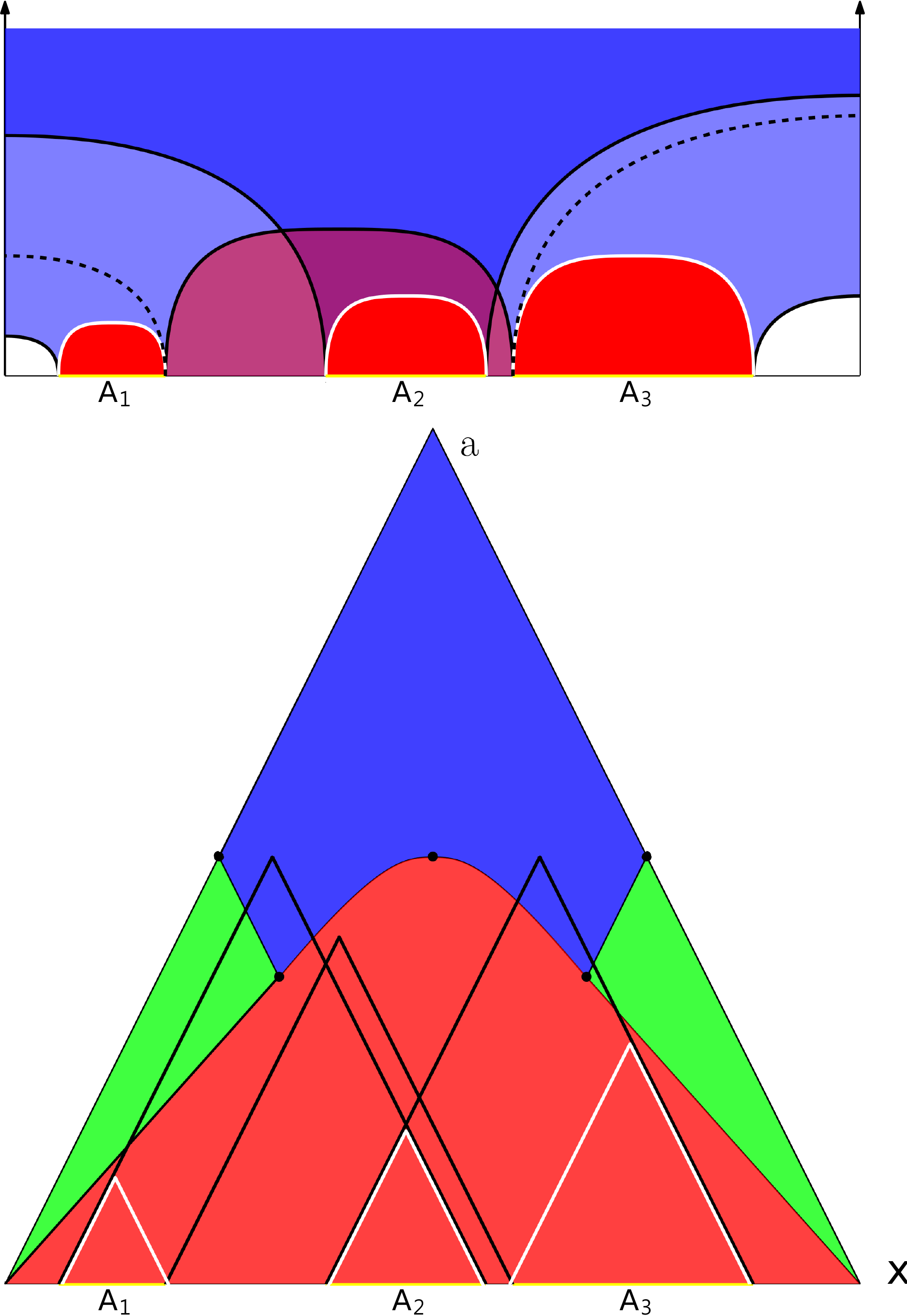} }\hspace{0.5cm}
  \subfloat[kksr]{
    \includegraphics[width=.2\linewidth]{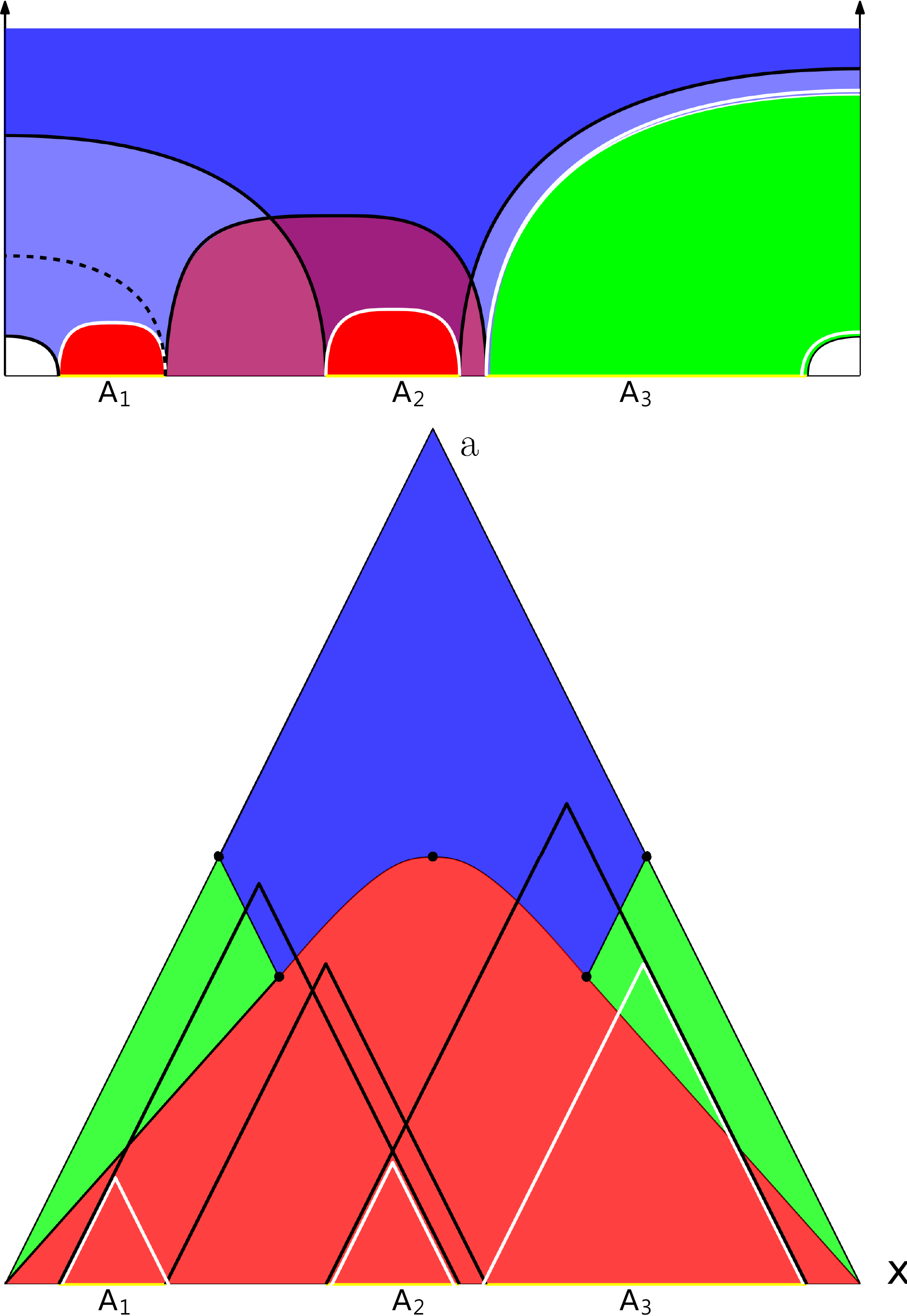} }\hspace{0.5cm}
  \subfloat[kkrs]{
    \includegraphics[width=.2\linewidth]{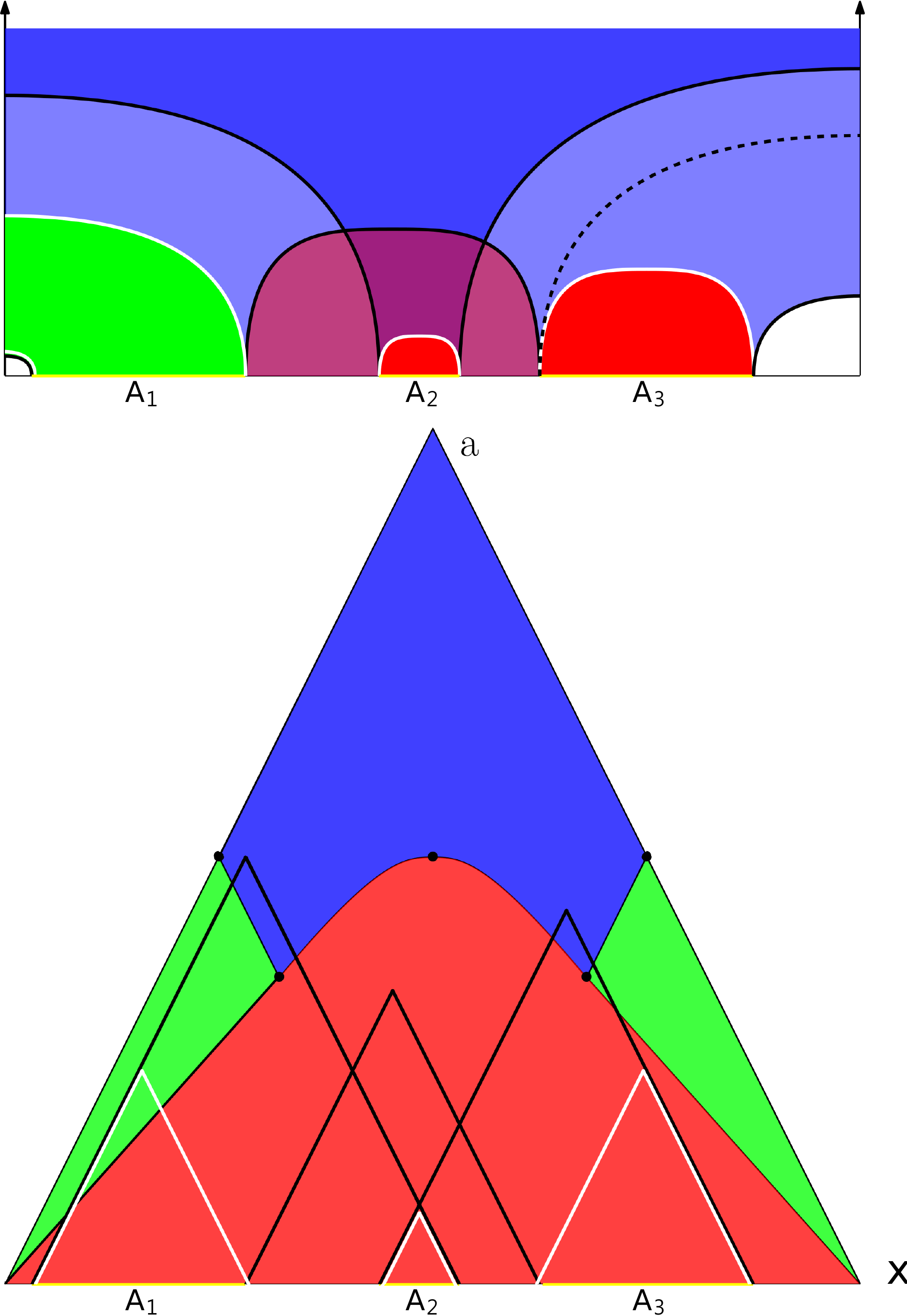} }
  \caption{The other 16 diagrams that need auxiliary curves.}
  \label{fig_srss}
\end{figure}

To prove the other 17 cases, we need to add auxiliary curves. For example, the sksr case is plotted in Fig.\ref{fig_ssss}(c) with the dashed black curve being the auxiliary curve. Besides the two joint curves in homology with $A_1$ and $A_2$, which are larger than their corresponding HEE's represented by the white curves, the third joint curve is larger than the dashed black curve. Furthermore, the dashed black curve is larger than the white quarter-circle-shaped curve anchored on the left boundary of the entangled region $A_3$ because we already assume that the HEE of $A_3$ is in the rainbow phase in this case. We thus proves the inequality Eq.(\ref{MMIc}) in the sksr case. The other 16 similar cases are shown in  Fig.\ref{fig_srss}.

Now let us consider the cases that one of $S_{ij}$ is not in the connected configuration, i.e. $S_{ij}=S_i+S_j$, it is easy to show that the MMI reduces to the SSA. On the other hand, if all $S_{ij}$ are in the connected configuration, since we have proved the MMI in the connected case (\ref{MMIc}), we can use  $S^c_{ij}\le S_i+S_j$ to show that
\begin{align}
S_{123}^{c} &\le S_{1}+S_{23}^{c} ,\\
S_{123}^{c} &\le S_{12}^{c}+S_{3},\\
S_{123}^{c} &\le S_{1}+S_{3}+S_{3},
\end{align}
which justifies that $S_{123}$ must be in the connected configuration. We thus complete the proof of the MMI in Eq.(\ref{MMI}) for BCFT.

Equivalently, we can define the non-positivity of the tripartite information,
\begin{align}
I_{3}&=S_{1}+S_{2}+S_{3}-\left(  S_{12}+S_{23}+S_{13}\right)
+S_{123} \leq0.
\end{align}

\section{Schwazschild-AdS Black Hole Background}

In the case of AdS black hole spacetime (\ref{AdS-Schw}), the size $a$ and the HEE for the entangled region $A$ can be expressed as the following \cite{1805.06117},
\begin{align}
a &  =2z_{0}\int_{0}^{1}\frac{v^{d}dv}{\sqrt{\left(  1-\left(  bv\right)
^{d+1}\right)  \left(  1-v^{2d}\right)  }},\\
S\left(  a\right)
&  =\frac{l_{AdS}^{d}}{2\left(  d-1\right)  G_{N}^{(d+2)}}\left(  \frac
{L}{\epsilon}\right)  ^{d-1}-\frac{aS_{BH}}{\left(  d-1\right)  Lb^{d}}%
+\frac{bz_{0}}{d-1}\int_{0}^{1}\frac{\left(  d-3\right)  v-\left(  d+3\right)
v^{2d+1}}{\sqrt{\left(  1-\left(  bv\right)  ^{d+1}\right)  \left(
1-v^{2d}\right)  }}dv. \label{SBH}
\end{align}
where $b=z_{0}/z_{H}$. The  first term in the HEE is divergent and is  proportional  to  the  boundary  of  the  entangled  region $A$.  The remaining terms are finite.

\begin{figure}
  \subfloat[Sunset]{
    \includegraphics[width=.3\linewidth]{image/BCFT/sun.pdf} }
  \subfloat[Sky]{
    \includegraphics[width=.3\linewidth]{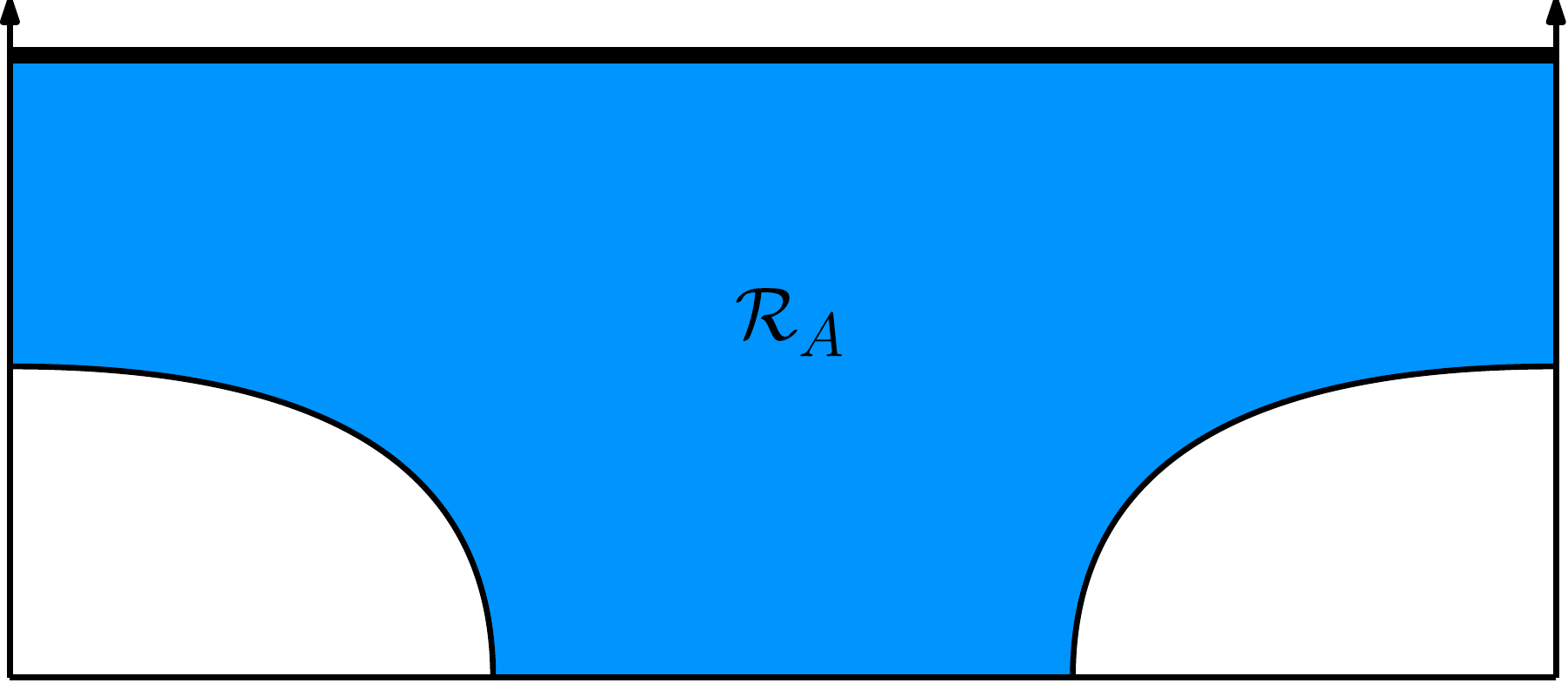} }
  \subfloat[Rainbow]{
    \includegraphics[width=.3\linewidth]{image/BCFT/rainbow.pdf} }
  \caption{The minimal surfaces in the Schwazschild AdS black hole spacetime. We fill the color red/blue/green in the entanglement wedge $\mathcal{R}_{ {A} }$ of the sunset/rainbow/sky phase.
  (a) $\mathcal{R}_{ {A} }$ is in the sunset phase when the entangled region $ {A} $ is relatively small.
  (b) $\mathcal{R}_{ {A} }$ is in the sky phase when the entangled region $ {A} $ is large enough. In this case, the RT surface includes the black hole horizon. (c) $\mathcal{R}_{ {A} }$ is in the rainbow phase when the entangled region $ {A} $ is close to the boundary.}
  \label{HEEshapesBH}
\end{figure}

As shown in \cite{1805.06117}, there are also three configurations of the entanglement wedges in the case with a black hole,
\begin{align}
\text{sunset}  &  :S^s_{A} =S_{A}\left(  a\right)  ,\label{sBH} \\
\text{rainbow}  &  :S^r_{A}  =\frac{1}{2}S_{A}\left(
l+a-2\left\vert {x}\right\vert \right)  +\frac{1}{2}S_{A}\left(  l-a-2
\left\vert {x}\right\vert \right) , \label{rBH} \\
\text{sky}  &  :S^k_{A}  =\frac{1}{2}S_{A}\left(  l-a+2
\left\vert {x}\right\vert \right)  +\frac{1}{2}S_{A}\left(  l-a-2\left\vert
{x}\right\vert \right)  +S_{BH}, \label{kBH}
\end{align}
which are shown in Fig.\ref{HEEshapesBH}.

In the black hole background, the interior part behind the horizon is removed from the spacetime so that the homology changes. In this case, the RT surface could be disconnected with one of them being enclosed on the horizon. This modifies the HEE in the sky phase by including the black hole entropy $S_{BH}$ as in Eq.(\ref{kBH}), and the HEE in the sunset or rainbow phase remains unchanged. Therefore, in the black hole background, we only need to reconsider the inequalities with terms involved the sky phase, otherwise the proof will be exactly the same as in the pure AdS background.

\begin{figure}
  \includegraphics[width=.4\linewidth]{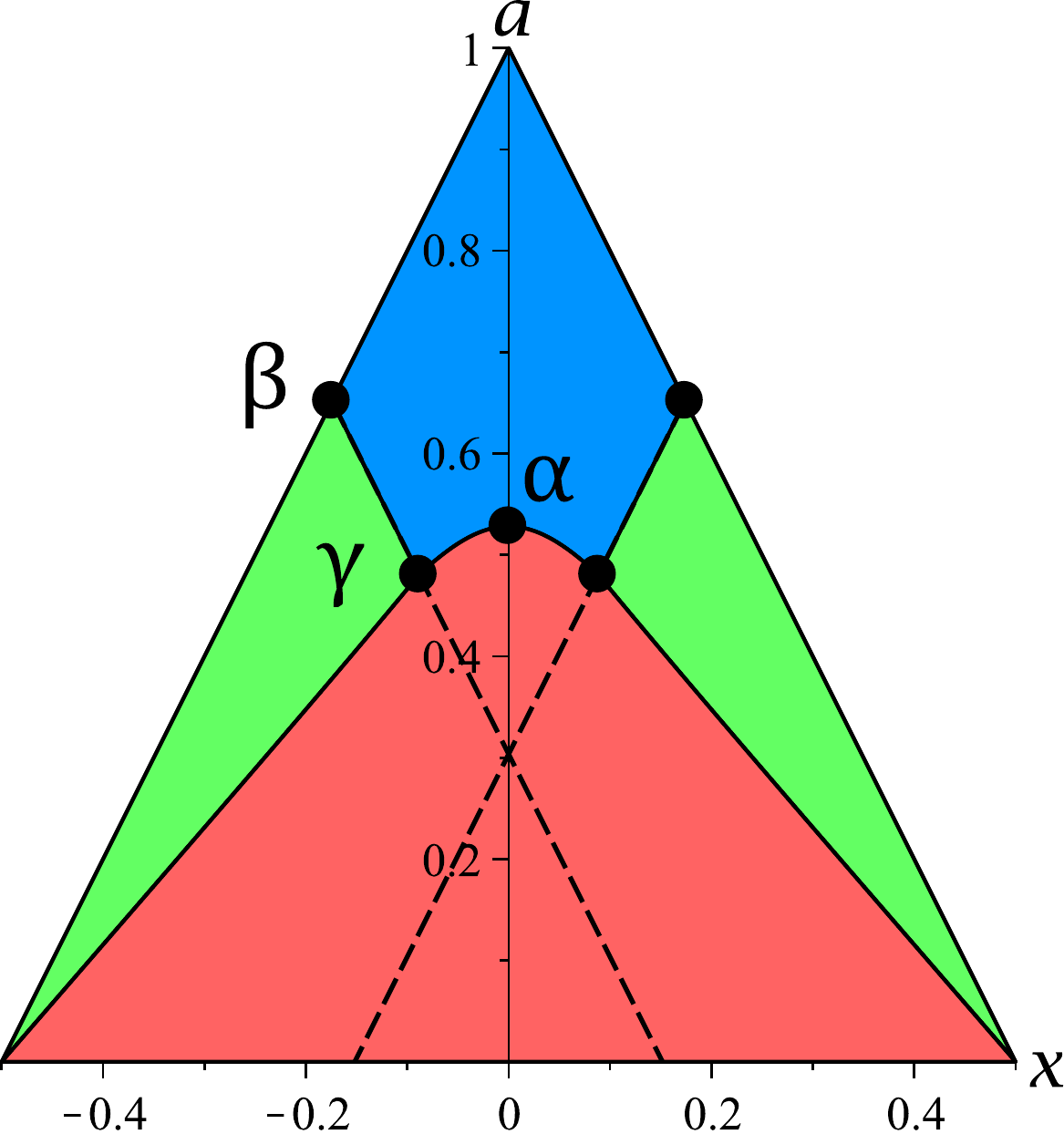}
  \caption{Phase Diagram of the holographic entanglement entropy in the AdS black hole background. The red/green/blue region represents the sunset/rainbow/sky phase.}
  \label{AdsBHPD}
\end{figure}

Fig.\ref{AdsBHPD} shows the phase diagram in the black hole background for the entangled region $A$ at the horizon $z_H=1.5$. The phase boundary between the sky and rainbow phases can be determined by $S_{ {A} }^{k}=S_{ {A} }^{r}$, which gives ${\left\vert
{x}\right\vert }=(a-x_0)/2$, where $x_0$ is determined by the equation,
\begin{equation}
S_{BH}    ={S\left(  {l+2}\left\vert {x}_{0}\right\vert \right)
-S}\left(  {l-2}\left\vert {x}_{0}\right\vert \right).
\end{equation}
We notice that the phase boundaries between the sky and rainbow phases in the black hole background are parallel to the corresponding edges of the equilateral triangle as in the pure AdS background, but their extensions do not intersect at the origin. This is because the extra contribution from black hole entropy $S_{BH}$ enlarges the HEE in the sky phase so that the region of the sky configuration in the phase diagram shrinks at finite temperature. In high temperature limit, the region of the sky configuration shrinks to zero eventually, we thus expect that only sunset phase survives. Since the HEE in Eq.(\ref{SBH}) is linearly dependent on the size of the entangled region $a$ in high temperature limit, the HEE in the sunset phase is always the minimum. Similarly, the phase boundaries between the sunset and sky/rainbow phases can be determined from
$S_{ {A} }^{s}=S_{ {A} }%
^{k}$ and $S_{ {A} }^{s}=S_{ {A} }^{r}$, respectively.

In the black hole background, the HEE of the entangled region $ {A} $ and its complement $ {A} ^{c}$ are different from the black hole entropy $\mathcal{E}_{ {A} ^{c}}=\mathcal{E}_{ {A} }\pm S_{BH}$.

\subsection{Strong subadditivity}
\begin{figure}
  \subfloat[sss]{
    \includegraphics[  width=.15\linewidth]{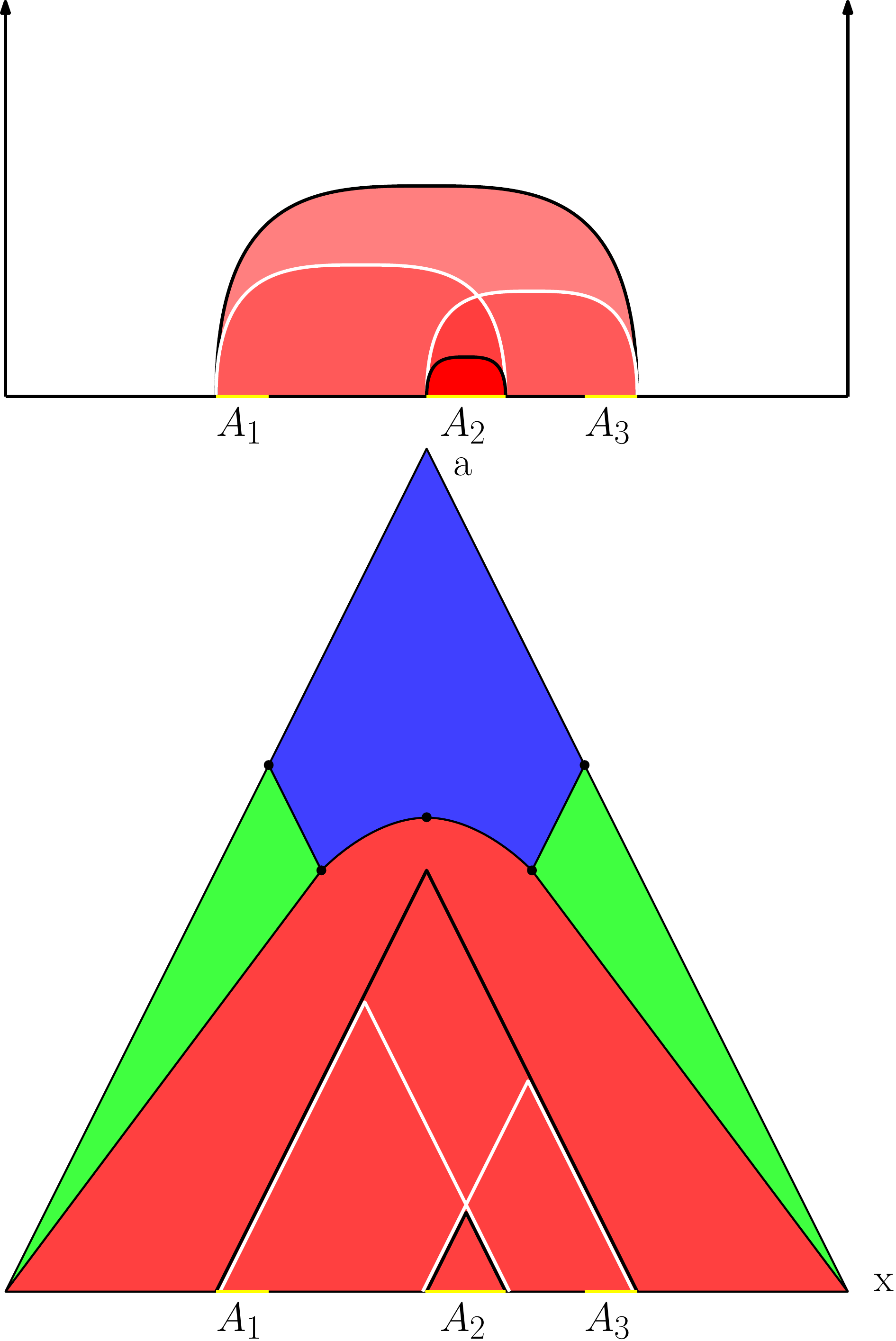} }\hspace{0.3cm}
  \subfloat[ssr]{
    \includegraphics[  width=.15\linewidth]{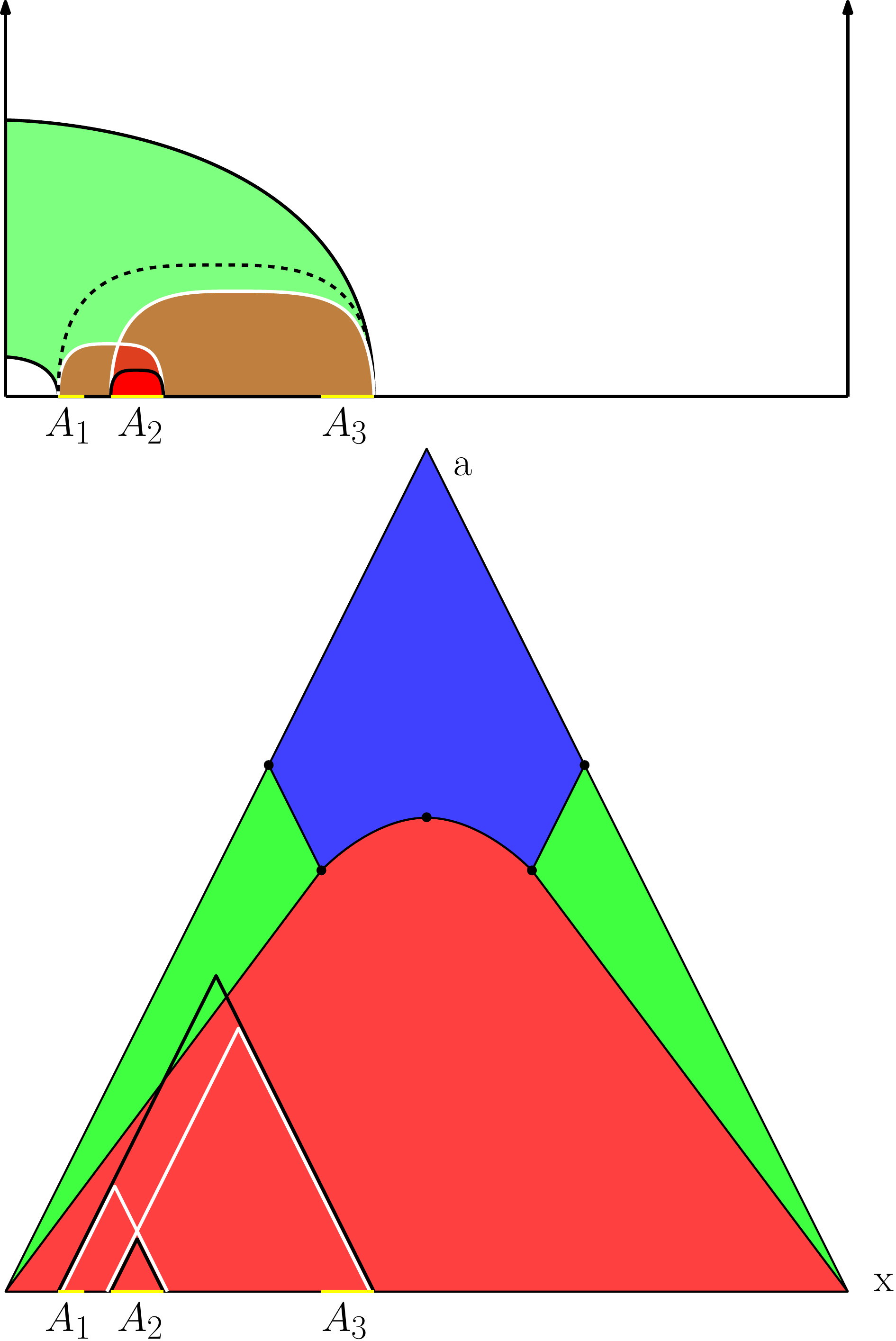} }\hspace{0.3cm}
  \subfloat[srr]{
    \includegraphics[  width=.15\linewidth]{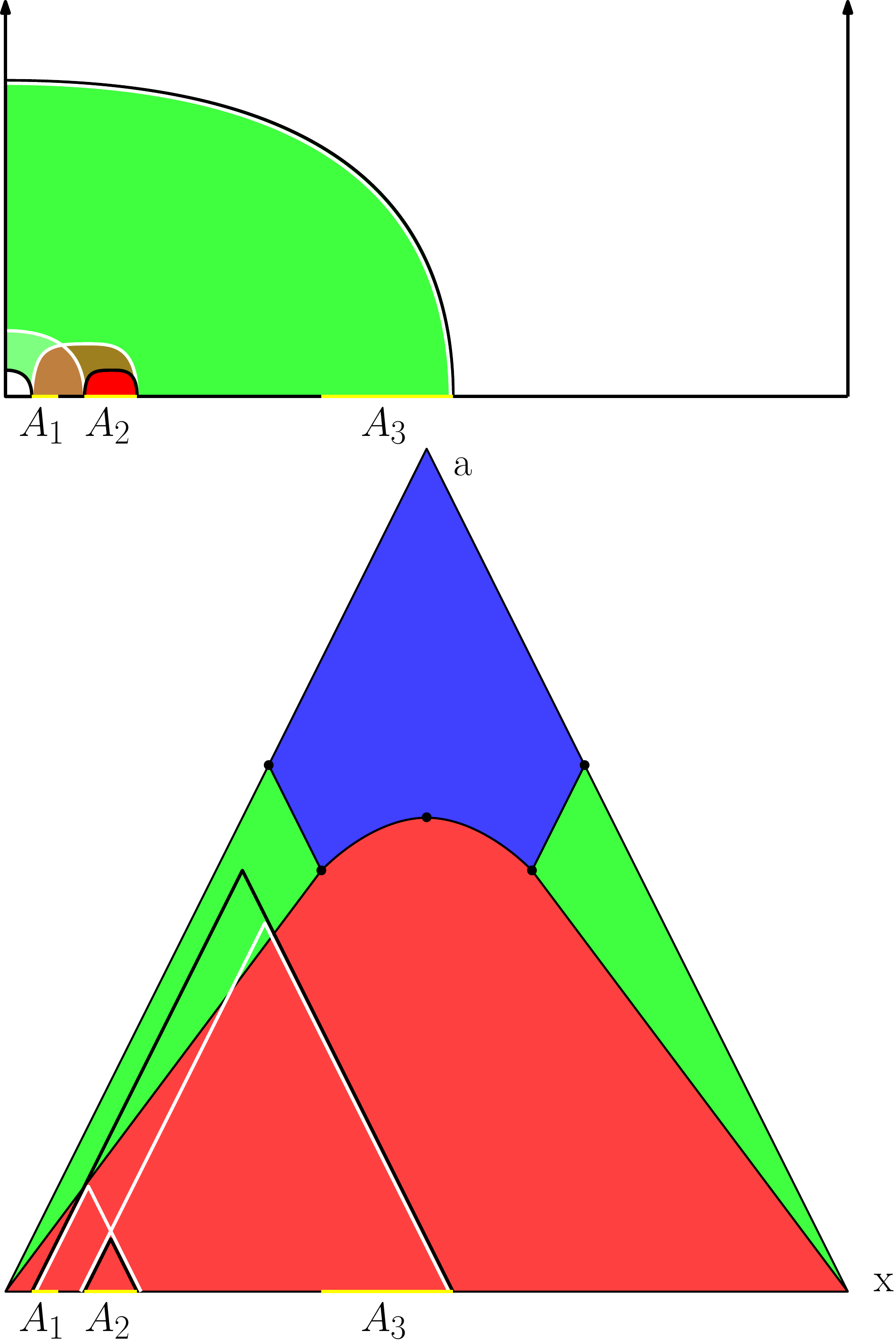} }\hspace{0.3cm}
  \subfloat[rsr]{
    \includegraphics[  width=.15\linewidth]{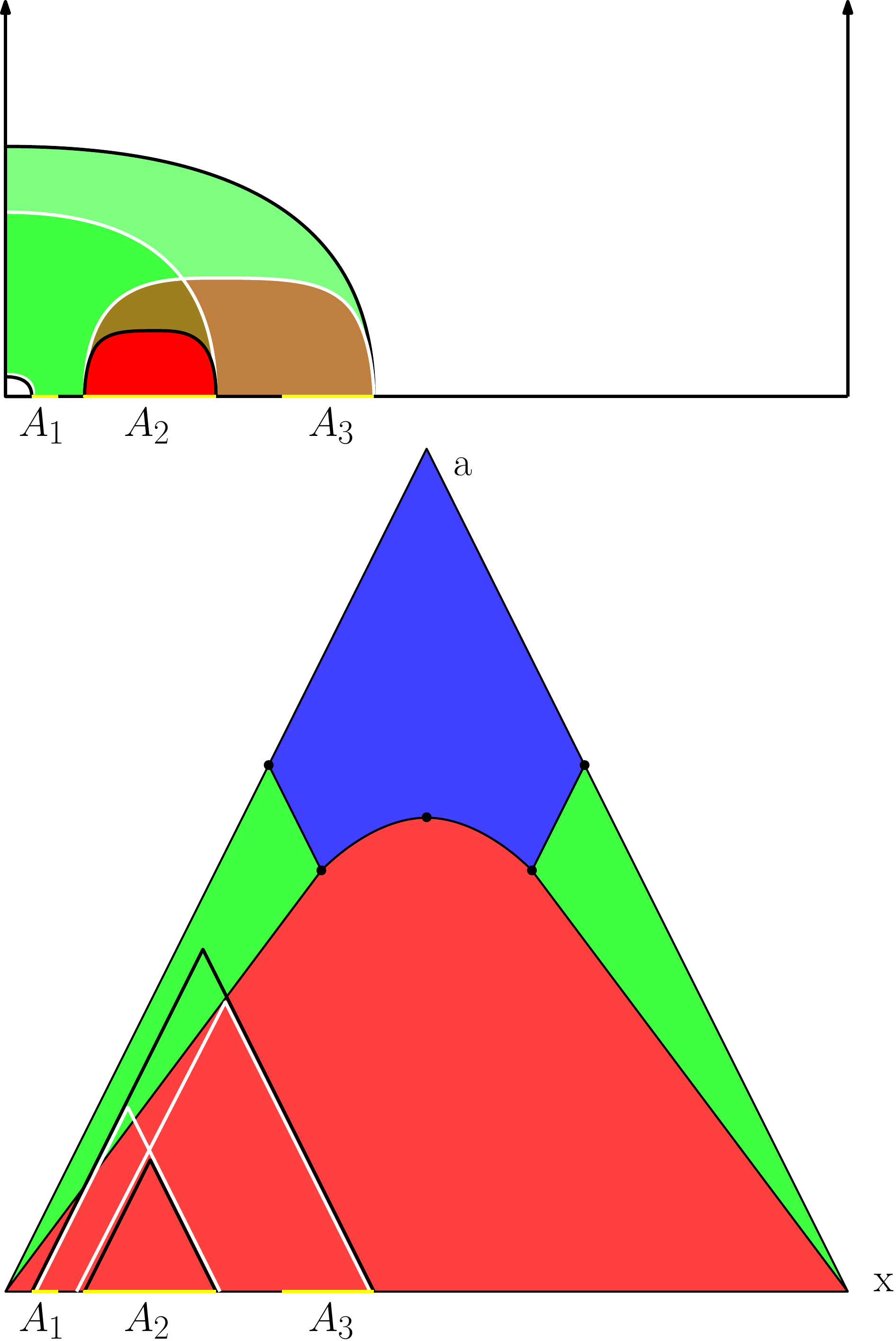} }\hspace{0.3cm}
  \subfloat[rrr]{
    \includegraphics[  width=.15\linewidth]{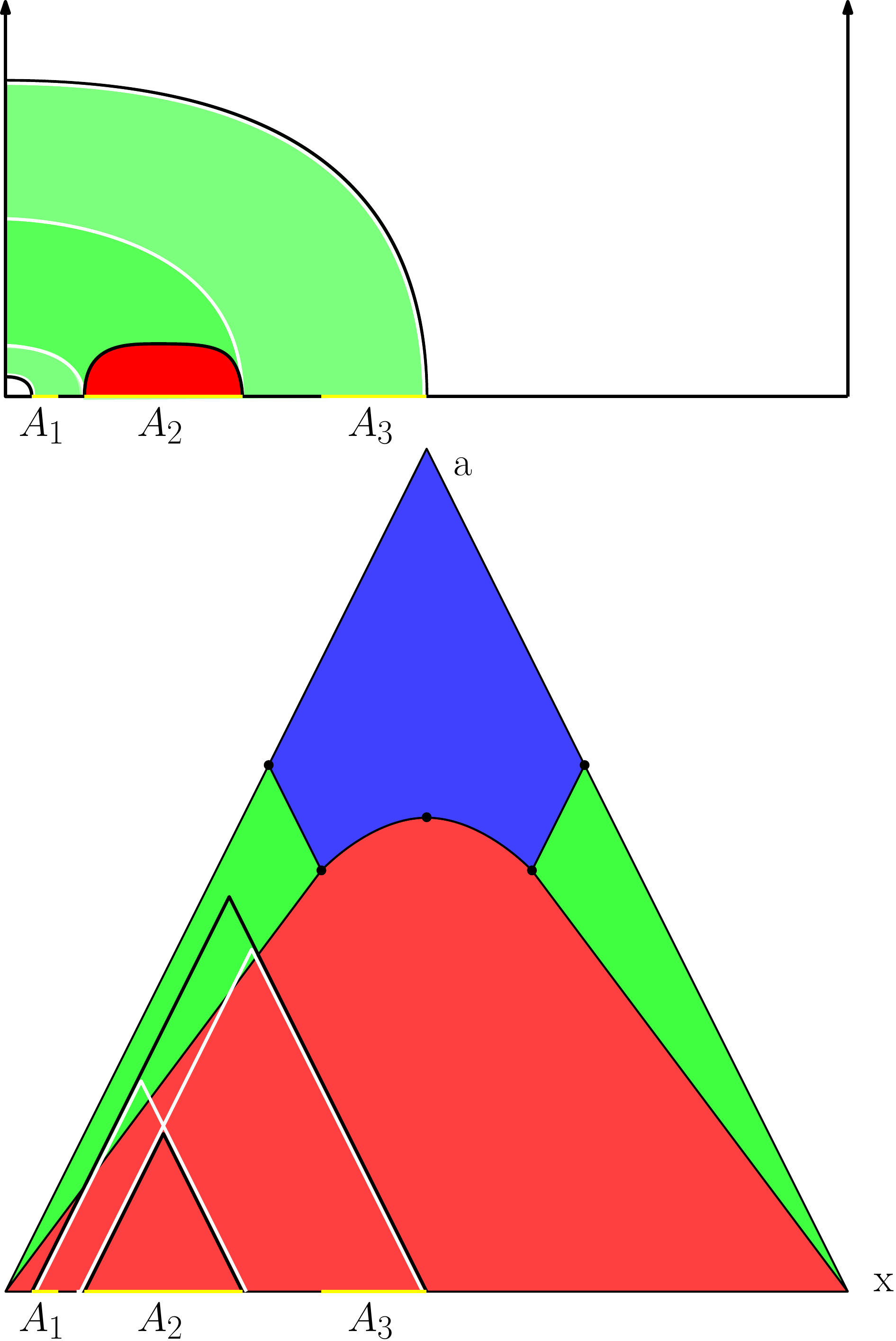} }\hspace{0.3cm}\\
  \subfloat[ssk]{
    \includegraphics[  width=.15\linewidth]{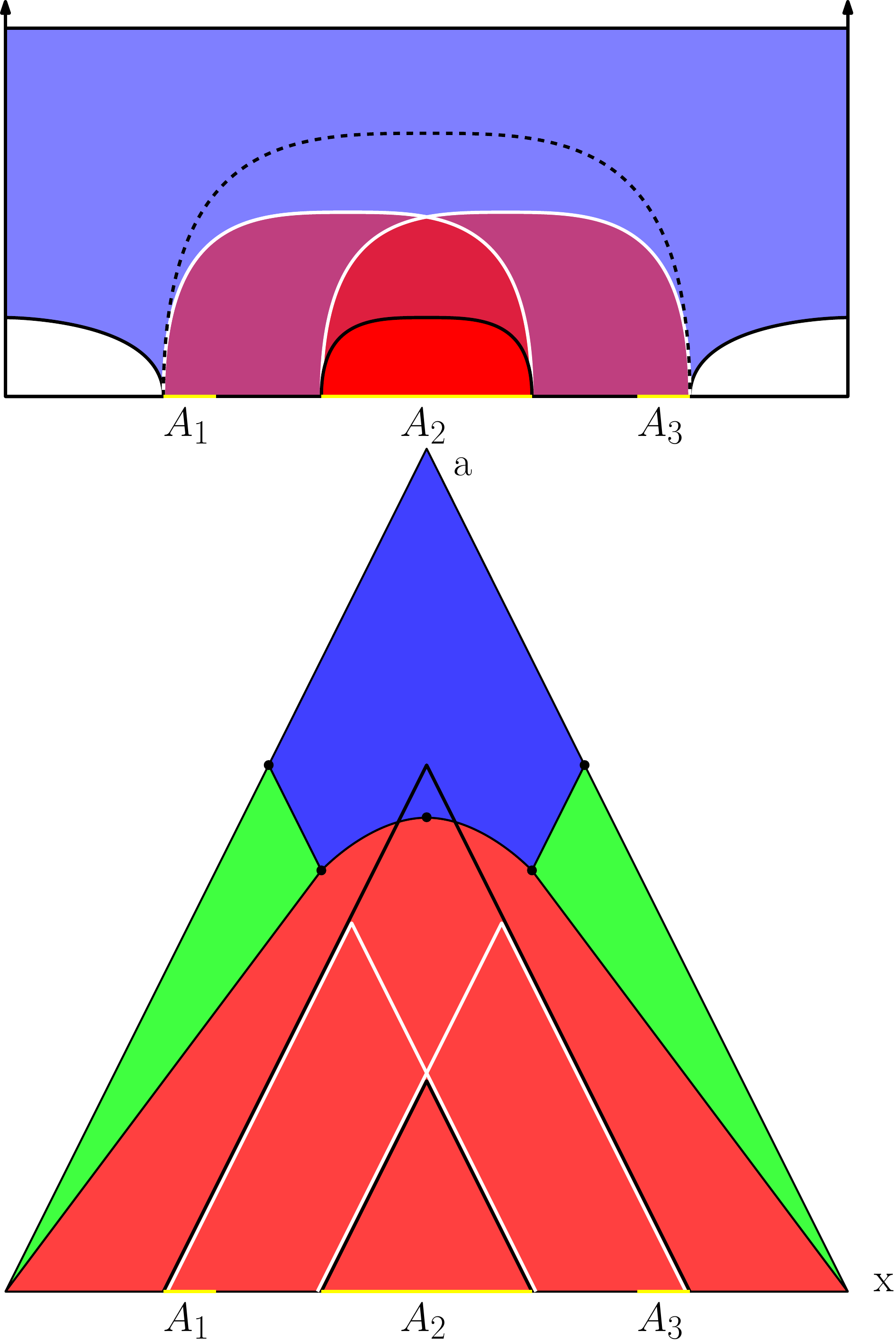} }\hspace{0.3cm}
  \subfloat[rsk]{
    \includegraphics[  width=.15\linewidth]{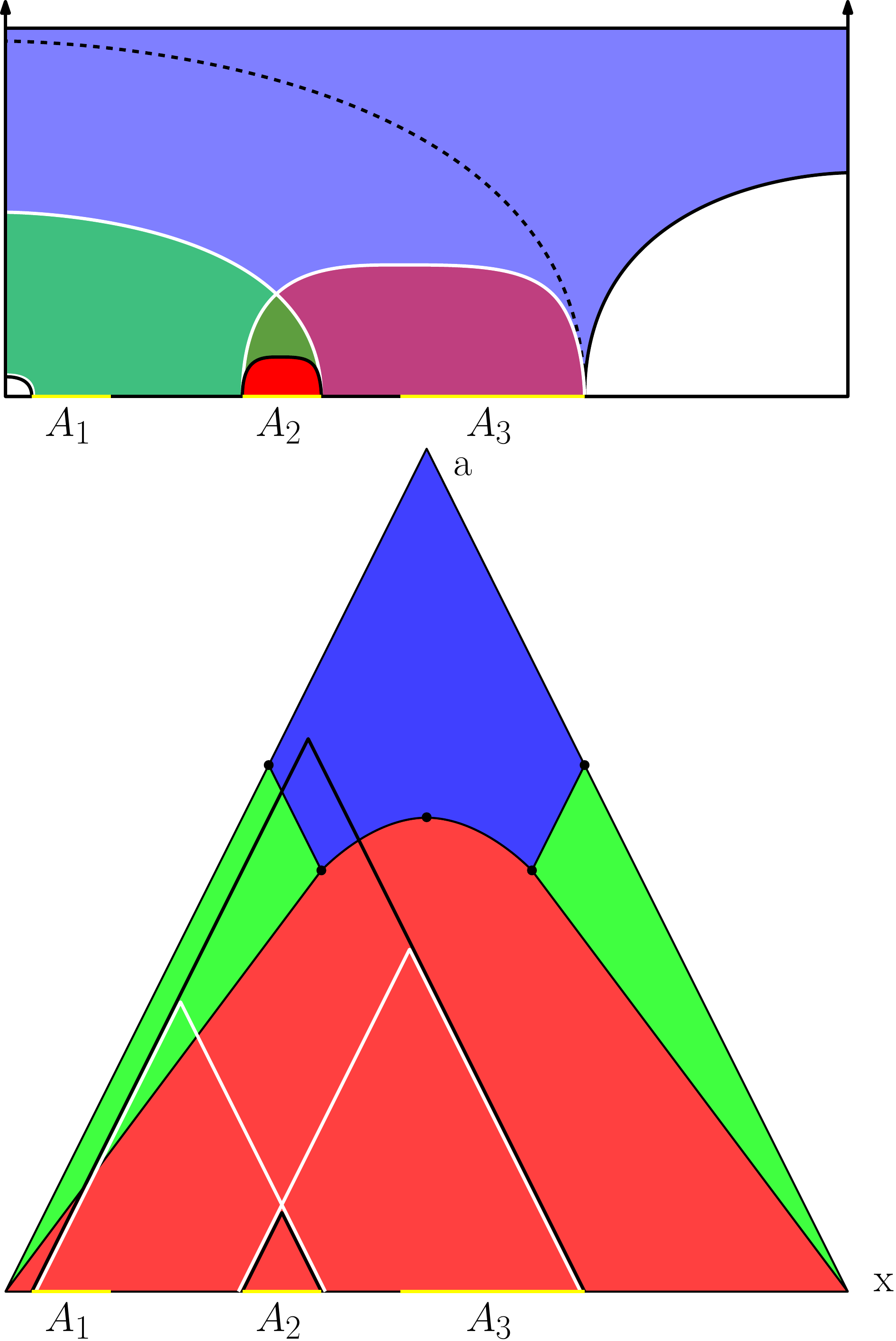} }\hspace{0.3cm}
  \subfloat[rkk]{
    \includegraphics[  width=.15\linewidth]{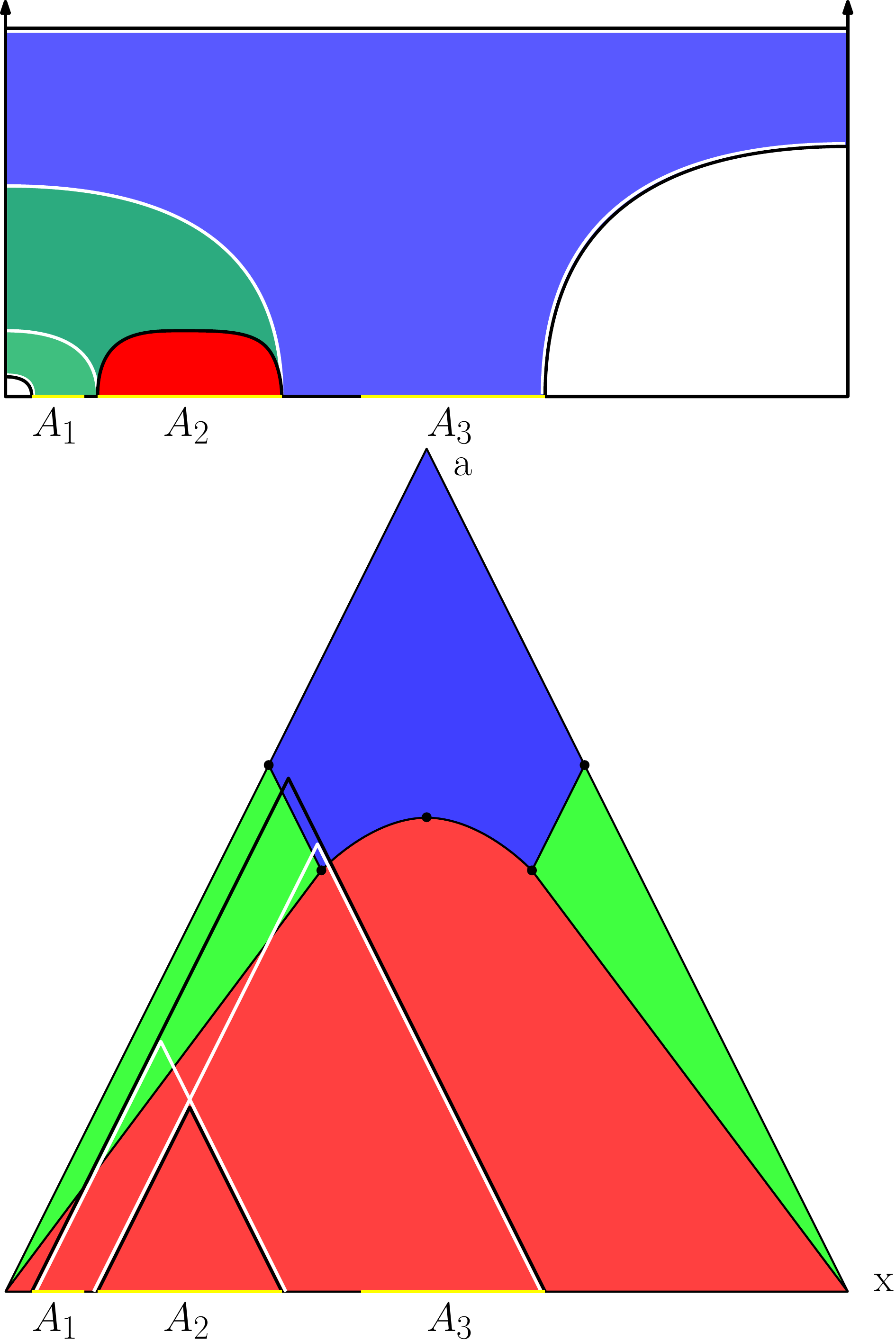} }\hspace{0.3cm}
  \subfloat[ksk]{
    \includegraphics[  width=.15\linewidth]{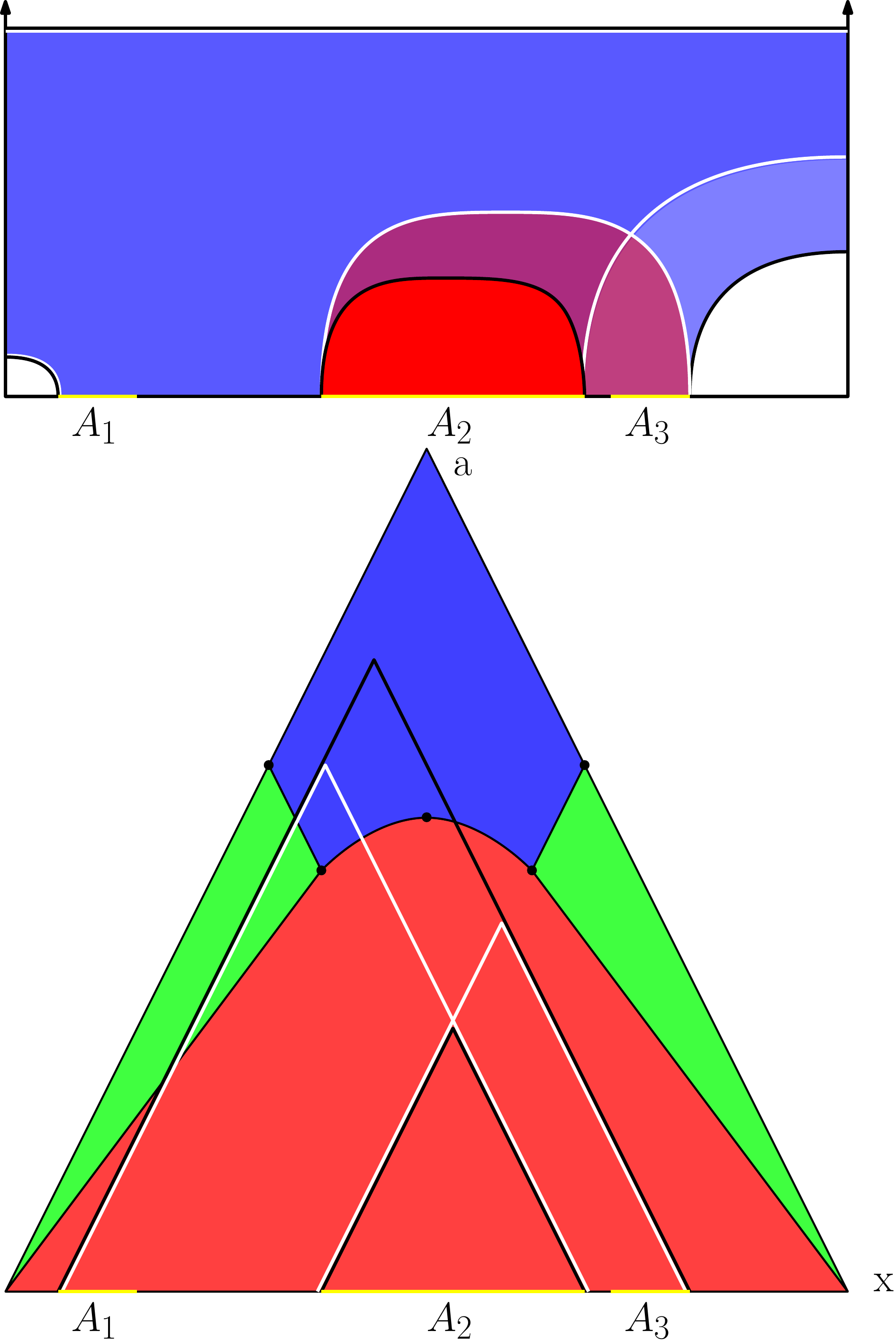} }\hspace{0.3cm}
  \subfloat[kkk]{
    \includegraphics[  width=.15\linewidth]{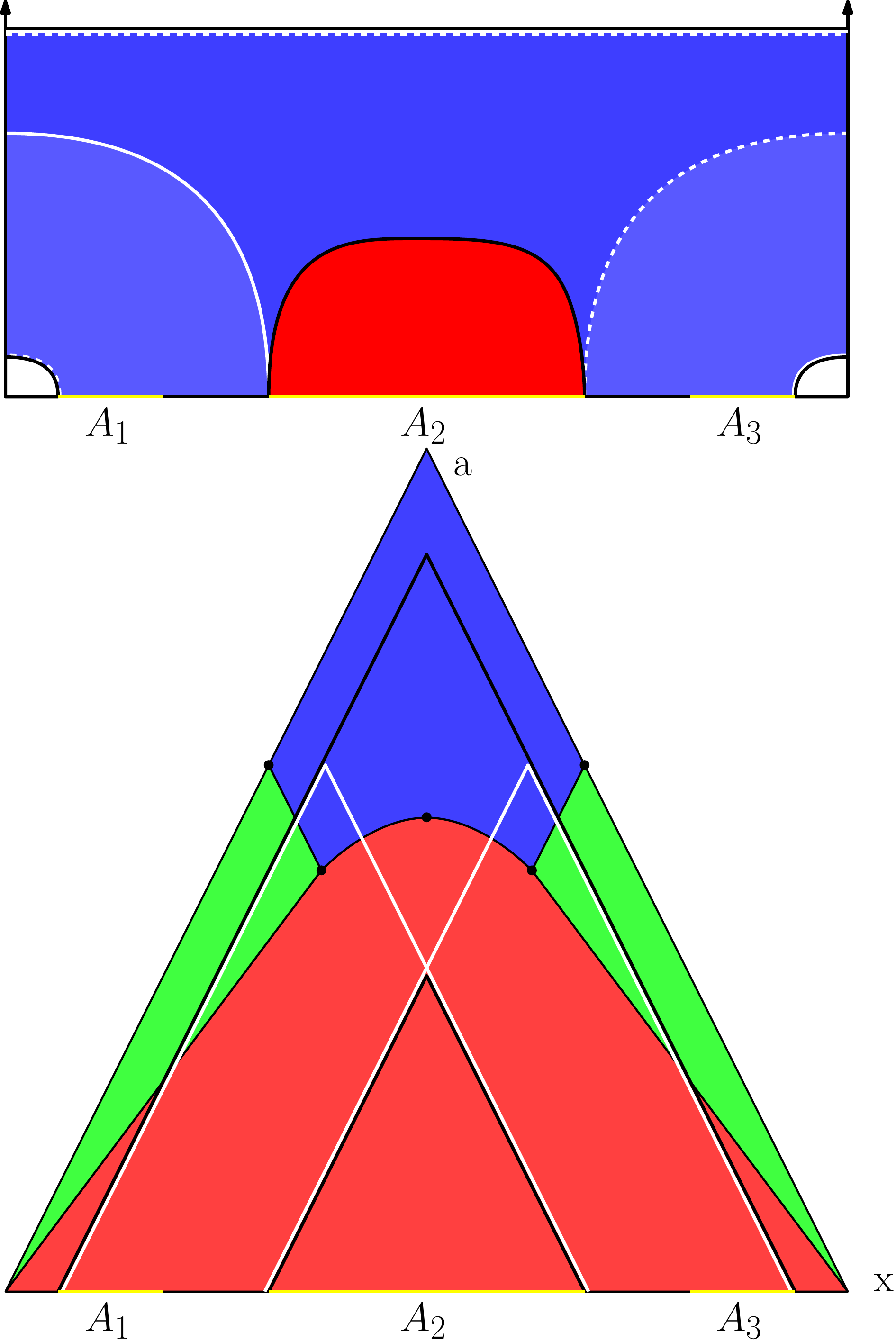} }
  \caption{The ten allowed cases for SSA in the black hole background corresponding to the ones in the pure AdS background.}
  \label{fig_SSAsame}
\end{figure}

\begin{figure}
  \includegraphics[width=.3\linewidth]{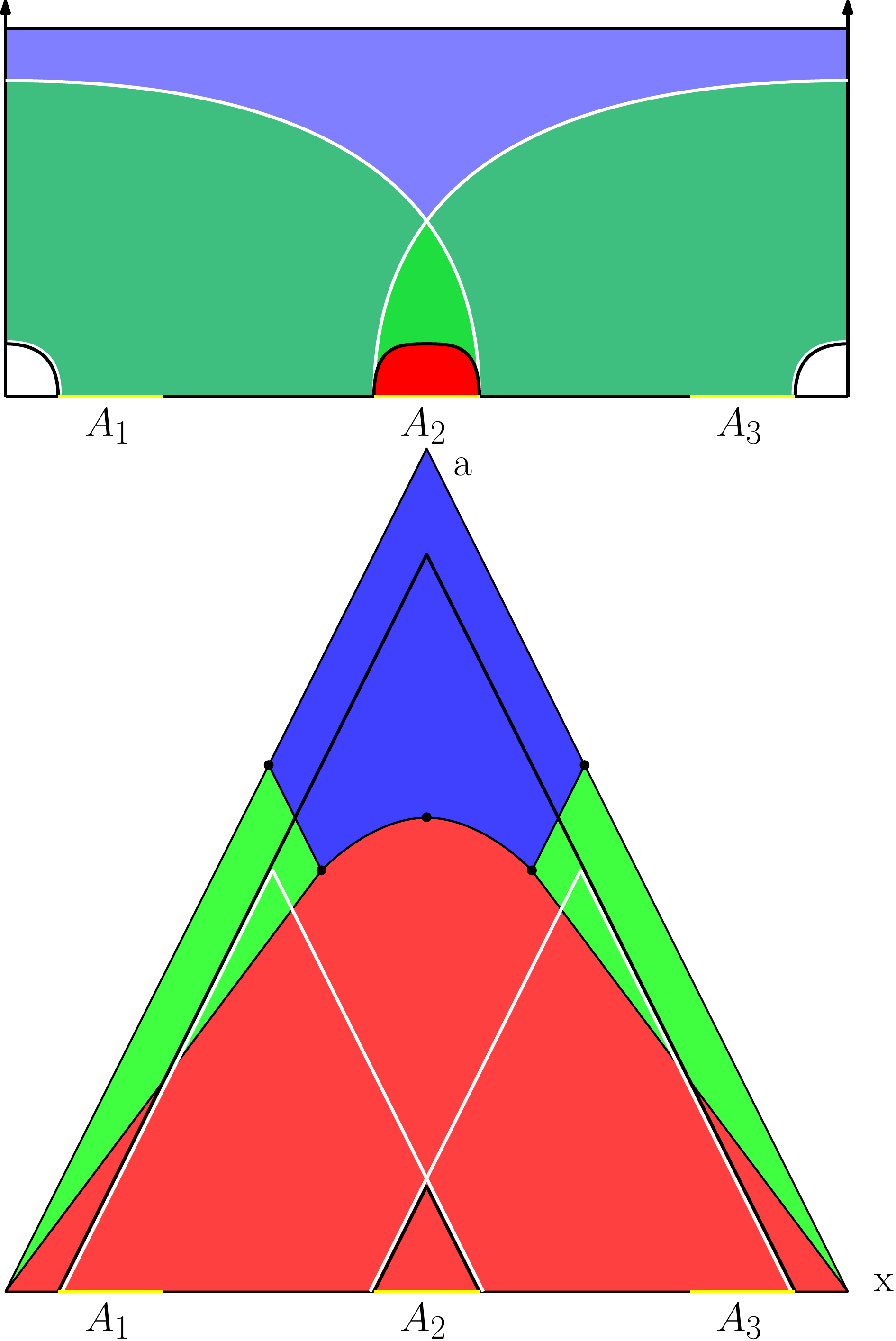}
  \caption{The new allowed case for SSA in the black hole background.}
  \label{fig_BH-new}
\end{figure}

In this section, we consider the strong subadditivity (SSA) Eq.(\ref{SSA}) in the black hole background. As in the pure AdS background, we first consider the case that $S_{12}$, $S_{23}$ and $S_{123}$ are all in the connected configuration as in Eq.(\ref{SSAf}) with $S_2$ being in the sunset phase. While the other three terms, $S_{[12]}$, $S_{[13]}$ and $S_{[23]}$ could be in any of the three phases. Besides the 10 cases listed in TABLE \ref{SSAtable} and plotted in Fig.\ref{fig_SSAsame}, there is a new allowed case in the black hole background, plotted in Fig.\ref{fig_BH-new}, due to the fact that the sky phase in the phase diagram is shrunk in the black hole background.

Among the ten cases in Fig.\ref{fig_SSAsame}, five of them, ($a$) - ($e$), do not involve the sky phase so that their RT surfaces are the same as the ones in the pure AdS background. While the other five cases, ($f$) - ($j$), involve the sky phase. By observing that the term involves the sky phase is always on the larger side of the inequality, we obtain that adding an extra black hole entropy $S_{BH}$ on this side does not change the inequality.

Finally, the new allowed case (rrk) is plotted in Fig.\ref{fig_BH-new}. By cutting and rejoining the two white curves and noticing the cancellation between the black and white curves around the two corners, it is easy to show that the sum of the white curves is larger than the sum of the black curves.

Furthermore, for $S_{12}$, $S_{23}$ and $S_{123}$ are not in the connected configuration, the similar argument as in the pure AdS background still holds now. Therefore, we prove the SSA in the black hole background.

\subsection{Monogamy of mutual information}
As in the pure AdS background, we first prove that $S_{12}$, $S_{23}$, $S_{13 }$ and $S_{123  }$ are all in the connected configurations Eq.(\ref{MMIc}), which reduces to Eq.(\ref{MMIcc}) by using the notation in Eqs.(\ref{S12c}) - (\ref{S123c}). Similarly, among the six terms in Eq.(\ref{MMIcc}), $S_{\left\langle
13\right\rangle }$ and  $S_2$ must be in the sunset phase. While the other four terms, $S_{[12]}$, $S_{[23]}$, $S_{1}$ and $S_{3}$, and $S_{\left[  23\right]  }$, could be in any of the three phases. By using the rules we discussed in the last section, we find total 28 allowed cases that are three more than the allowed cases in the pure AdS background.

\begin{figure}
  \subfloat[ssss]{
    \includegraphics[  width=.15\linewidth]{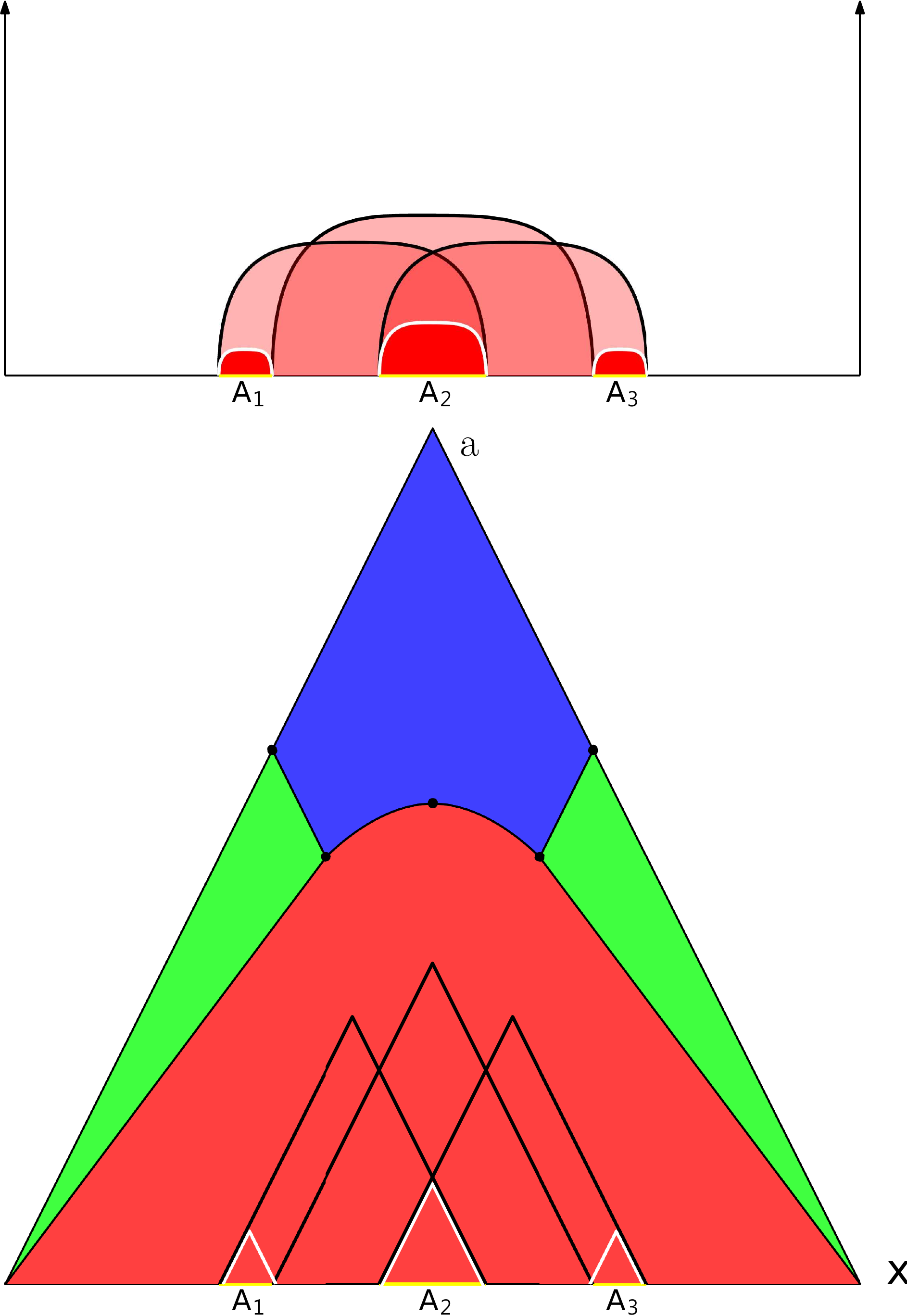} }\hspace{0.3cm}
  \subfloat[srss]{
    \includegraphics[  width=.15\linewidth]{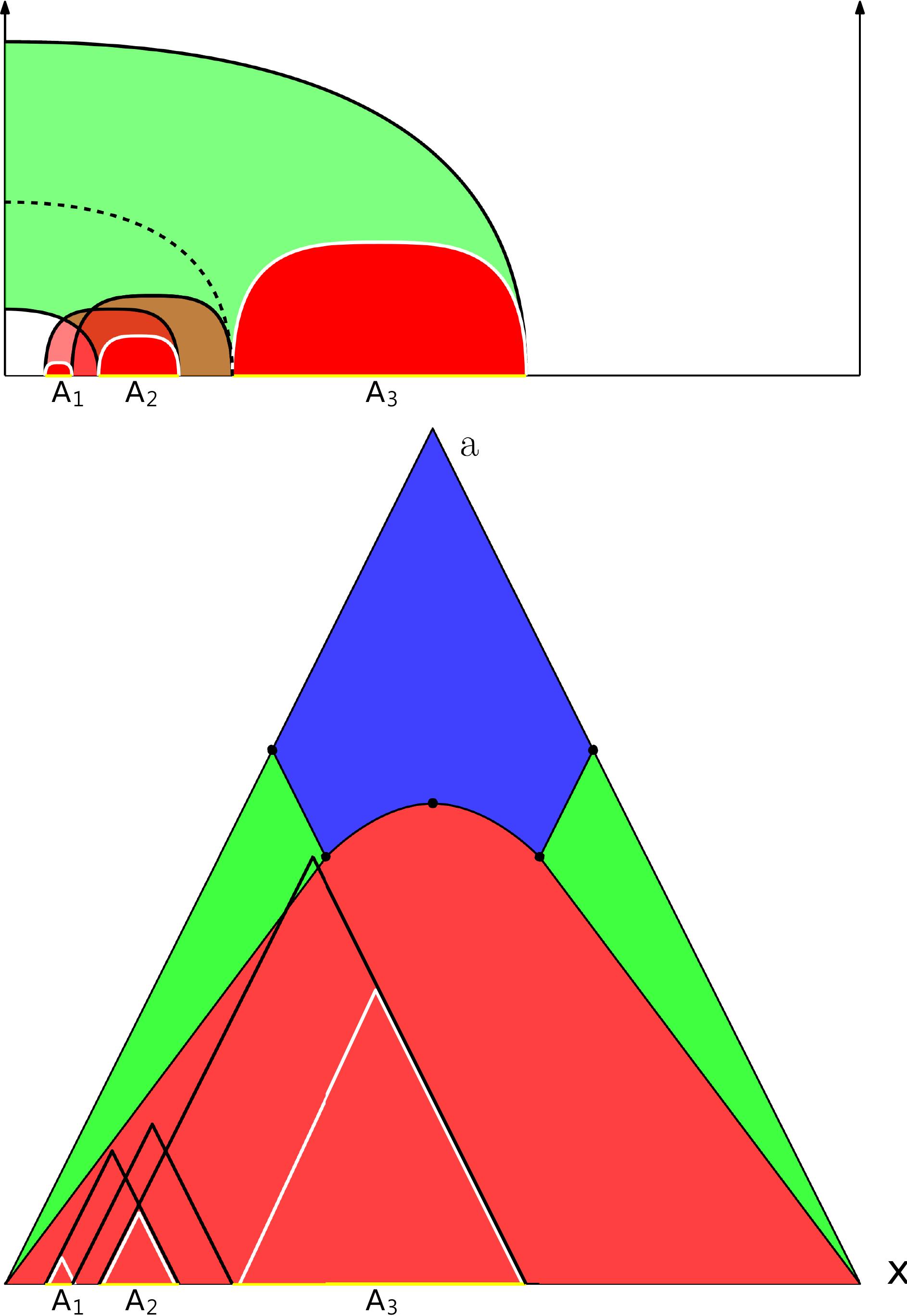} }\hspace{0.3cm}
  \subfloat[srsr]{
    \includegraphics[  width=.15\linewidth]{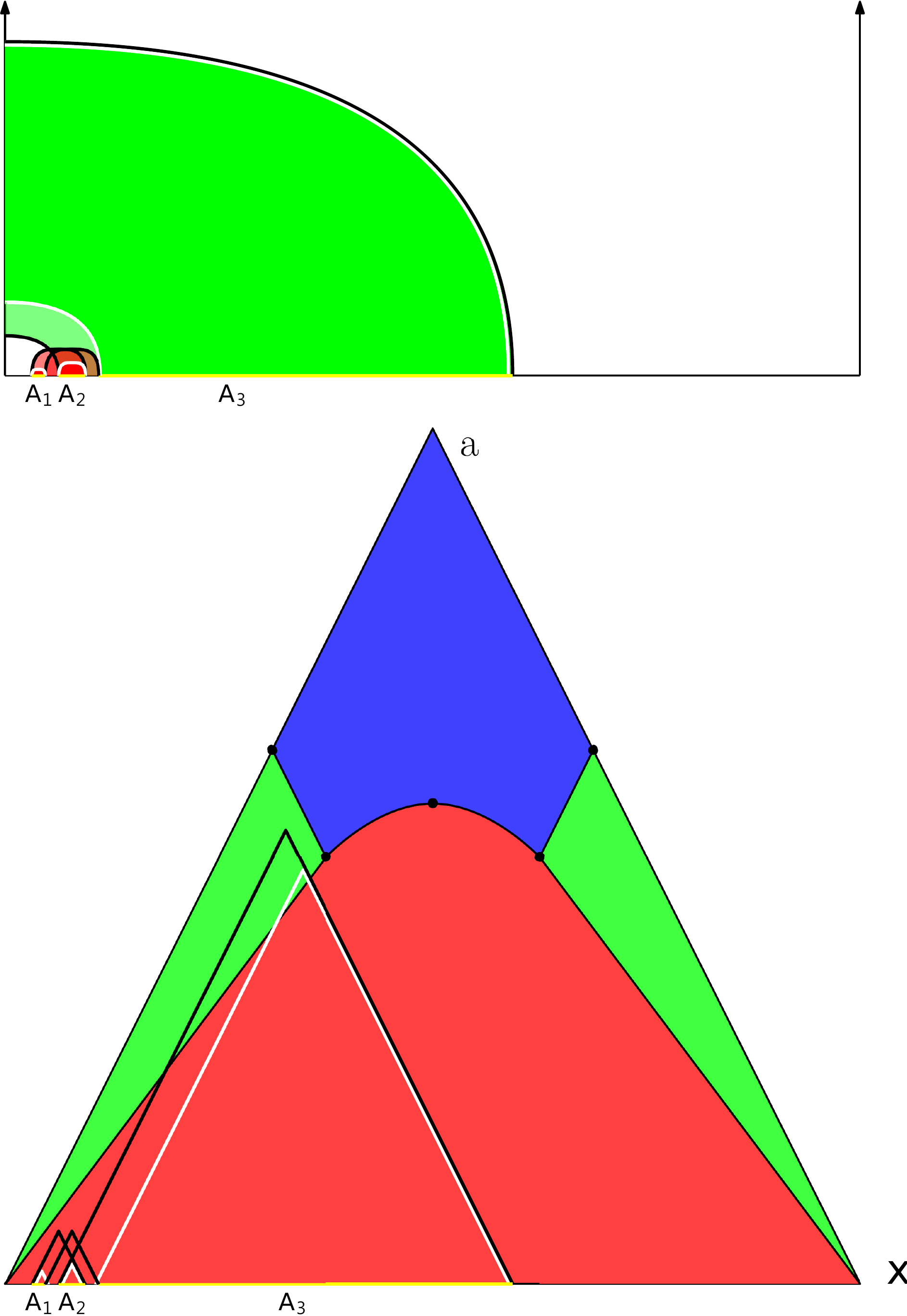} }\hspace{0.3cm}
  \subfloat[rsss]{
    \includegraphics[  width=.15\linewidth]{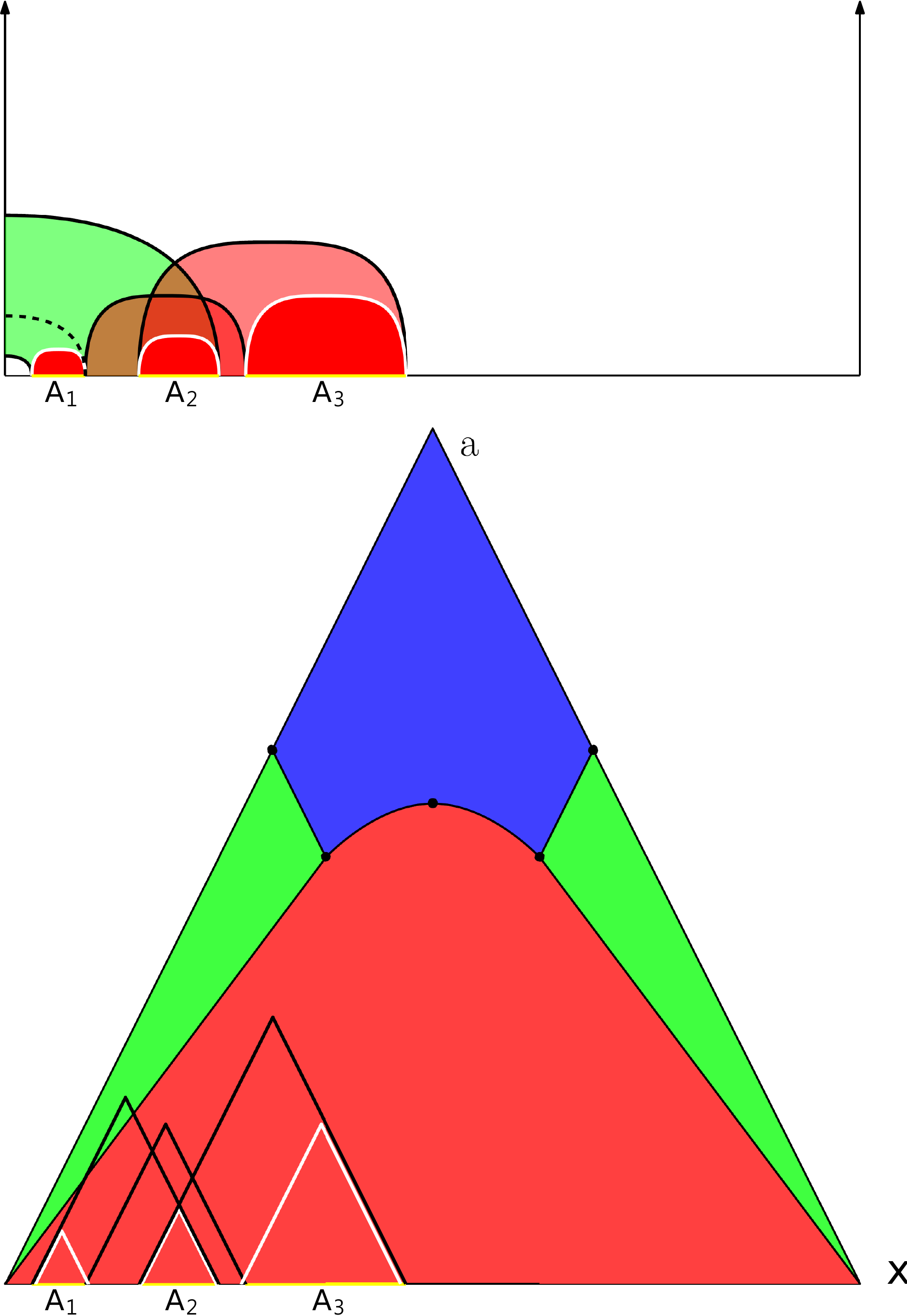} }\hspace{0.3cm}
  \subfloat[rsrs]{
    \includegraphics[  width=.15\linewidth]{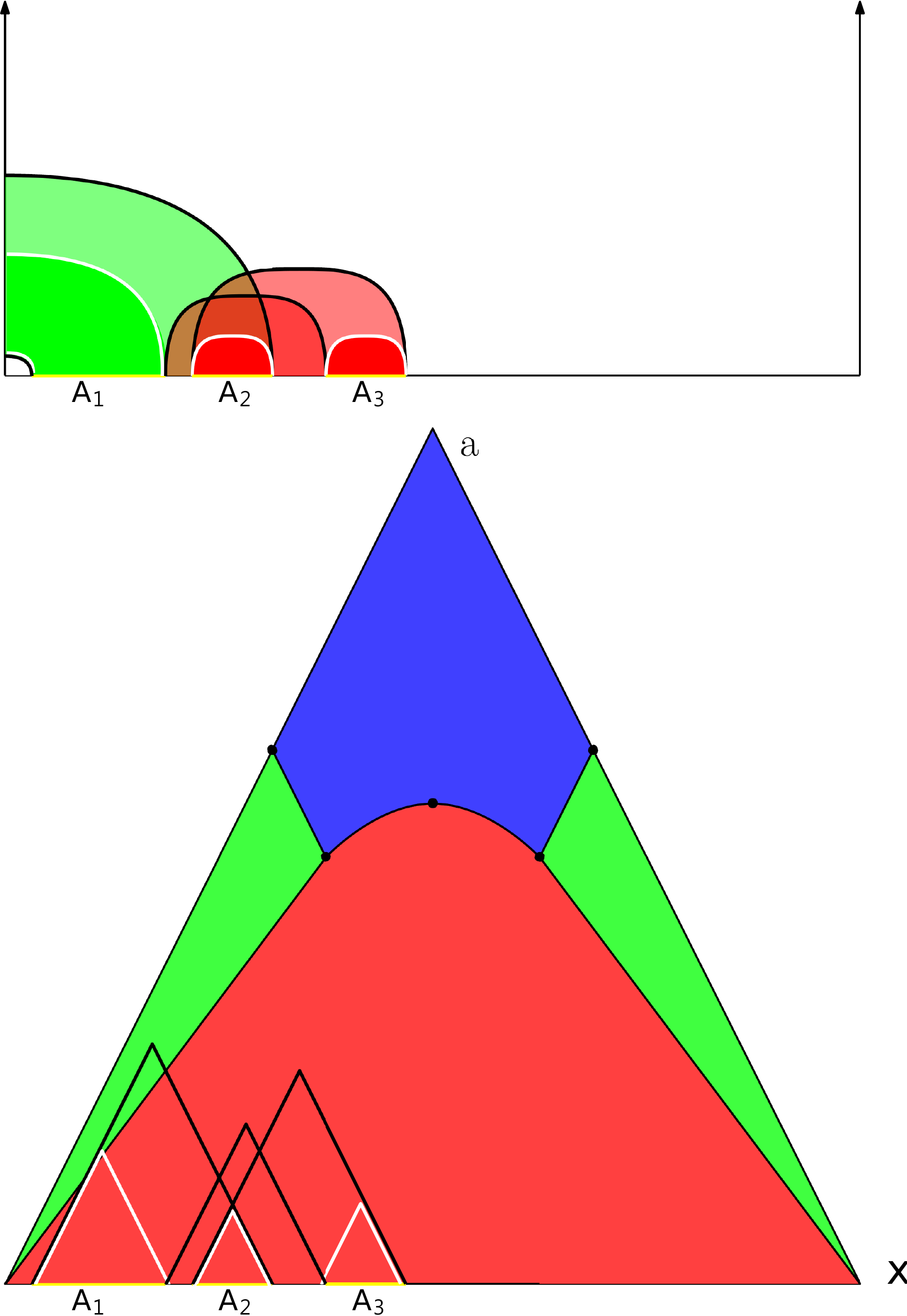} }\\
  \subfloat[rrss]{
    \includegraphics[  width=.15\linewidth]{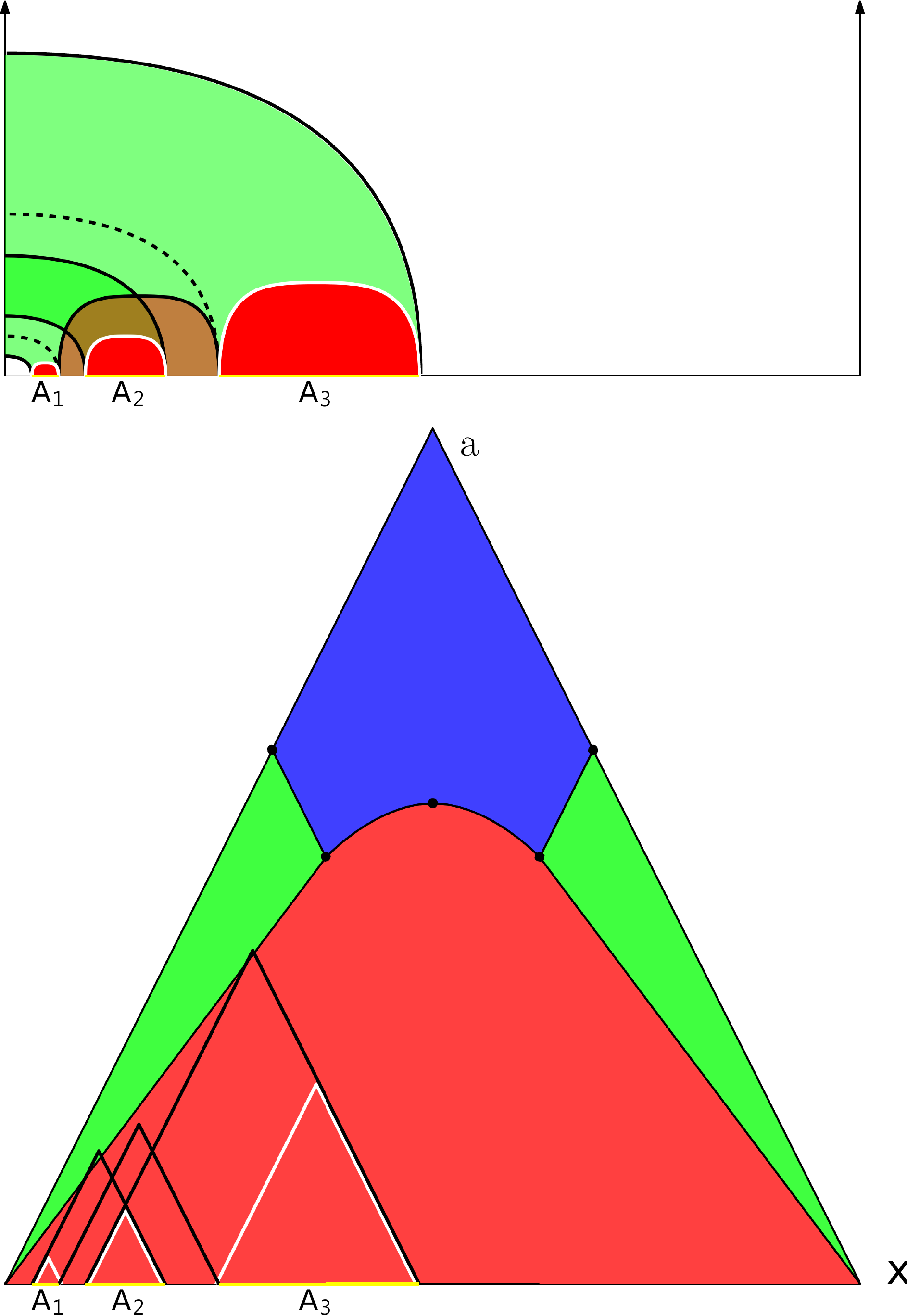} }\hspace{0.3cm}
  \subfloat[rrsr]{
    \includegraphics[  width=.15\linewidth]{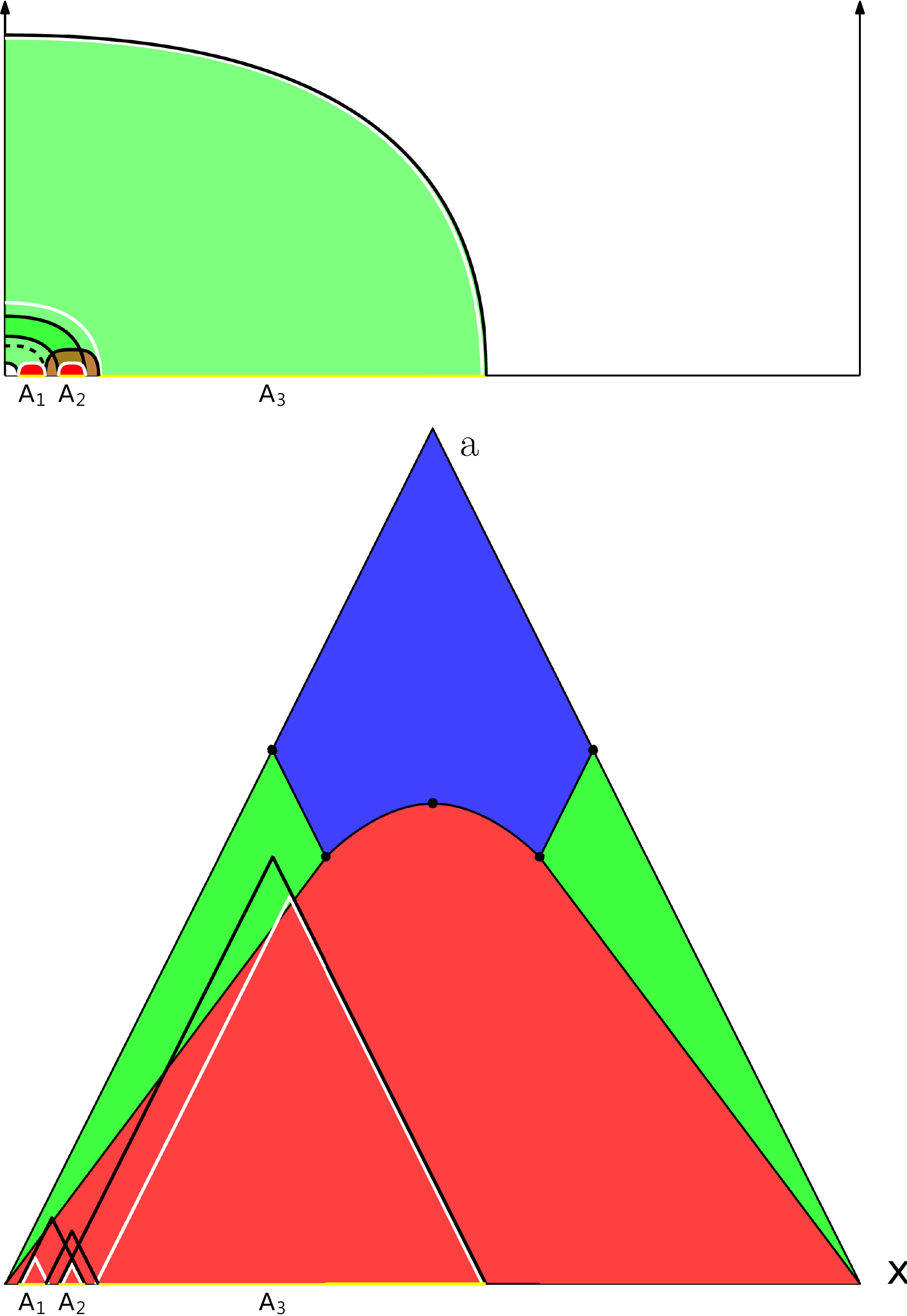} }\hspace{0.3cm}
  \subfloat[rrrs]{
    \includegraphics[  width=.15\linewidth]{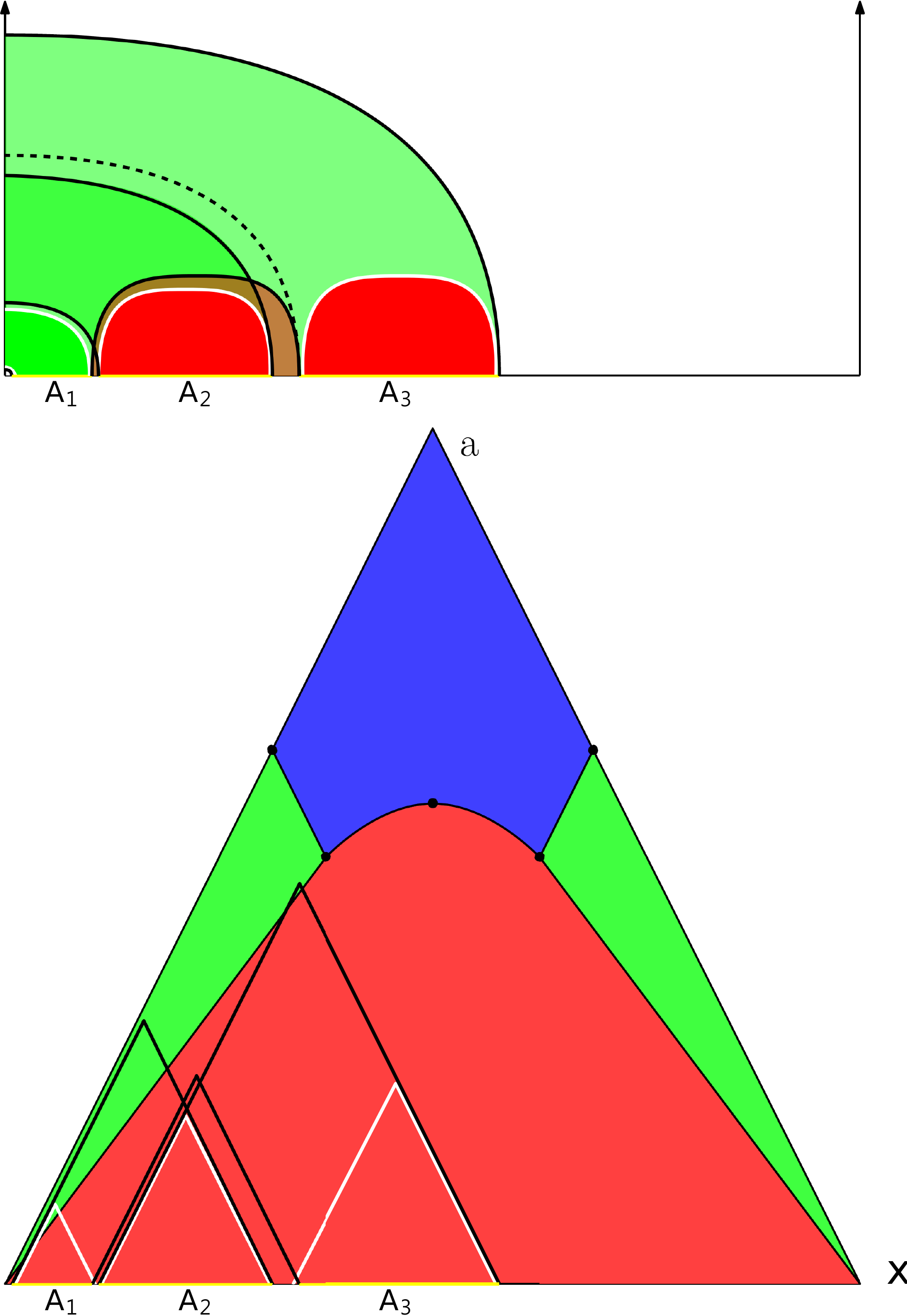} }\hspace{0.3cm}
  \subfloat[rrrr]{
    \includegraphics[  width=.15\linewidth]{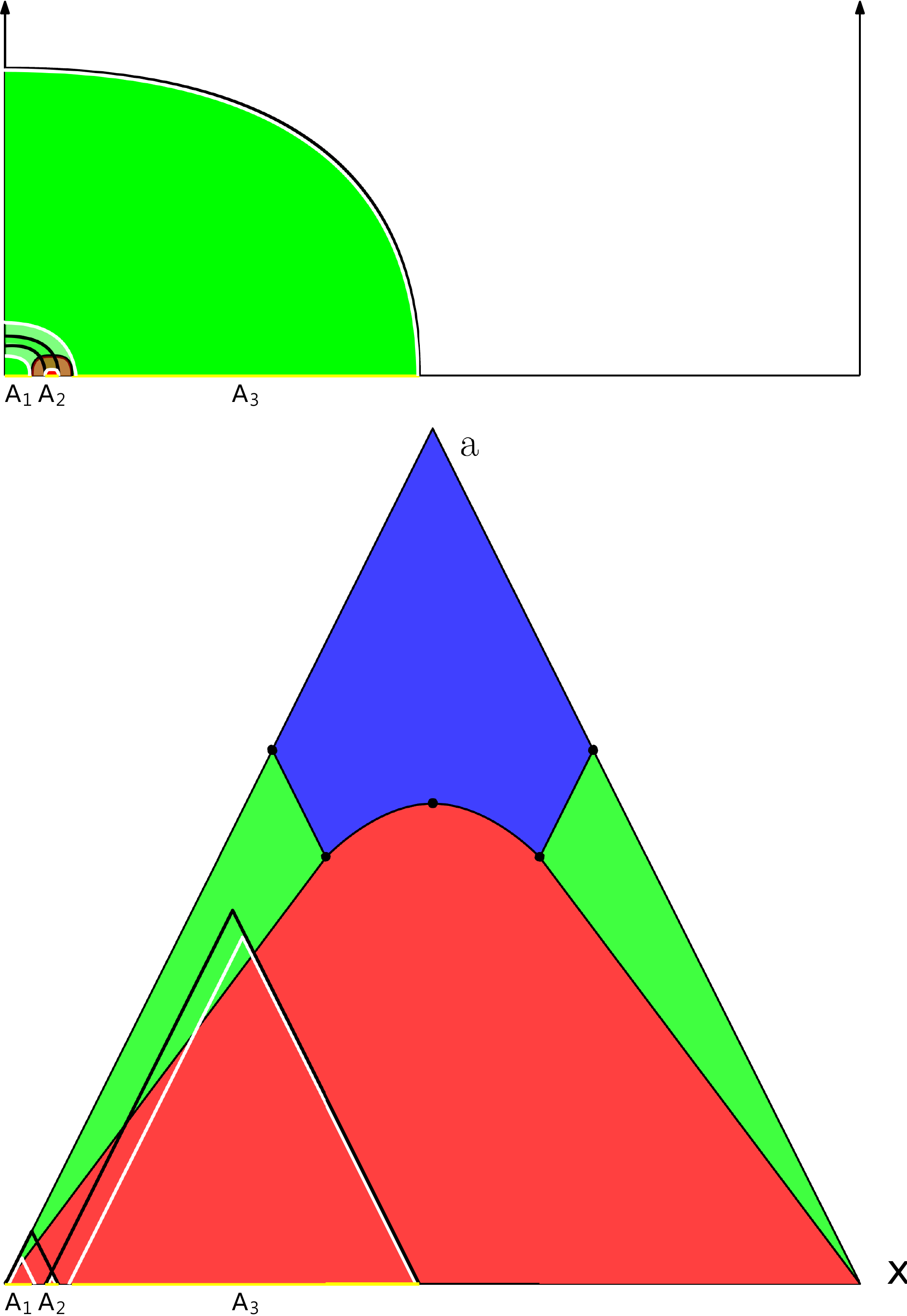} }\hspace{0.3cm}
  \subfloat[skss]{
    \includegraphics[  width=.15\linewidth]{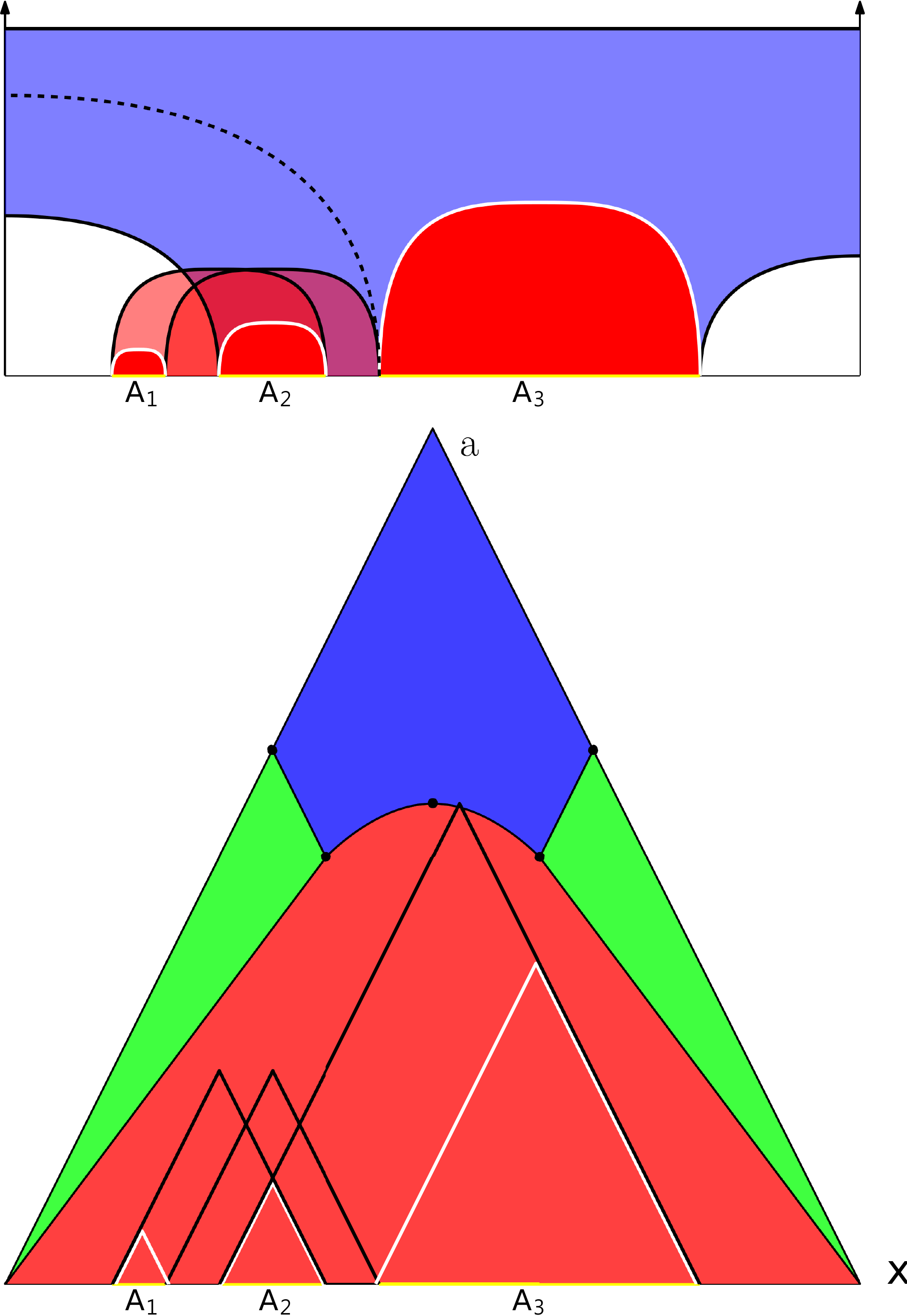} }\\
  \subfloat[sksr]{
    \includegraphics[  width=.15\linewidth]{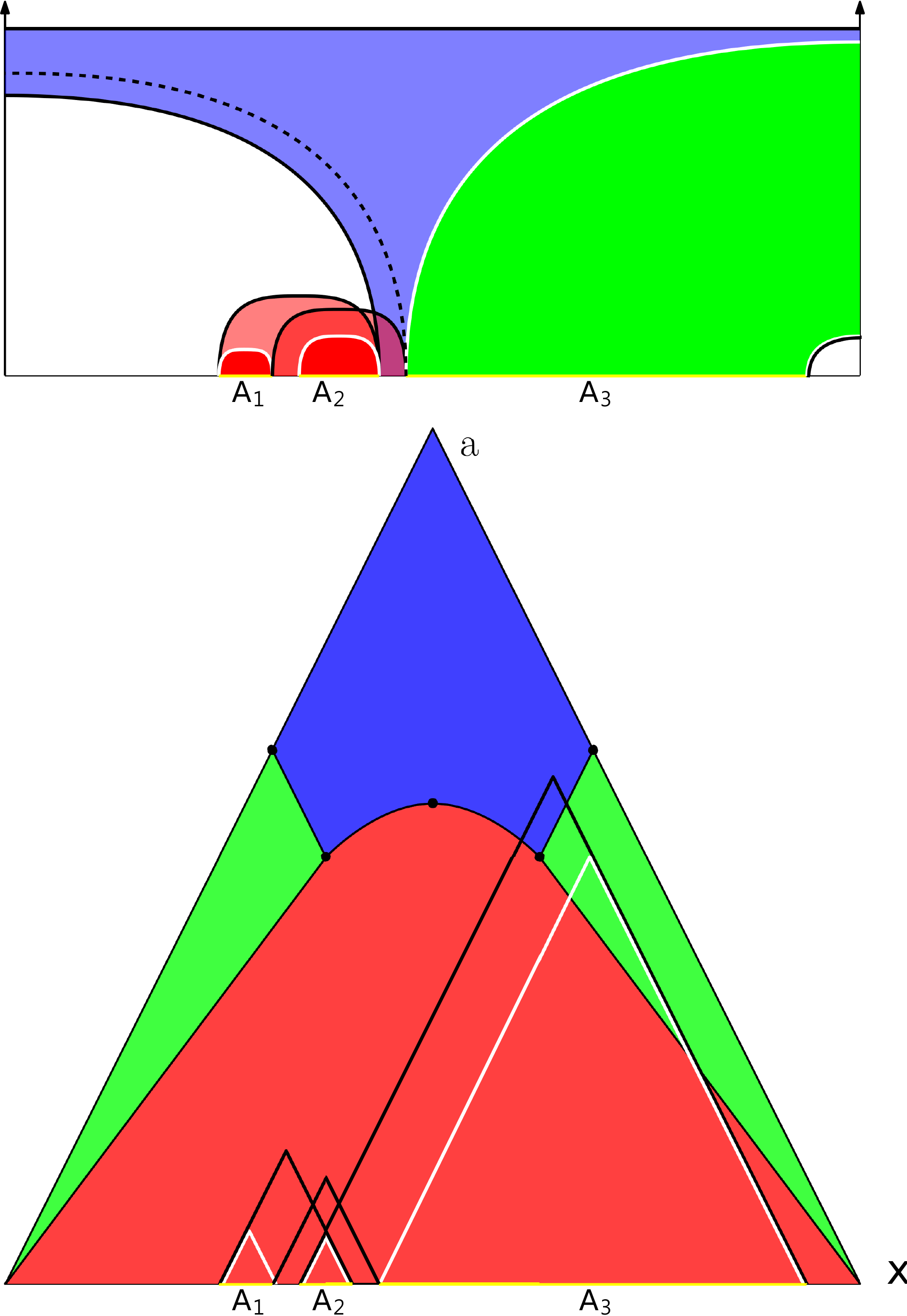} }\hspace{0.3cm}
  \subfloat[rkss]{
    \includegraphics[  width=.15\linewidth]{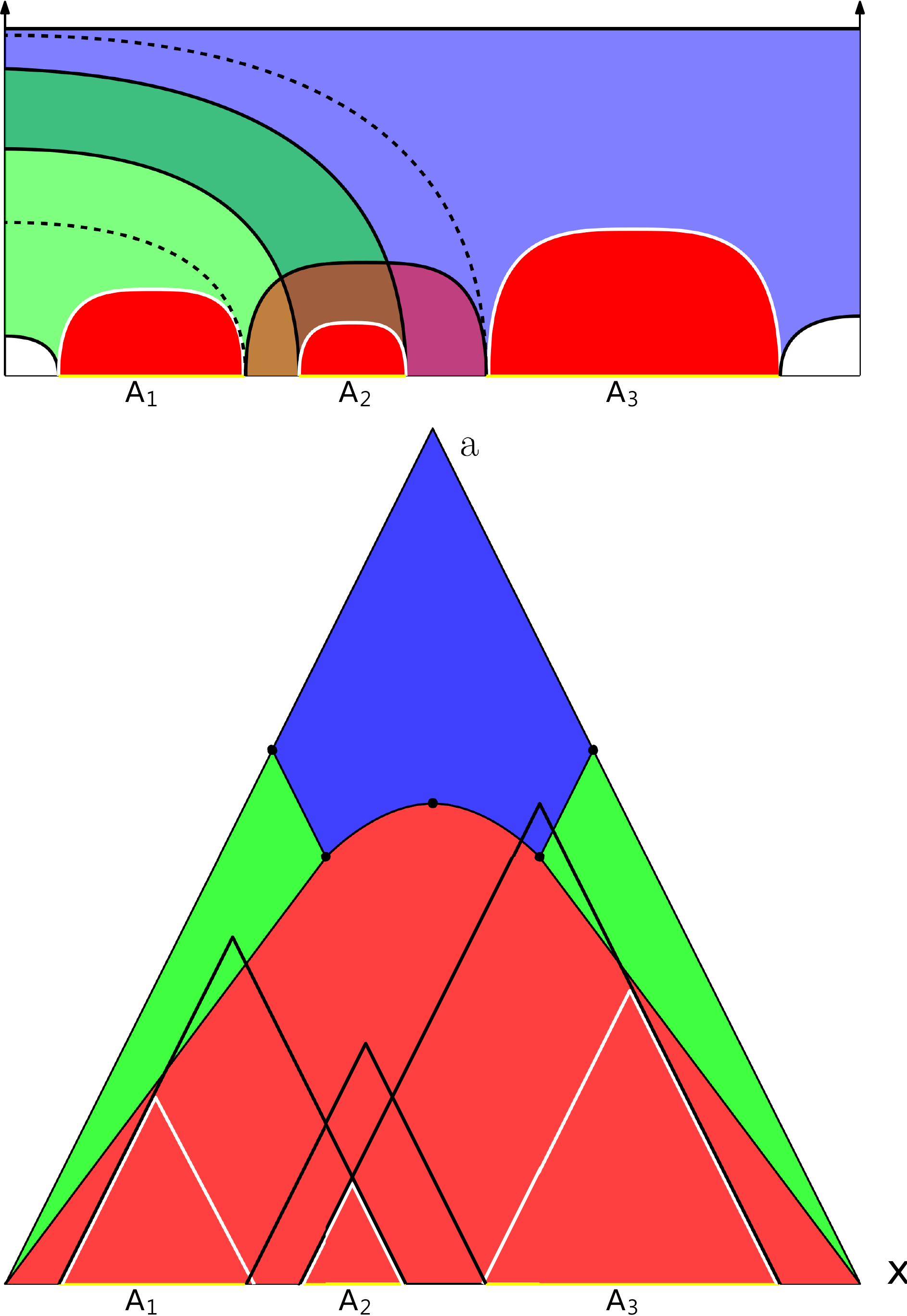} }\hspace{0.3cm}
  \subfloat[rksr]{
    \includegraphics[  width=.15\linewidth]{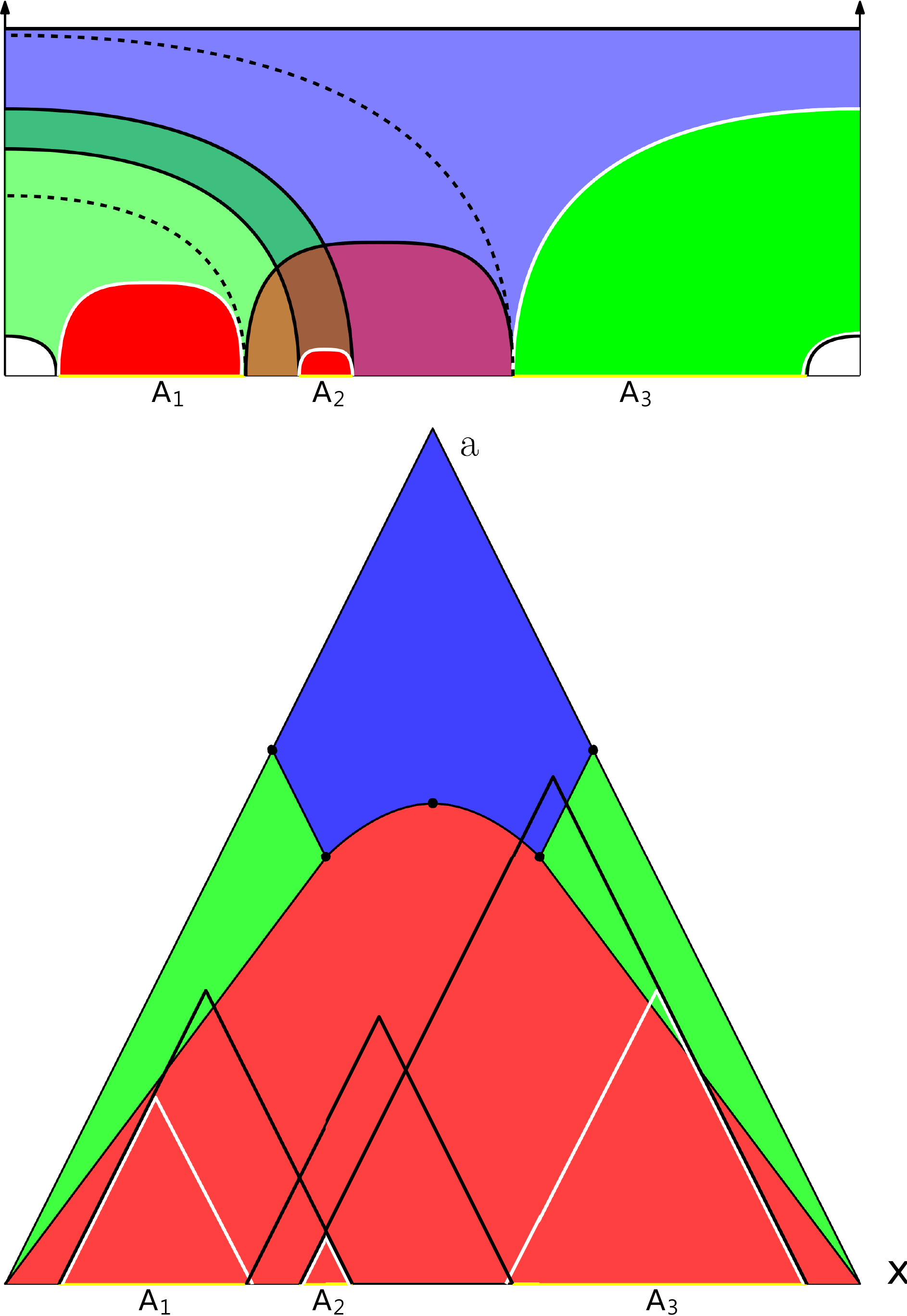} }\hspace{0.3cm}
  \subfloat[rkrs]{
    \includegraphics[  width=.15\linewidth]{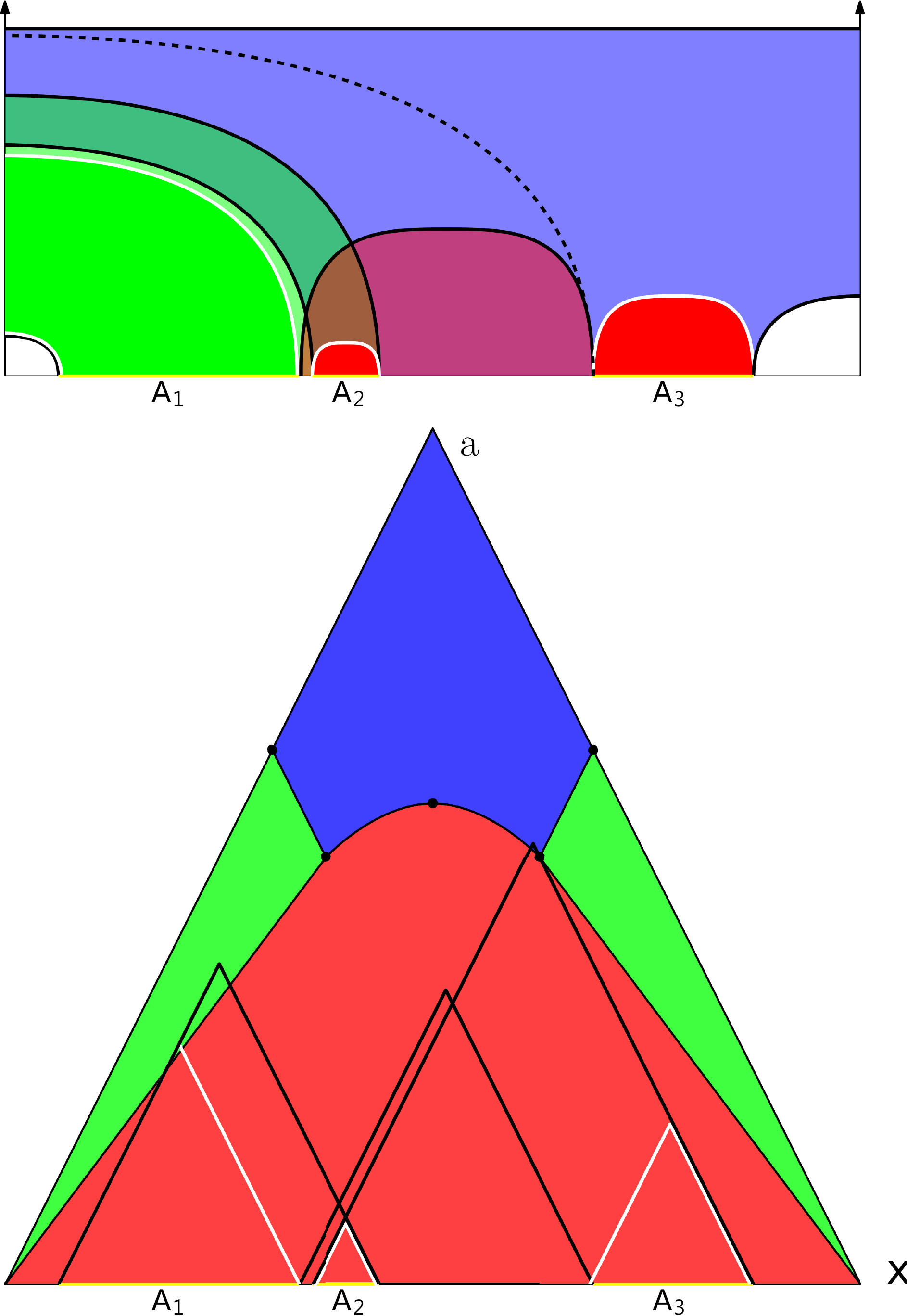} }\hspace{0.3cm}
  \subfloat[rkrr]{
    \includegraphics[  width=.15\linewidth]{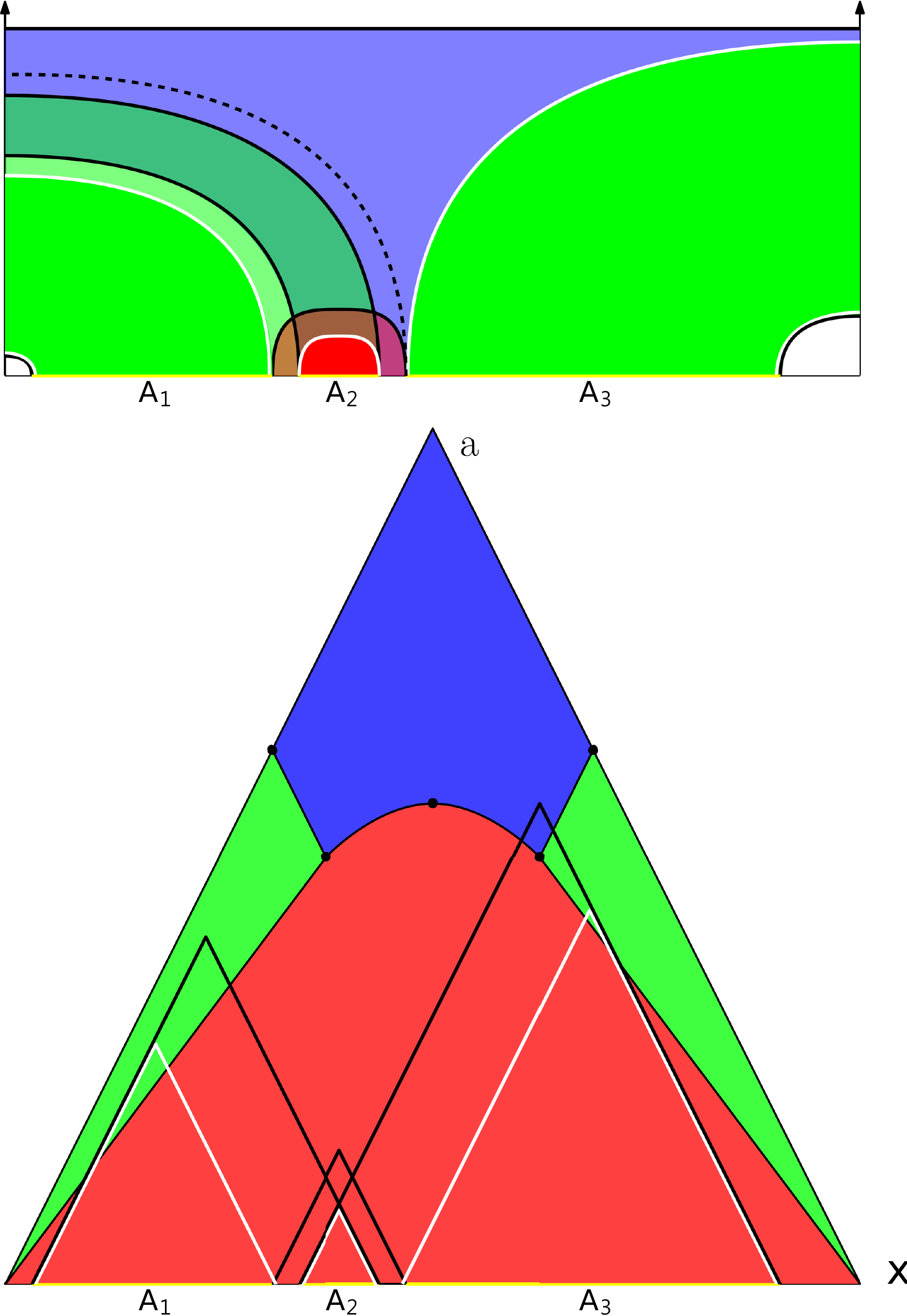} }\\
  \subfloat[ksss]{
    \includegraphics[  width=.15\linewidth]{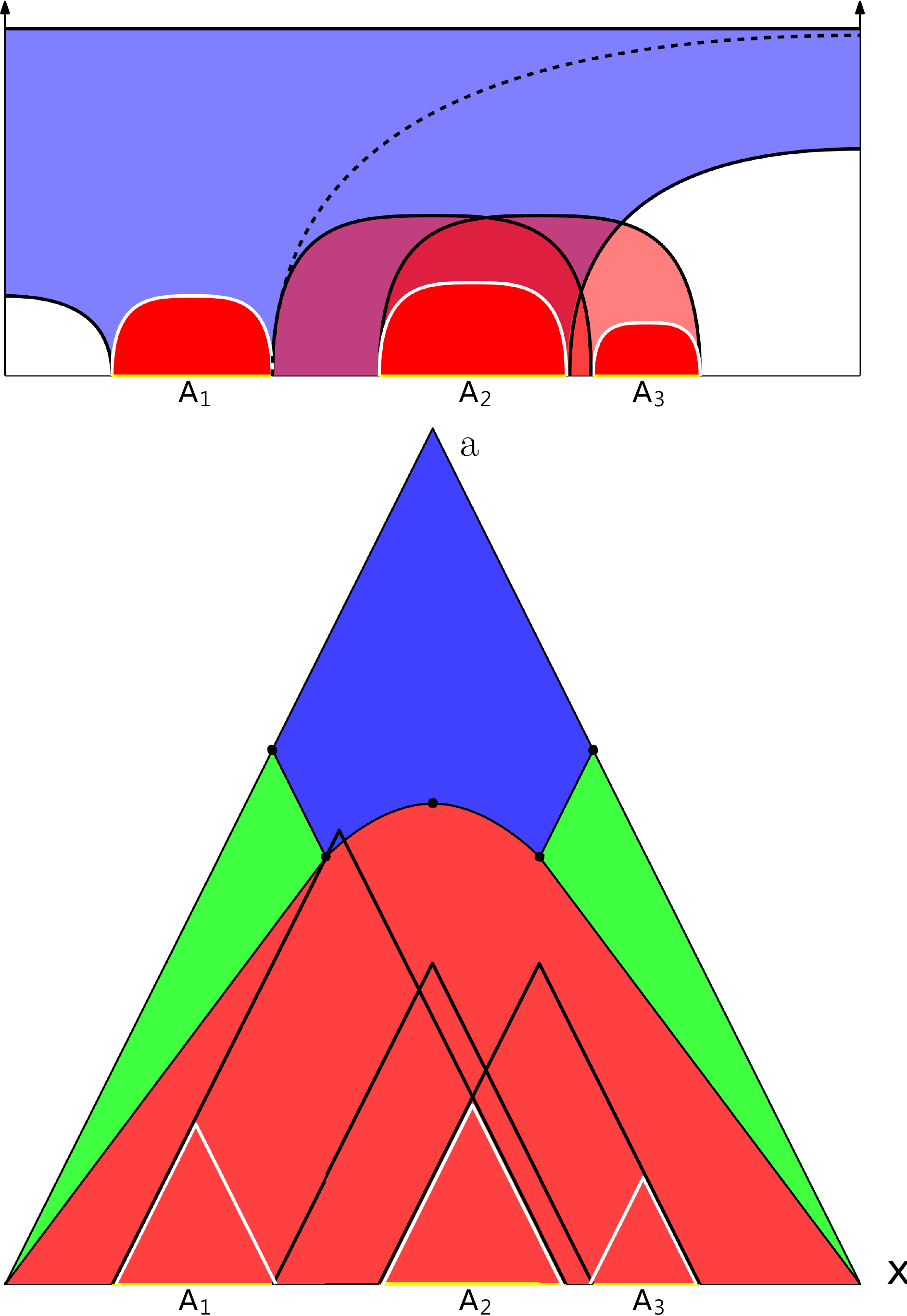} }\hspace{0.3cm}
  \subfloat[ksrs]{
    \includegraphics[  width=.15\linewidth]{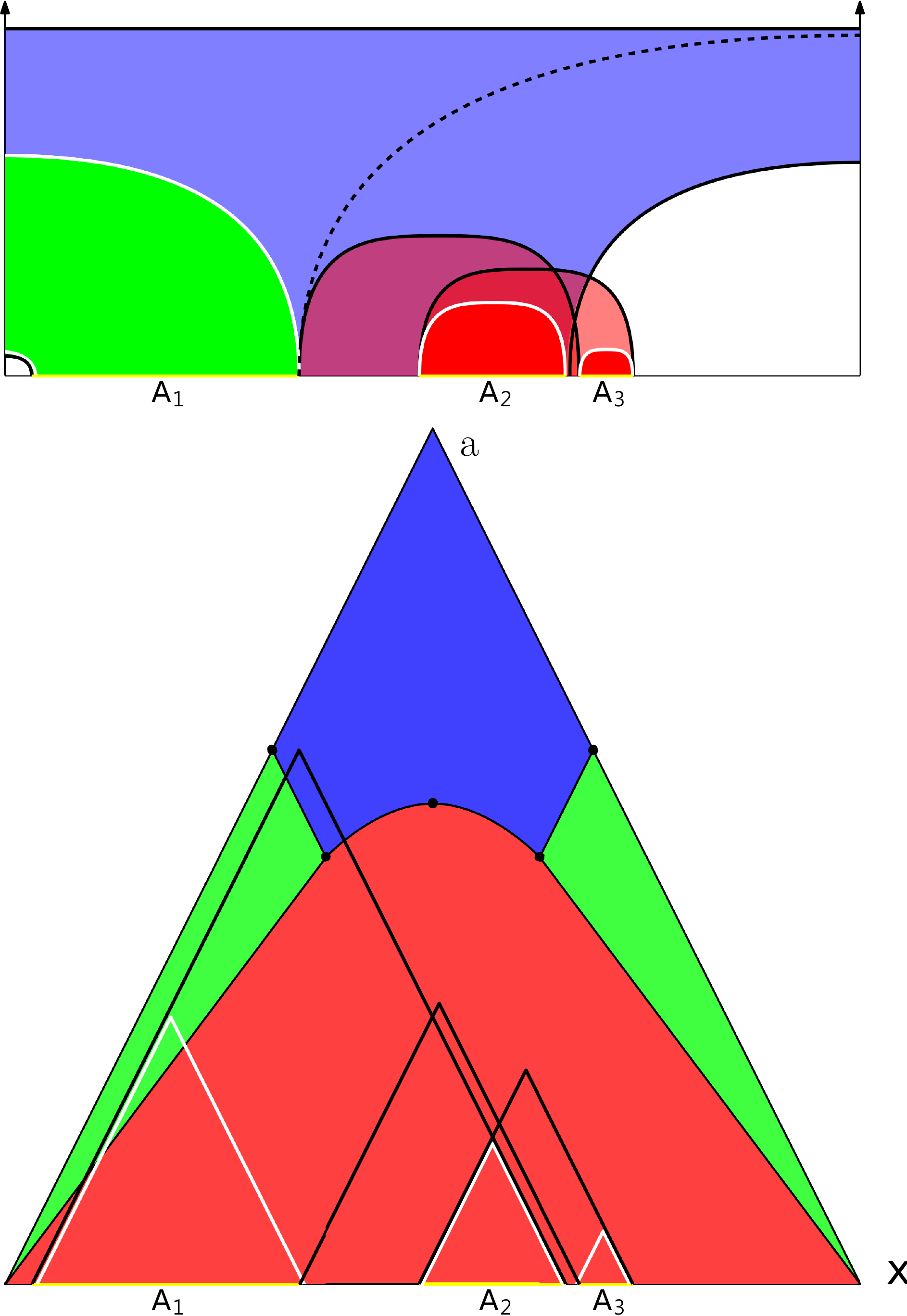} }\hspace{0.3cm}
  \subfloat[sksk]{
    \includegraphics[  width=.15\linewidth]{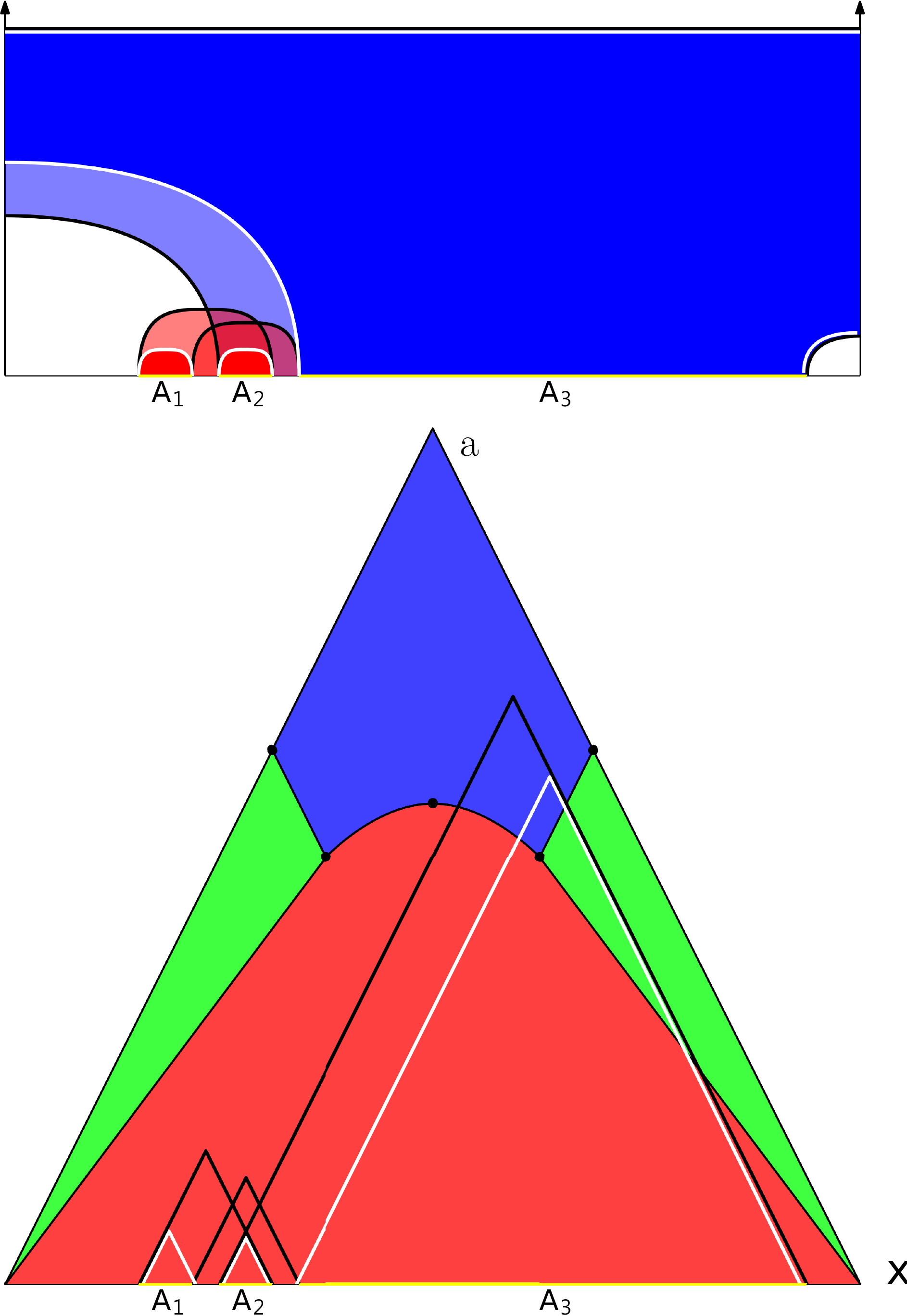} }\hspace{0.3cm}
  \subfloat[rksk]{
    \includegraphics[  width=.15\linewidth]{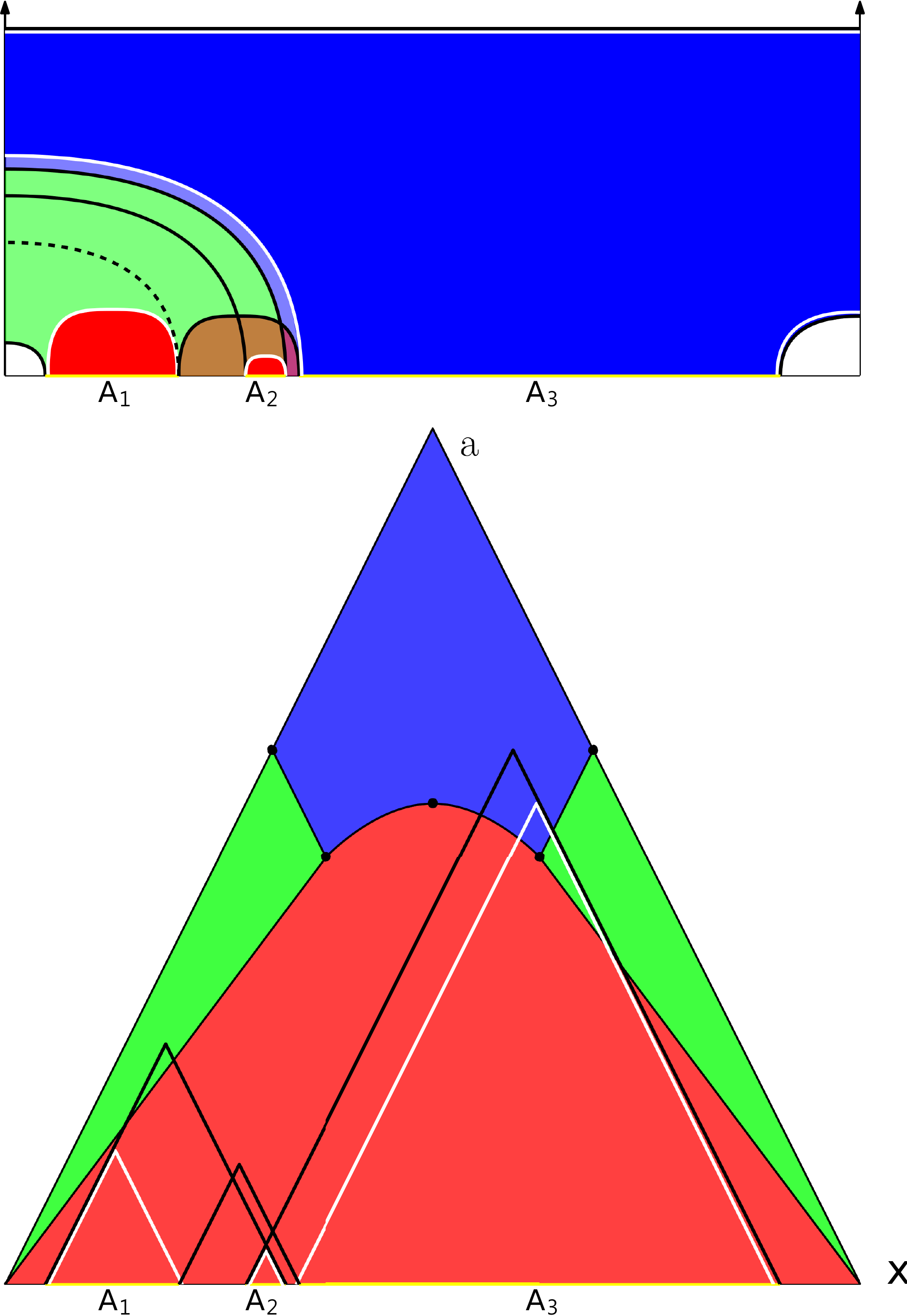} }\hspace{0.3cm}
  \subfloat[rkrk]{
    \includegraphics[  width=.15\linewidth]{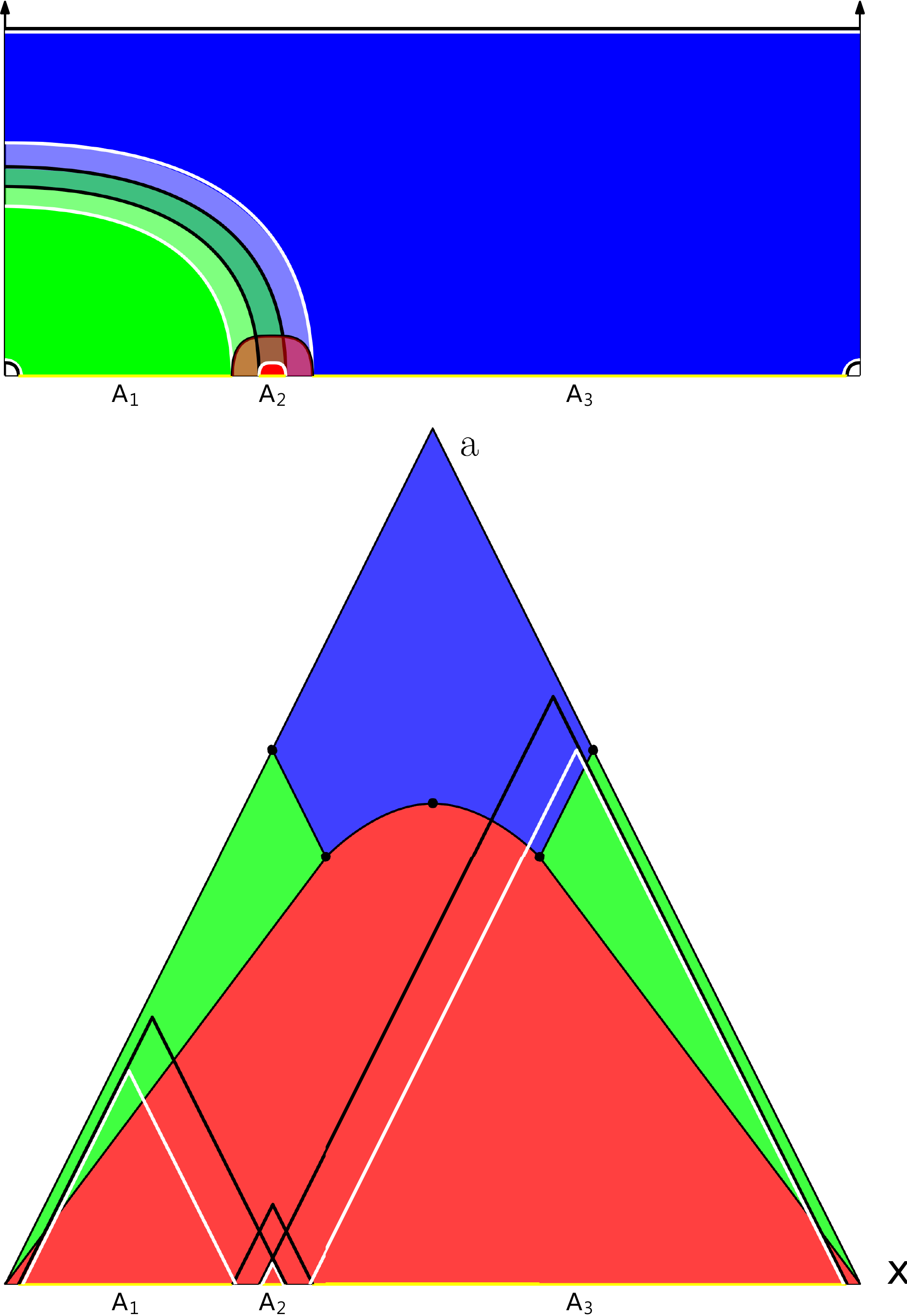} }\\
  \subfloat[ksks]{
    \includegraphics[  width=.15\linewidth]{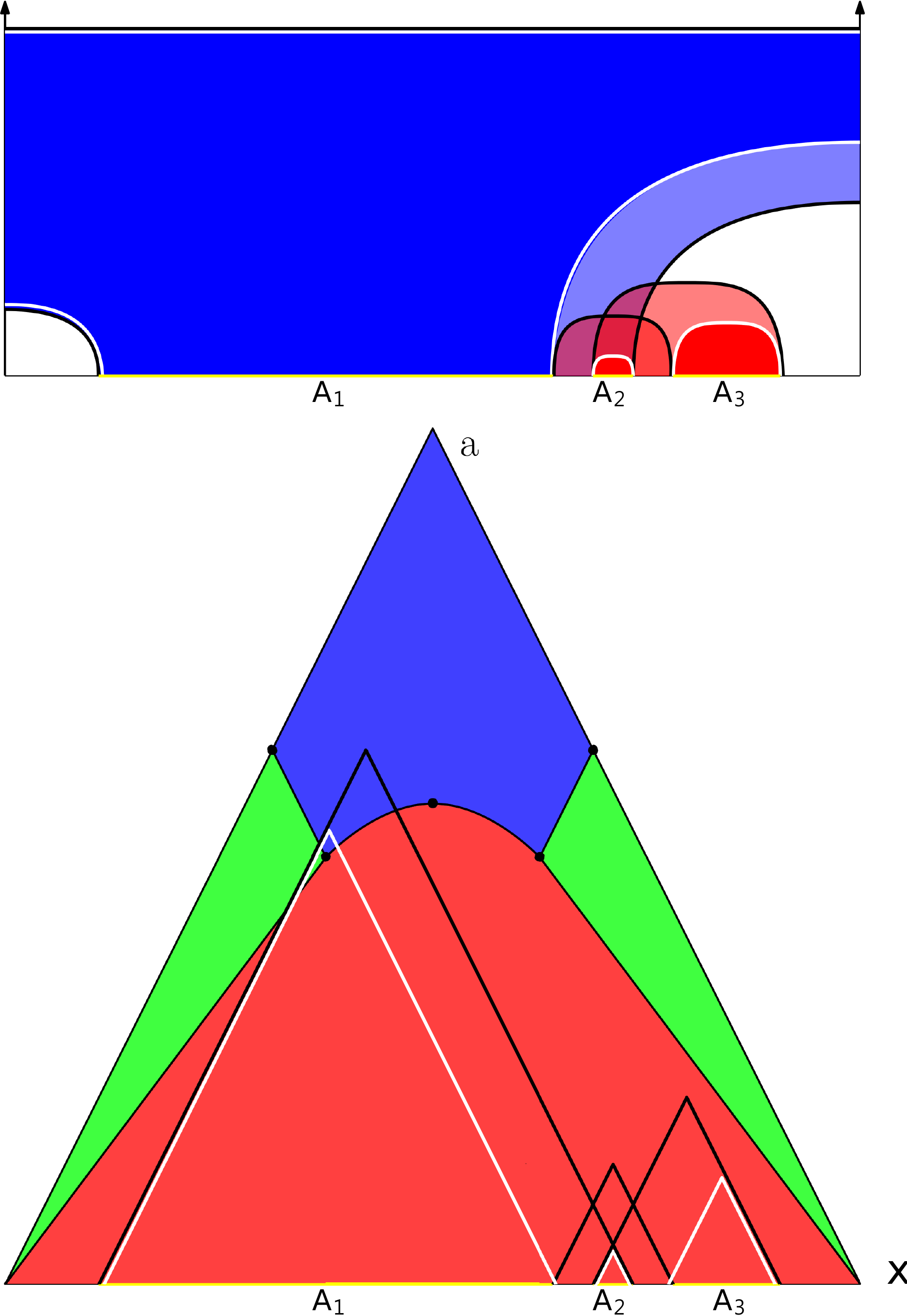} }\hspace{0.3cm}
  \subfloat[kkss]{
    \includegraphics[  width=.15\linewidth]{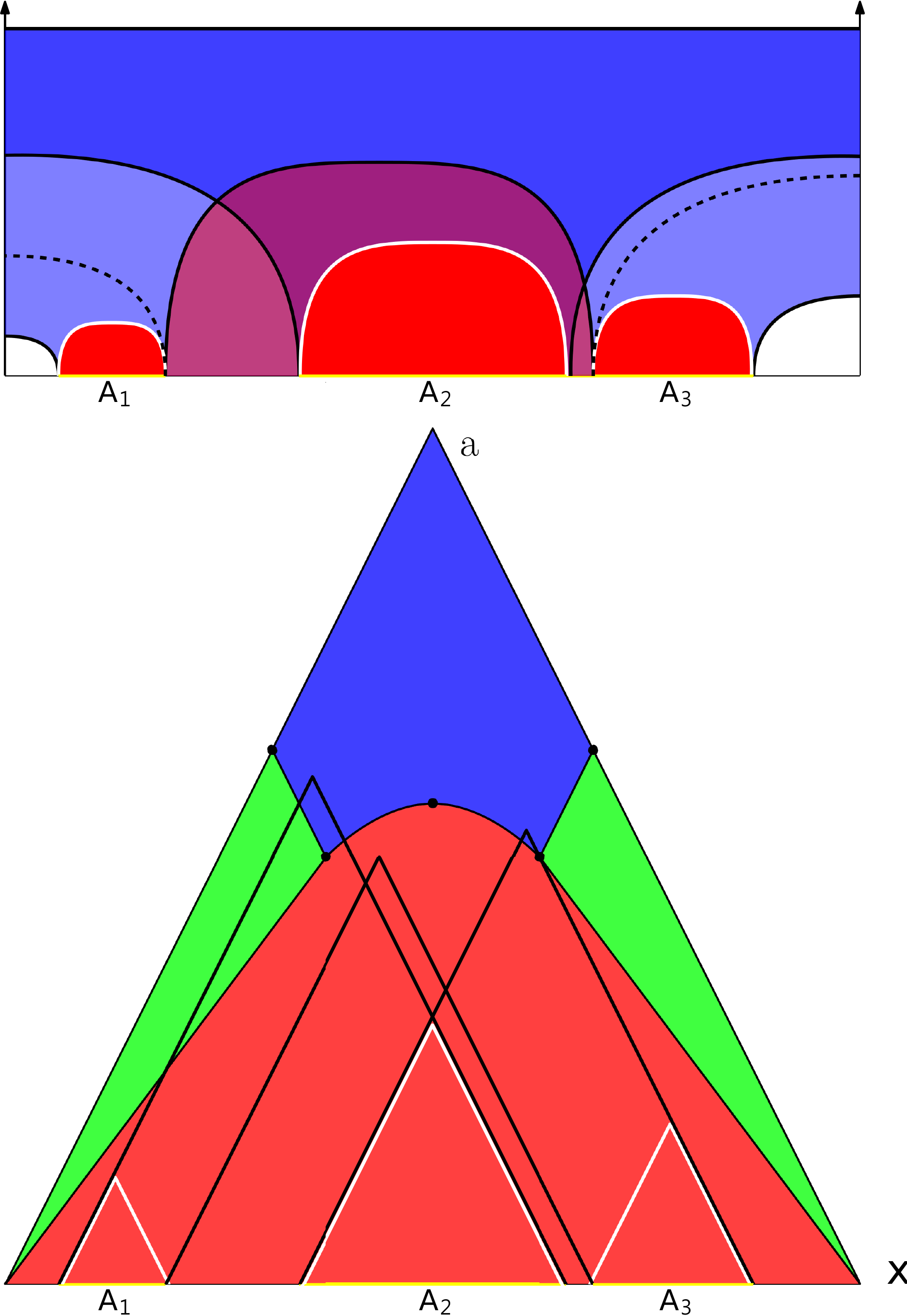} }\hspace{0.3cm}
  \subfloat[kksr]{
    \includegraphics[  width=.15\linewidth]{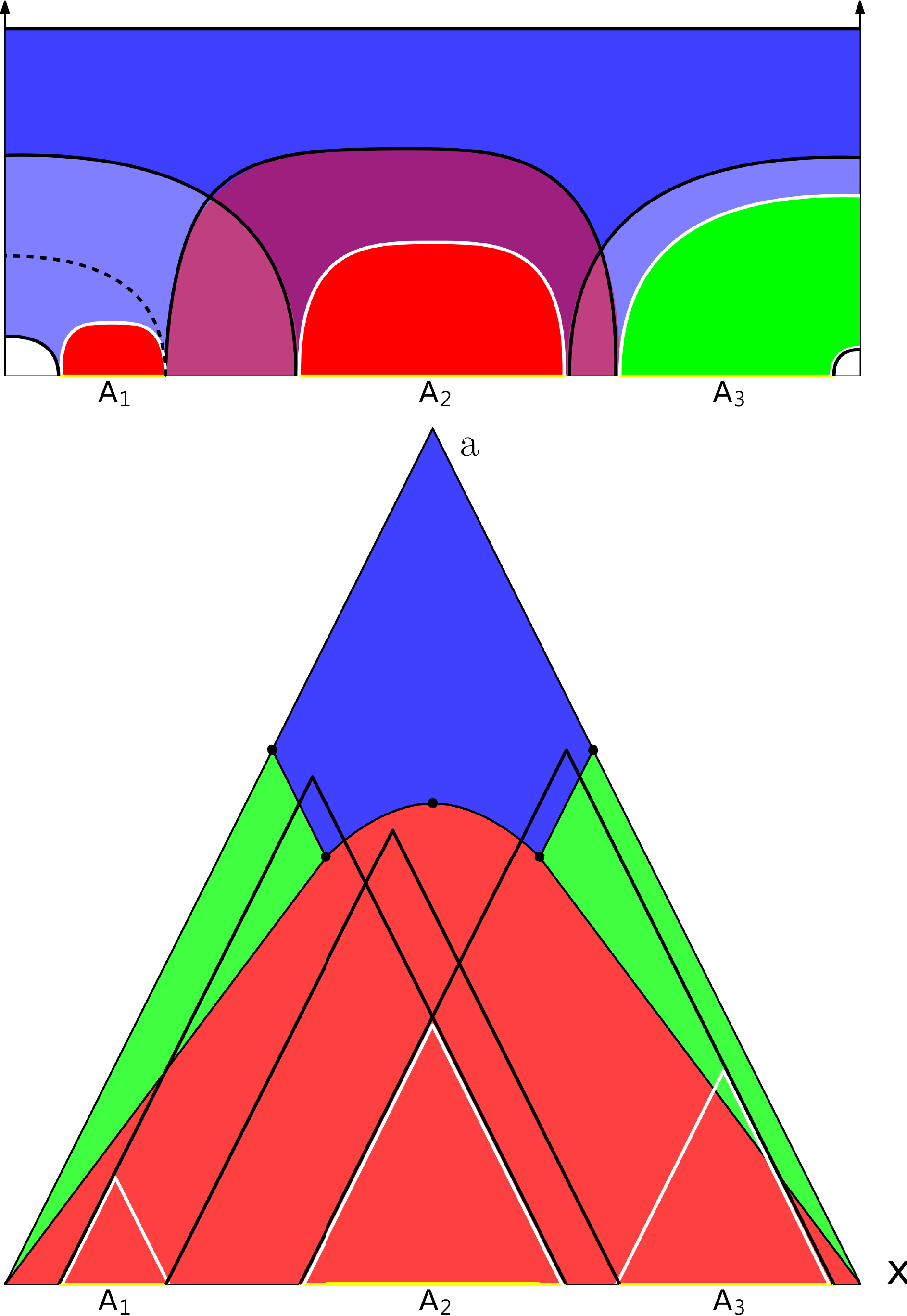} }\hspace{0.3cm}
  \subfloat[kkrs]{
    \includegraphics[  width=.15\linewidth]{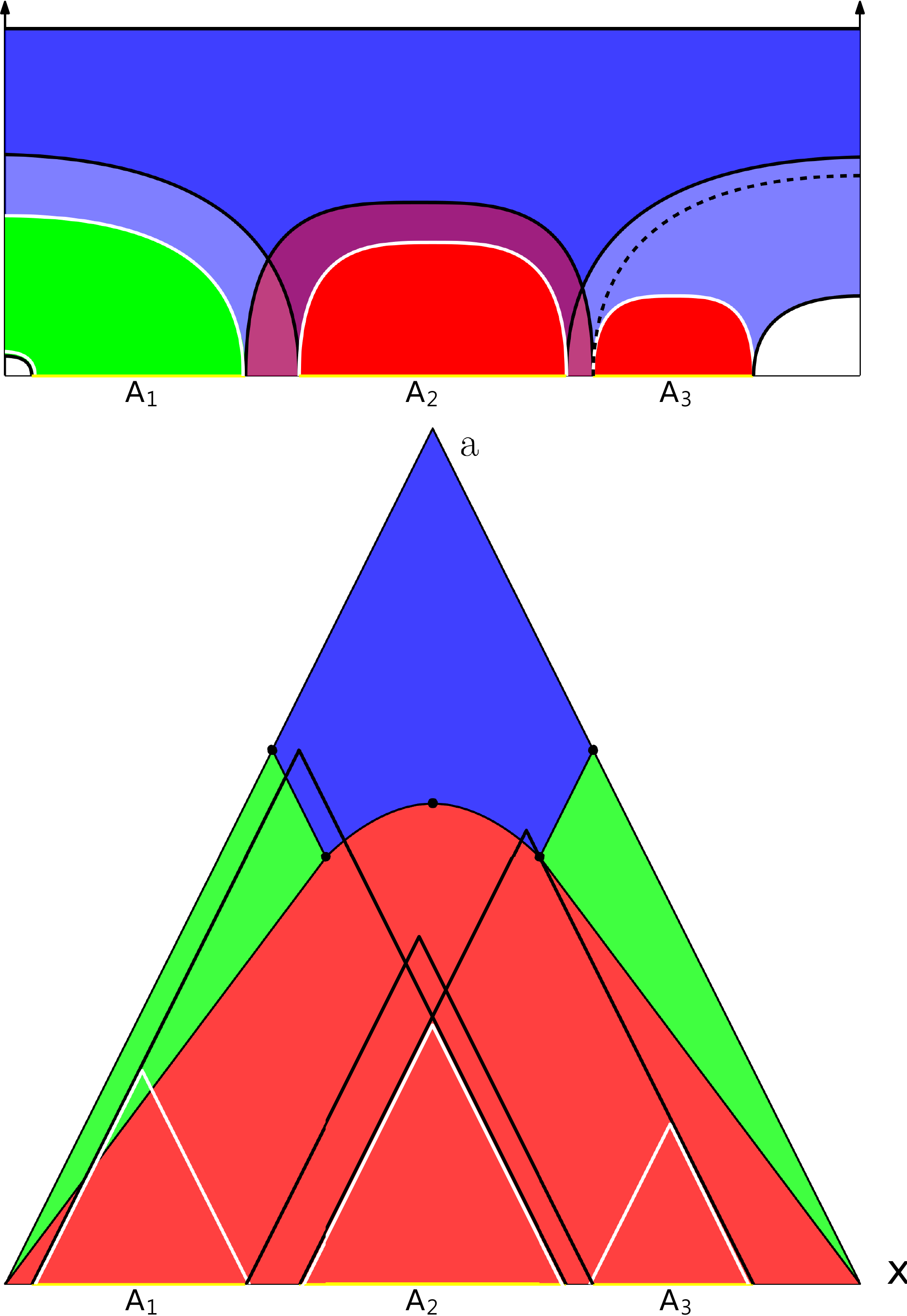} }\hspace{0.3cm}
  \subfloat[kkrr]{
    \includegraphics[  width=.15\linewidth]{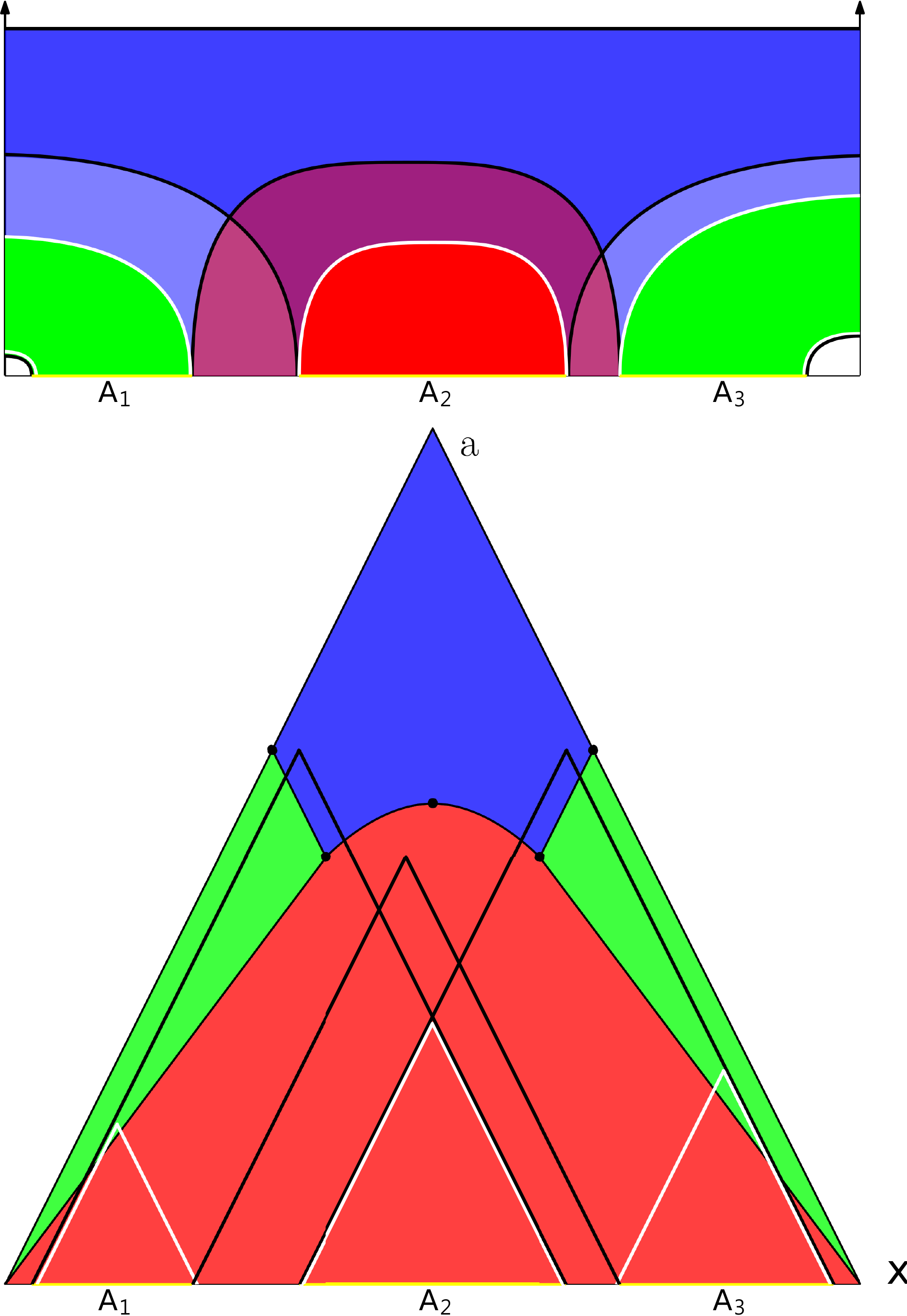} }
  \caption{The 25 cases for MMI without changes in the black hole background.}
  \label{fig_same}
\end{figure}

Among the 28 allowed cases, 25 of them correspond to the cases in the pure AdS background listed in TABLE \ref{MMItable}. These 25 cases are plotted in Fig.\ref{fig_same} and can be verified accordingly by the same arguments as the SSA.

\begin{figure}[h]
  \subfloat[rrrr]{
    \includegraphics[width=.25\linewidth]{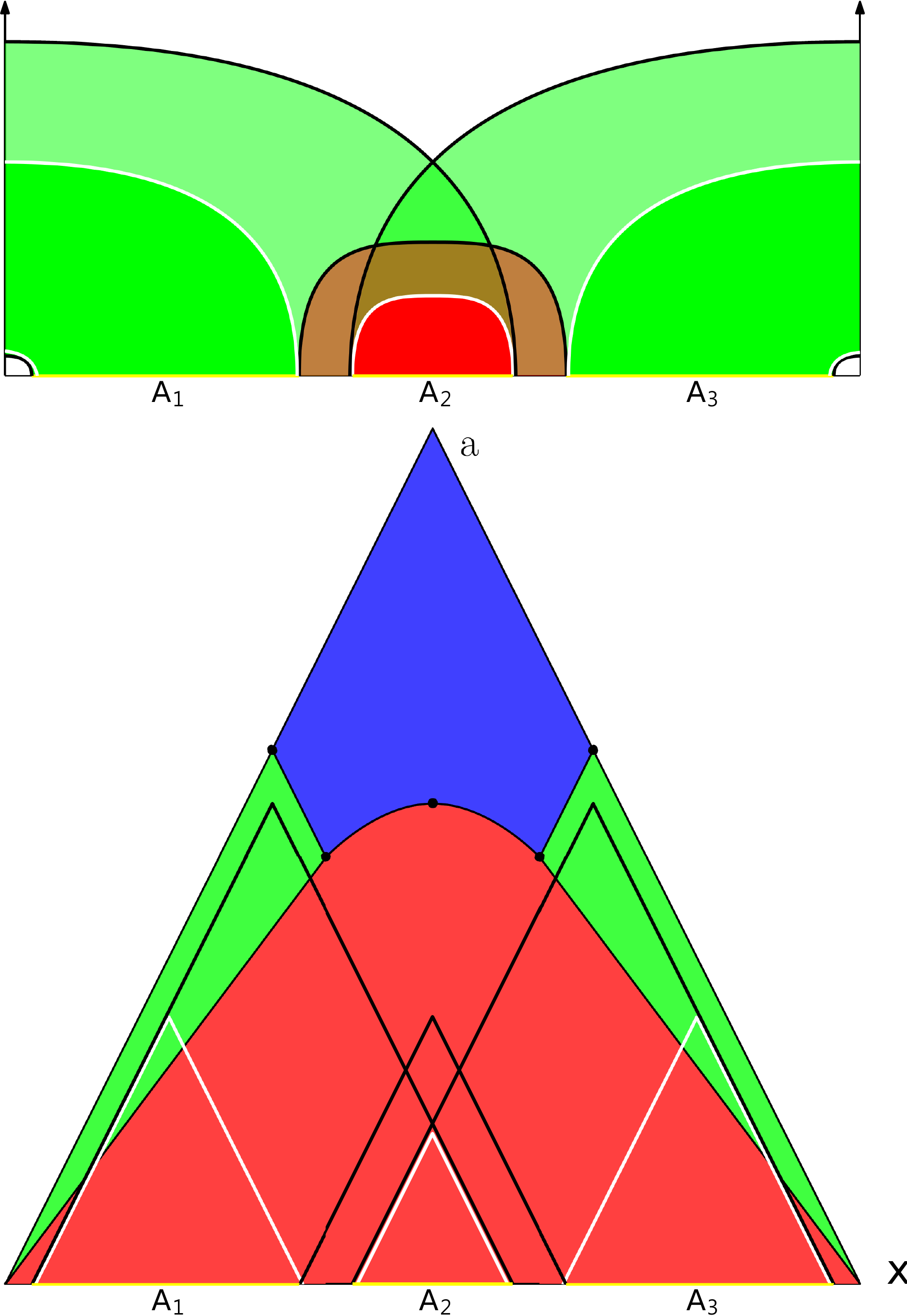} }\hspace{0.5cm}
  \subfloat[rrrs]{
    \includegraphics[width=.25\linewidth]{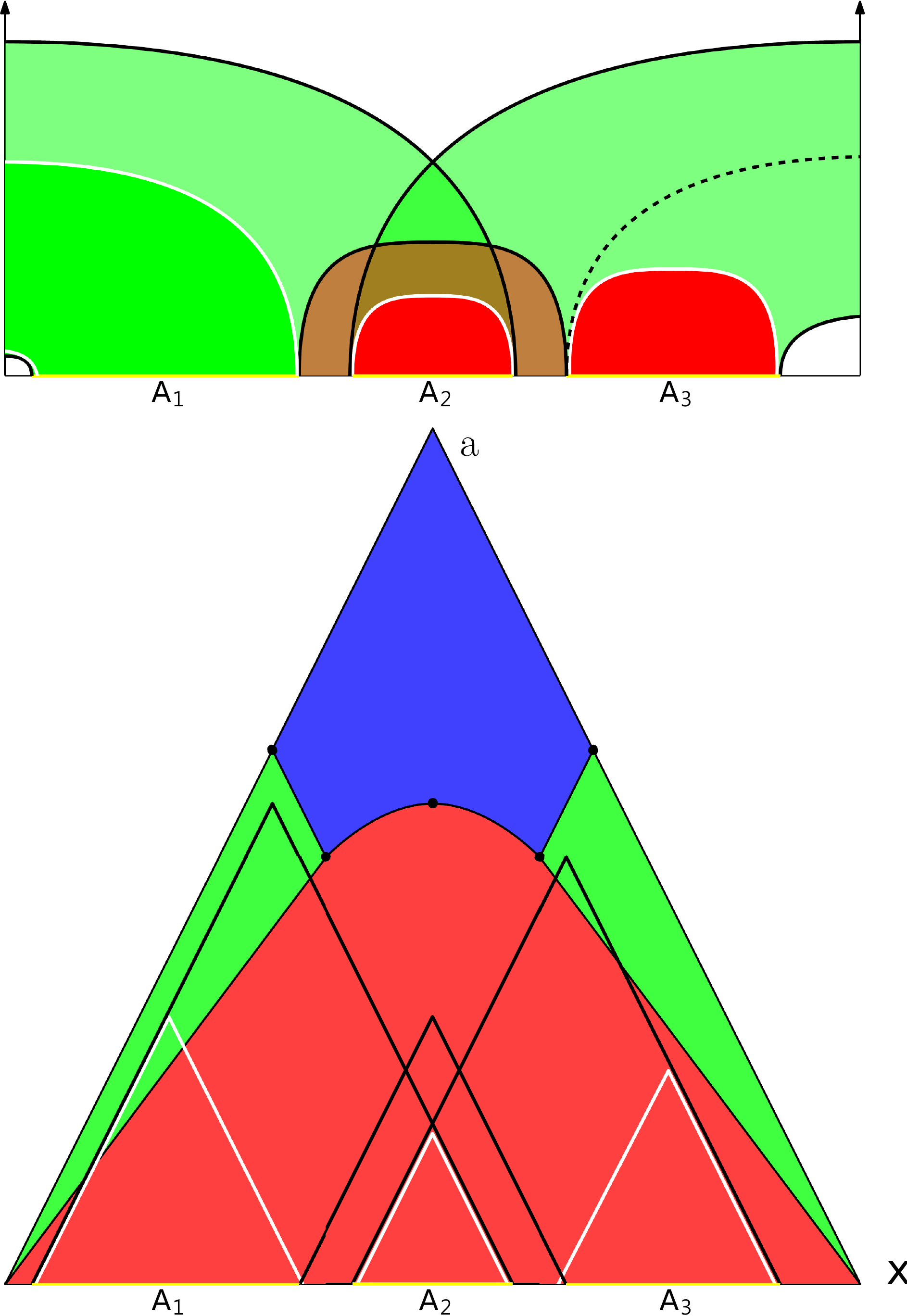} }\hspace{0.5cm}
  \subfloat[rrss]{
    \includegraphics[width=.25\linewidth]{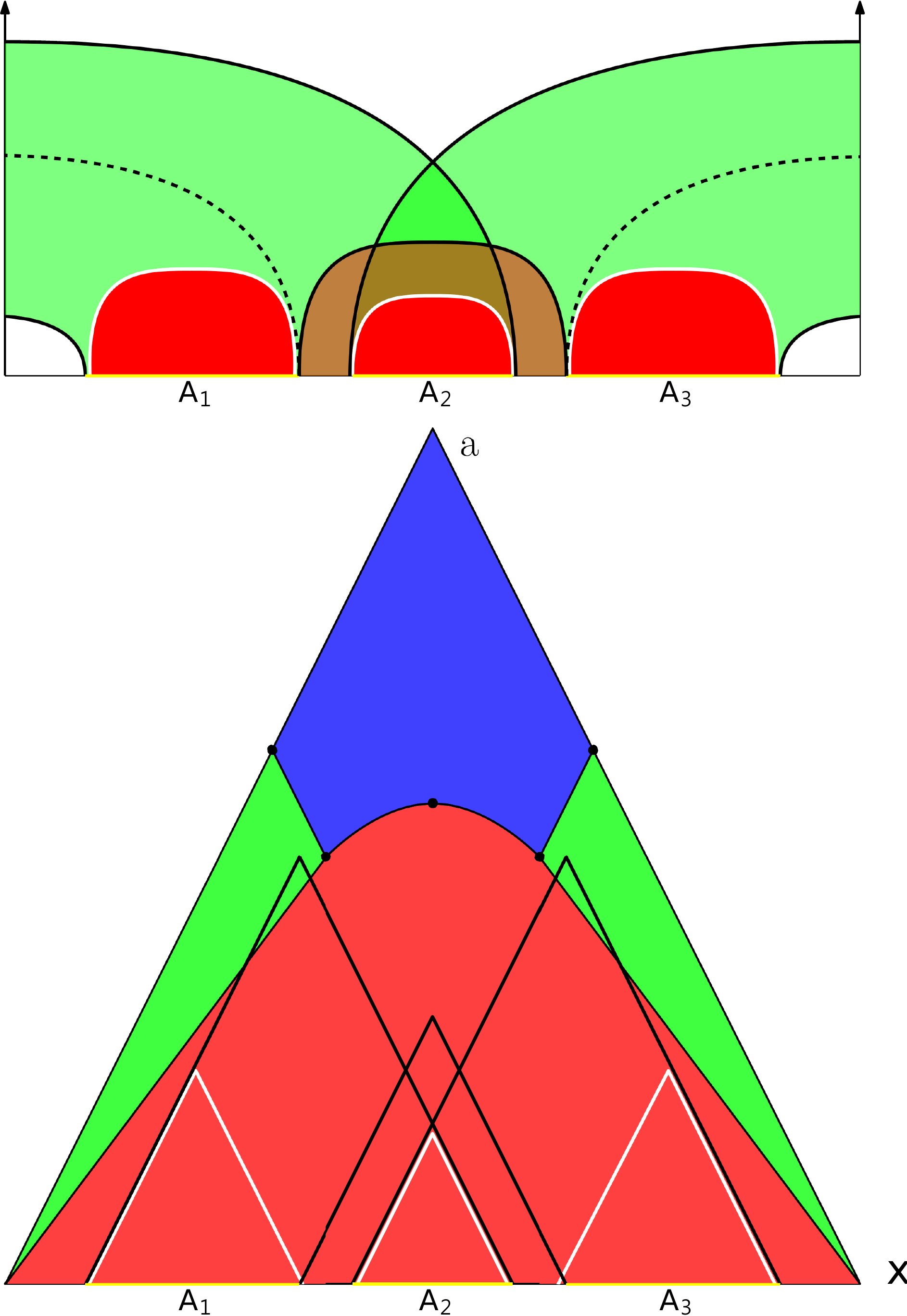} }
  \caption{The three additional diagrams of MMI in the black hole background.}
  \label{fig_extra}
\end{figure}

In addition, there are three new allowed cases in the black hole background as plotted in Fig.\ref{fig_extra}. In these three cases, there is no term involved the sky phase. However, because the phase diagram is modified in the black hole background, the HEE for the entangled region $A_{[23]}$ is allowed in the right part of the rainbow phase. We prove these MMI by cutting the three black curves at their intersections and rejoining them to three joint curves. The joint curve in homology with $A_2$ must be larger than its corresponding HEE's represented by the white curves. Moreover, it can be shown that the other two joint curves must be larger than the white quarter-circle-shaped curves anchored on the right/left boundary of the entangled region $A_1$/$A_3$ with or without the help of the auxiliary line (dashed black curve).

We therefore show that the sum of the black curves is larger than the sum of the white curves, which proves the MMI in the three new cases. Similar argument verifies the cases that not all of $S_{12}$, $S_{23}$, $S_{13 }$ and $S_{123  }$ are in the connected configurations.  Hence, we prove the MMI in the black hole background.

\section{Summary and Outlook}

In this paper, we studied the HEE inequalities in a (d+1) dimensional holographic BCFT at both zero and finite temperatures. We proved the strong subadditivity (SSA) and the monogamy of mutual information (MMI) in the tripartite systems.

In the presence of boundaries, HEE could be in the different phases with different shapes of the RT surfaces. We first considered the phase diagrams of HEE in the bipartite system. The phase diagrams provide the constraints for the configurations of the RT surfaces.  We derived the rules for the allowed configurations.

We then applied the rules to the tripartite system to classify all allowed configurations for the SSA and MMI. In each allowed configuration, we proved the SSA and MMI at both zero and finite temperatures by using the pure AdS and AdS black hole gravity duals, respectively.

In this work, we introduced the new quantities $ {A} _{\left\langle
ij\right\rangle }$ and $ {A} _{[
ij]}$ in Eqs.(\ref{small region}, \ref{large region}), which greatly simplify the classification of the RT surfaces. This encourages us to generalize the method to the holographic entropy cone in our future work.

Furthermore, the entanglement entropy can be used to characterize the quantum phase transition in condensed matter physics. Many important condensed matter systems have nontrivial effects on boundaries, such as quantum Hall effect, chiral  magnetic  effect,  topological  insulator. Thus, we are going to apply the inequalities of the entanglement entropy in BCFT to some condensed matter systems with boundary, and study their phase transitions at both zero and finite temperature near quantum critical points.

\section*{Acknowledgements}

We would like to thank Chong-Sun Chu and Rong-Xin  Miao for useful discussions. This work is supported by the Ministry of Science and Technology (MOST 109-2112-M-009-005), R.O.C.

\end{document}